\documentclass[a4paper,11pt]{article}
\pdfoutput=1 
\usepackage{jcappub} 
\usepackage[utf8]{inputenc}
\usepackage{ulem} 
\usepackage{rotating}
\usepackage{multirow}
\usepackage{array}
\newcolumntype{?}{!{\vrule width 1pt}}
\newcommand{\upchi}{\protect\raisebox{2.5pt}{$\chi$}}
\newcommand*{\blue}{\textcolor{blue}}

\DeclareUnicodeCharacter{2212}{\textendash}

\title{\boldmath Observational constraints and predictions of the interacting dark sector with field-fluid mapping}

\author{Joseph P Johnson,}
\author{Archana Sangwan}
\author{ and S. Shankaranarayanan}
\affiliation{Department of Physics, Indian Institute of Technology Bombay,\\ Mumbai 400076, India}

\emailAdd{josephpj@iitb.ac.in}
\emailAdd{arch06san@gmail.com}
\emailAdd{shanki@phy.iitb.ac.in}

\abstract{We consider an interacting field theory model that describes the dark energy - dark matter interaction. Only for a specific interaction term, this interacting field theory description has an equivalent interacting fluid description. For inverse power law potentials and linear interaction function, we show that the interacting dark sector model with field-fluid mapping is consistent with \textit{four cosmological data sets} --- Hubble parameter measurements (Hz), Baryonic Acoustic Oscillation data (BAO), Supernova Type Ia data (SN),  and High redshift HII galaxy measurements (HIIG). More specifically, these data sets prefer a negative value of interaction strength in the dark sector and lead to consistent best-fit values of Hubble constant and other cosmological parameters. Having established that this interacting field theory model is consistent with cosmological observations, we obtain quantifying tools to distinguish between the interacting and non-interacting dark sector scenarios. We focus on the variation of the scalar metric perturbed quantities as a function of redshift related to structure formation, weak gravitational lensing, and the integrated Sachs-Wolfe effect. We show that the difference in the evolution becomes significant for $z < 20$, for all length scales, and the difference peaks at smaller redshift values $z <  5$. We then discuss the implications of our results for the upcoming missions.}

\begin{document}
\maketitle
\flushbottom

\section{Introduction}

Cosmological observations suggest that the energy budget of the Universe is dominated by dark energy and dark matter~\cite{1998-Riess.Others-Astron.J.,1999-Perlmutter.Others-Astrophys.J.,2007-Spergel.Others-Astrophys.J.Suppl.,2018-Scolnic.Others-Astrophys.J.,2018-Akrami.Others-Astron.Astrophys.,2020-Aghanim.others-Astron.Astrophys.}. $\Lambda$CDM model provides the simplest description of the Universe dominated by dark energy and dark matter while being highly successful in describing various cosmological observations and phenomena like the cosmic microwave background (CMB) and nucleosynthesis~\cite{2000-Padmanabhan-TheoreticalAstrophysicsVolume,2005-Mukhanov-PhysicalFoundationsCosmology,2008-Weinberg-Cosmology,2011-Gorbunov.Rubakov-IntroductionTheoryEarly}. But with the availability of high precision cosmological observational data, there have been some inconsistencies in the values of cosmological parameters estimated using the $\Lambda$CDM model, with the most prominent of them being the difference in the value of the Hubble's constant estimated from the local distance measurements and CMB observations~\cite{2013-Marra.etal-Phys.Rev.Lett.,2013-Verde.etal-Phys.DarkUniv.,2014-Bennett.etal-Astrophys.J.,2016-Riess.others-Astrophys.J.,2019-Riess.etal-Astrophys.J.,2020-Aghanim.others-Astron.Astrophys.}. These inconsistencies point towards the limitations of the $\Lambda$CDM model and the need for modifications to the standard model of cosmology.

Apart from the gravitational interaction, we know very little about the properties of dark matter and dark energy. $\Lambda$CDM model assumes that dark energy is constant in time. The quintessence model provides a more general time-varying dark energy represented by a scalar field \cite{1988-Ratra.Peebles-Phys.Rev.D,2006-Copeland.etal-Int.J.Mod.Phys.}. A quintessence dark energy model can be further generalized by introducing a non-gravitational interaction between dark energy and dark matter, which is not ruled out by cosmological observations ~\cite{2000-Amendola-Mon.Not.Roy.Astron.Soc.,2000-Amendola-Phys.Rev.D,2000-Billyard.Coley-Phys.Rev.D,2005-Olivares.etal-Phys.Rev.D,2007-Amendola.etal-Phys.Rev.D,2008-Olivares.etal-Phys.Rev.D,2008-Boehmer.etal-Phys.Rev.D,2009-CalderaCabral.etal-Phys.Rev.D,2008-He.Wang-JCAP,2008-Pettorino.Baccigalupi-Phys.Rev.D,2008-Quartin.etal-JCAP,2010-Boehmer.etal-Phys.Rev.D,2011-Beyer.etal-Phys.Rev.D,2010-LopezHonorez.etal-Phys.Rev.D,2012-Avelino.Silva-Phys.Lett.B,2015-Pan.etal-Mon.Not.Roy.Astron.Soc.,2013-Salvatelli.etal-Phys.Rev.D,2013-Chimento.etal-Phys.Rev.D,2014-Amendola.etal-Phys.Rev.D,2016-Wang.etal-Rept.Prog.Phys.,2016-Marra-Phys.DarkUniv.,2017-Bernardi.Landim-Eur.Phys.J.C,2017-Pan.Sharov-Mon.Not.Roy.Astron.Soc.,2018-VanDeBruck.Mifsud-Phys.Rev.D,2018-CarrilloGonzalez.Trodden-Phys.Rev.D,2019-Barros.etal-JCAP,2019-Landim-Eur.Phys.J.C}. Recently, it has been shown that the dark matter-dark energy interaction can reconcile the tensions in the estimated values of
Hubble constant $H_0$ ~\cite{2017-DiValentino.etal-Phys.Rev.D,2017-Kumar.Nunes-Phys.Rev.D,2018-Yang.etal-Phys.Rev.Da,2018-Yang.etal-JCAP,2019-Pan.etal-Phys.Rev.D,2020-DiValentino.etal-Phys.Rev.D,2020-GomezValent.etal-Phys.Rev.D}. Hence it is important to develop the analytical and numerical tools to detect the interaction between dark energy and dark matter. For this purpose, we need a theoretical framework that provides a comprehensive description of the interacting dark sector.  

In Ref.\cite{2021-Johnson.Shankaranarayanan-Phys.Rev.D}, two of the current authors, have explicitly constructed such a framework starting from a classical field theory action that describes interacting dark sector. The authors showed that: (i) A one-to-one mapping between the field theory description and the fluid description of the interacting dark sector exists for a unique interaction term. (ii) This class of interacting dark sector models has an attractor solution describing the accelerated expansion of the Universe. The establishment of such a mapping enables us to analyze the background and perturbed evolution of the Universe with dark energy - dark matter interaction. 

To constrain the model parameters, especially the interaction strength, and to make testable predictions, one needs to specify the scalar field potential and the interaction function. In this work, we look at an inverse power law potential~\cite{2013-Pavlov.etal-Phys.Rev.D} $U(\phi) \sim 1/\phi^{n}$ where $(n = 1, 2)$ and a linear interaction function $\alpha(\phi) \sim C \phi$ where $C \in [-1, 1]$. We constrain the cosmological and model parameters
using Hubble parameter measurements (Hz) \cite{2017-Farooq.etal-Astrophys.J.,2005-Simon.etal-Phys.Rev.D,2010-Stern.etal-Astrophys.J.Suppl., 2012-Moresco.others-JCAP,2016-Moresco.etal-JCAP,2015-Moresco-Mon.Not.Roy.Astron.Soc.,2014-Zhang.etal-Res.Astron.Astrophys.,2017-Ratsimbazafy.etal-Mon.Not.Roy.Astron.Soc.}, high redshift HII Galaxy (HIIG) data \cite{2019-GonzalezMoran.etal-Mon.Not.Roy.Astron.Soc.,2014-Chavez.etal-Mon.Not.Roy.Astron.Soc.,2006-Erb.etal-Astrophys.J.,2014-Masters.others-Astrophys.J.,2014-Maseda.others-Astrophys.J.,2015-Terlevich.etal-Mon.Not.Roy.Astron.Soc.}, Baryon acoustic oscillation (BAO) data \cite{2017-Alam.others-Mon.Not.Roy.Astron.Soc., 2011-Beutler.etal-Mon.Not.Roy.Astron.Soc.,2015-Ross.etal-Mon.Not.Roy.Astron.Soc.,2018-Ata.others-Mon.Not.Roy.Astron.Soc.,2017-Bautista.others-Astron.Astrophys.} and the Type Ia supernovae (SN) observations \cite{2014-Betoule.others-Astron.Astrophys.}. The key results are: 
\begin{enumerate}
\item Although both negative and positive values of interaction strength $(C)$ are allowed, observations show a preference for negative interaction strength $(C < 0$). 
\item For our interacting dark sector model, the constraint on the Hubble constant from the combined data set is $H_0 = 69.79^{0.29}_{0.52}$ km s$^{-1}$Mpc$^{-1}$. This value lies between the value of Hubble constant reported by Planck is $H_0=67.5\pm0.5$ km s$^{-1}$Mpc$^{-1}$ which uses base $\Lambda$CDM cosmology~\cite{2020-Aghanim.others-Astron.Astrophys.} and the distance ladder estimates of Hubble constant is $H_0=73.48\pm 1.66$ km s$^{-1}$ Mpc$^{-1}$( from SH0ES data \cite{2018-Riess.others-Astrophys.J., 2016-Riess.others-Astrophys.J.}), and $H_0=74.03\pm1.42$ km s$^{-1}$Mpc$^{-1}$ ( measurements of LMC Cepheids \cite{2019-Riess.etal-Astrophys.J.}). We also see that the constraint obtained from the individual data sets are consistent with each other. 
\item The constraints on $\Omega_m$ obtained by the model are consistent with $\Omega_m=0.31\pm0.007 $ reported in \cite{2020-Aghanim.others-Astron.Astrophys.} (for latest constraints on the cosmological parameters see  \cite{2018-Ryan.etal-Mon.Not.Roy.Astron.Soc.,2018-Ooba.etal-Astrophys.J.,2020-Liu.etal-JCAP,2018-Zhang.Li-,2018-Sangwan.etal-,2020-Khadka.Ratra-Mon.Not.Roy.Astron.Soc.,2020-UrenaLopez.Roy-Phys.Rev.D,2020-Khadka.Ratra-Mon.Not.Roy.Astron.Soc.a,2021-Cao.etal-} and the references therein). 
\end{enumerate}

Our analysis shows that, with respect to the  low-redshift background observations, there is a \emph{strong degeneracy} between the interacting and non-interacting dark sector models.
To distinguish between the two scenarios, we need to go beyond the background evolution. In this work, we identify three specific tools that we can obtain by studying the difference in the evolution of cosmological perturbations in both of these scenarios~\cite{2005-Mukhanov-PhysicalFoundationsCosmology,2011-Gorbunov.Rubakov-IntroductionTheoryEarly}: Structure formation, Weak gravitational lensing, and Integrated Sachs-Wolfe effect. More specifically, we look at the evolution of the density perturbation ($\delta_m$), the Bardeen potential, and its derivative ($\Phi$ and $\Phi'$ respectively) for the inverse power law potential $U(\phi) \sim 1/\phi^{n}$ where $(n = 1, 2)$ and linear interaction function with negative interaction strength $(C < 0)$.  We evolve all the perturbed quantities in the redshift range $1500 \lesssim z < 0$. We see a significant difference in the evolution of the relevant perturbed quantities in the interacting and non-interacting scenarios, at all length scales, for $z < 20$. The maximum difference in the evolution is around $z \sim  5$. We thus explicitly show that it is possible to detect and constrain the interaction between dark energy and dark matter from cosmological observations.

In Sec.~\ref{sec:IDSmodel}, we introduce the interacting dark sector model we have used for the analysis. In Sec.~\ref{sec:bgevo} we discuss the background evolution in the model and the numerical analysis using various observational data sets to obtain the parameter constraints. The evolution of the cosmological perturbations and their observational consequences are discussed in Sec.~\ref{sec:pertevo}. In Sec.~\ref{sec:conclusion}, we briefly discuss the results and discuss the implications of our analysis. Appendices ~\ref{sec:wlambda} -~\ref{app:Sound} contain additional details.

In this work, we use the natural units where $m_{\rm Pl}^2 = G^{-1}$, and the metric signature (-,+,+,+). Greek letters denote the four-dimensional space-time coordinates, and Latin letters denote the three-dimensional spatial coordinates. Unless otherwise specified, \textit{dot} represents derivative with respect to cosmic time and \textit{prime} denotes derivative with respect to number of e-foldings $N \equiv \ln a(t)$.

\section{Interacting dark sector with field-fluid mapping: The model}
\label{sec:IDSmodel}

In this work, we consider the model described by the action \cite{2021-Johnson.Shankaranarayanan-Phys.Rev.D}, 
\begin{equation}
\label{eq:Scde}
S = \int d^4x \sqrt{-g}\left(\dfrac{1}{16 \pi G}R-\dfrac{1}{2}g^{\mu \nu}\nabla_{\mu}\phi \nabla_{\nu}\phi - U(\phi)-\dfrac{1}{2} e^{2\alpha(\phi)}g^{\mu \nu}\nabla_{\mu}\upchi \nabla_{\nu}\upchi -e^{4 \alpha(\phi)} V(\upchi) \right).
\end{equation}
where $\phi$ corresponds to the dark energy and $\upchi$ corresponds to the dark matter. 
The dark matter fluid in a homogeneous and isotropic Universe can be mapped to these scalar fields by defining the four velocity $u_{\mu}$
\begin{equation}
\label{eq:dm4v}
u_{\mu} = -\left[-g^{\alpha \beta} \nabla_{\alpha}\upchi \nabla_{\beta}\upchi \right]^{-\frac{1}{2}} \nabla_{\mu} \upchi \, ,
\end{equation}
the  energy density ($\rho_m$) and pressure ($p_m$) of the dark matter fluid
\begin{equation}
\label{eq:dmrhop}
p_{m}  = -\dfrac{1}{2}e^{2 \alpha}\left[g^{\mu \nu} \nabla_{\mu} \upchi \nabla_{\nu} \upchi + e^{2\alpha}V(\upchi) \right], \quad
\rho_{m}  = -\dfrac{1}{2}e^{2 \alpha}\left[g^{\mu \nu} \nabla_{\mu} \upchi \nabla_{\nu} \upchi - e^{2\alpha}V(\upchi) \right].
\end{equation} 
In this description, we can rewrite Einstein's equation in terms of dark energy scalar field and dark matter fluid:
\begin{equation}
\label{eq:flfeq}
 G_{\mu \nu} = 16 \pi G \left[\nabla_{\mu}\phi \nabla_{\nu}\phi - \dfrac{1}{2}g_{\mu \nu} \nabla^{\sigma}\phi \nabla_{\sigma}\phi -g_{\mu \nu} V(\phi)+ p_m g_{\mu \nu} + (\rho_m + p_m) u_{\mu} u_{\nu} \right] \, ,
\end{equation}
where the energy-momentum tensor for the dark matter fluid is given by
\begin{equation}
T^{(m)\mu}_{\nu}=p_m g_{\mu \nu} + (\rho_m + p_m) u_{\mu} u_{\nu} \, .
\end{equation}
The interaction between the dark energy and the dark matter fluid is described by:
\begin{equation}
\label{eq:interaction02}
\nabla_{\mu} T^{(m)\mu}_{\nu} = Q_{\nu}^{\rm (F)} \, ,
\end{equation}
where the interaction term is given by
\begin{equation}
\label{eq:interactionterm}
Q_{\nu}^{\rm (F)} =   -e^{2\alpha(\phi)} \alpha_{,\phi}(\phi) \nabla_{\nu} \phi \left[ \nabla^{\sigma} \upchi \nabla_{\sigma} \upchi + 4 e^{2\alpha(\phi)} V(\upchi) \right] =  -\alpha_{,\phi}(\phi) \nabla_{\nu}\phi (\rho_m  - 3 p_m) .
\end{equation}
Identifying $T^{(m)}=T^{(m) \mu}_{\mu} = -(\rho_m - 3 p_m ) $, we get
\begin{equation}
\label{eq:traceinter}
Q_{\nu}^{\rm (F)}  = T^{(m)} \nabla_{\nu}\alpha(\phi) \, .
\end{equation}
{The time component of $Q_{\nu}^{\rm (F)}$ represents the energy transfer between dark energy and dark matter. It is important to know that the field-fluid mapping in Ref.~\cite{2021-Johnson.Shankaranarayanan-Phys.Rev.D} is valid \emph{only} for the above form of $Q_{\nu}^{\rm (F)}$.  For easy reading, we denote  $Q_{0}^{\rm (F)}$ as $Q$.  $Q$ will be split into the background and perturbed parts given by $Q = \overline{Q} + \delta Q$}.

To study the cosmological evolution and obtain predictions and constraints, we need to consider a specific form of scalar field potential $U(\phi)$ and the interaction function $\alpha(\phi)$. In this work, we focus on the 
quintessence dark energy model with an inverse power law potential~\cite{2013-Pavlov.etal-Phys.Rev.D} and a linear interaction function 
\begin{equation}
\quad U(\phi) \sim \dfrac{1}{\phi^n}, \quad \alpha(\phi) \sim \phi \, ,
\end{equation}
where $n=1, 2$.  
The inverse power-law potential provides a
self-consistent phenomenological description of DE whose
density decreases as the Universe expands, but decreases
less rapidly than the nonrelativistic (cold dark, and baryonic) matter density in a spatially flat universe~\cite{2013-Pavlov.etal-Phys.Rev.D}. The above interaction function is the simplest form for obtaining the interacting dark sector considered in this work from a field theory action.

\section{Background evolution and observational constraints}
\label{sec:bgevo}

We consider a spatially flat universe governed by Friedmann equations
\begin{equation*}
    \left(\dfrac{\dot{a}}{a}\right)^2=\dfrac{8\pi G}{3}\rho_{tot}, \quad \dfrac{\ddot{a}}{a}=-\dfrac{4\pi G}{3}(\rho_{tot}+3P_{tot})
\end{equation*}
where $\rho_{tot}$ and $P_{tot}$ denote the total energy density and pressure of the universe at a given time. At late times, the contribution of the relativistic matter density ($\rho_r$) is negligible as compared to the dark (non-relativistic) matter ($\rho_m$) and dark energy density ($\rho_{\phi}$). Hence, for the analysis in this section, we neglect $\rho_r$ and total density is $\rho_{tot}=\rho_m+\rho_{\phi}$.

The dynamics of the scalar field is governed by
\begin{equation*}
    (\ddot{\phi}+3H\dot{\phi}+U_{,\phi})\dot{\phi}=\overline{Q},
\end{equation*}
where, $\overline{Q}$ is the background interaction term.  Here,  $\phi$ is in the units of $m_{\rm Pl}=G^{-1/2}$.  The scalar field potential is assumed to be 
\begin{equation}
\label{eq:Potentialdef}
U(\tilde{\phi})=\dfrac{\kappa}{2} \, m_{\rm Pl}^2\tilde{\phi}^{-n} \, 
\end{equation}
where $\kappa$ is of the order of unity.  
To make the analysis simpler, we rescale the scalar field $\phi$ to $\tilde{\phi} = \sqrt{16 \pi G} \, 
\phi $. Note that $\tilde{\phi}$ is dimensionless. 

The evolution of non-relativistic matter density is given by
\begin{equation*}
\dot{\rho}_m+3H\rho_m= -\overline{Q},
\end{equation*}
where we have considered a pressureless dark matter fluid, $p_m=0$.
For the interaction term, $\overline{Q}=-\alpha_{,\phi}\dot{\phi}\rho_m$, the above equation gives,  $\rho_m=\rho_{m_0}e^{\alpha(\phi)-\alpha(\phi_0)}a^{-3}$, which we use for the analysis in this section\footnote{The constant factor $e^{-\alpha(\phi_0)}$ can be absorbed in $\rho_{m_0}$}.

In terms of dimensionless scalar field variable ($\tilde{\phi}$), the Friedmann equations and the field equation are:
\begin{eqnarray}
\label{coup_pr1}
\left(\dfrac{\dot{a}}{a}\right)^2 = H_0^2 \Omega_m a^{-3} e^{C(\tilde{\phi}-\tilde{\phi}_0)} + \dfrac{\dot{\tilde{\phi}}^2}{12} + \frac{\kappa m_{Pl}^2}{12} \tilde{\phi}^{-n}\\   \label{coup_pr2}
\ddot{\tilde{\phi}} + 3H\dot{\tilde{\phi}} + U_{,\tilde{\phi}}(\tilde{\phi})= -6 H_0^2 C \Omega_m a^{-3}e^{C(\tilde{\phi}-\tilde{\phi}_0)},
\end{eqnarray}
where we have assumed $\alpha(\tilde{\phi})$ to be a linear function of $\tilde{\phi}$, i. e. $\alpha(\tilde{\phi}) = C \, \tilde{\phi}$, giving $\alpha_{,\tilde{\phi}}=C$. The parameter
$C$ is dimensionless and defines the strength of interaction between dark energy and dark matter. In our analysis, we obtain the constraint on $C$ by keeping it as a free parameter with  $C \in [-1,1]$.

\subsection{Observational data}
\label{sec::observations}

We analyze four different observational data sets to constrain the model parameters in the interacting dark sector model. More specifically, we use  Hubble parameter measurements (Hz) \cite{2005-Simon.etal-Phys.Rev.D,2010-Stern.etal-jcap,2009-Gaztanaga.etal-mnras,2012-Moresco.others-JCAP,2014-Zhang.etal-Res.Astron.Astrophys.,2017-Farooq.etal-Astrophys.J.}, high redshift HII Galaxy (HIIG) data \cite{2019-GonzalezMoran.etal-Mon.Not.Roy.Astron.Soc.,2014-Chavez.etal-Mon.Not.Roy.Astron.Soc.,2006-Erb.etal-Astrophys.J.,2014-Masters.others-Astrophys.J.,2014-Maseda.others-Astrophys.J.,2015-Terlevich.etal-Mon.Not.Roy.Astron.Soc.}, Baryon acoustic oscillation (BAO) data \cite{2017-Alam.others-Mon.Not.Roy.Astron.Soc.,2011-Beutler.etal-Mon.Not.Roy.Astron.Soc.,2015-Ross.etal-Mon.Not.Roy.Astron.Soc.,2018-Ata.others-Mon.Not.Roy.Astron.Soc.,2017-Bautista.others-Astron.Astrophys.} and the joint lightcurve analysis (JLA) sample of Type Ia supernovae (SN) observations \cite{2014-Betoule.others-Astron.Astrophys.,2018-Sako.others-Publ.Astron.Soc.Pac.,2011-Conley.others-Astrophys.J.Suppl.,2018-Balland.others-Astron.Astrophys.,2010-Guy.others-Astron.Astrophys.}.

\noindent {\bf Hubble Parameter Measurement (H(z)) data:}
The Hubble parameter measurements (abbreviated as Hz) at different redshifts is an
effective tool to constrain the cosmological parameters \cite{2012-Moresco.others-JCAP, 2017-Farooq.etal-Astrophys.J.}.  Hz observations are useful in constraining the cosmological parameters as it uses the model parameters directly without having an integral term that might obscure or cover valuable information. In the literature, two different techniques are employed to measure the Hubble parameter: a) Differential age method \cite{2010-Stern.etal-jcap} and b) Radial BAO method \cite{2009-Gaztanaga.etal-mnras}. In this work, we use the \emph{differential age method}, where the Hubble rate as a function of redshift is evaluated by using the expression:
\begin{equation}
    H(z) = -\frac{1}{\left(1 + z\right)}\frac{dz}{dt},
\end{equation}
where $t$ denotes the age of the Universe when the observable photon is emitted.  In the differential method,  we can obtain a direct estimate of the expansion rate by taking the derivative of redshift with respect to time. Hubble parameter obtained through this method does not depend on the cosmological model but on the age-redshift relation of cosmic chronometers.
So very carefully, the selection of passively evolving early galaxies as cosmic chronometers is made depending upon a galaxy's star formation history and its metallicity.

In this work, we consider the Hz data points obtained through the cosmic chronometric technique and use the data points compiled in Ref.~\cite{2017-Farooq.etal-Astrophys.J.}.  In this compilation, 
the authors dropped older Hubble parameter estimates from SDSS galaxy clustering \cite{2013-Chuang.Wang-Mon.Not.Roy.Astron.Soc.} and Ly-$\alpha$ forest measurement \cite{2013-Busca.others-Astron.Astrophys.} and added new data sets. Out of the $38$ data points reported in Ref.~\cite{2017-Farooq.etal-Astrophys.J.}, in this analysis, we only use $31$ independent measurements of the Hubble parameter (H(z)). More specifically, we use 9 data points from Ref.~\cite{2005-Simon.etal-Phys.Rev.D}, 2 points from Ref.~\cite{2010-Stern.etal-Astrophys.J.Suppl.}, 8 points from Ref.~\cite{2012-Moresco.others-JCAP}, 5 points from Ref.~\cite{2016-Moresco.etal-JCAP}, 2 points from Ref.~\cite{2015-Moresco-Mon.Not.Roy.Astron.Soc.}, 4 points from Ref.~\cite{2014-Zhang.etal-Res.Astron.Astrophys.},  and one point from Ref.~\cite{2017-Ratsimbazafy.etal-Mon.Not.Roy.Astron.Soc.}. Note that the three points reported in Ref.~\cite{2012-Blake.others-Mon.Not.Roy.Astron.Soc.} and another three points in Ref.~\cite{2017-Alam.others-Mon.Not.Roy.Astron.Soc.} are also used in the BAO observations, hence removed from this data set. 

\noindent {\bf BAO:} Baryon Acoustic Oscillations (BAO) are fluctuations in the correlation function of large-scale structures that appear as overdense regions in the distribution of the visible, baryonic matter. This is the consequence of acoustic waves set up in the primordial plasma because of competing forces of radiation pressure and gravity.
These acoustic waves travel within the plasma. However, they are frozen at the time of recombination when the plasma cooled down enough to make the cosmos neutral. The distances where the waves stall are imprinted as overdense regions and are used as a standard ruler to measure cosmological distances. 

The characteristic angular scale of the acoustic peak is given in terms of sound horizon at drag epoch, $r_s(z_d)$, as $\theta_A=r_s(z_d)/D_V(z)$, 
where $D_V$ is the effective distance ratio given in terms of angular diameter distance $D_A$:
\begin{equation}
    D_V(z) = \left[(1+z)^2D_A(z)^2\frac{cz}{H(z)}\right]^{1/3} \, , \qquad
    r_s(z_d) = \int^{\infty}_{z_d} \frac{c_s(z')dz'}{H(z')}.\\
\end{equation}
In order to use the BAO data, the knowledge of the sound horizon scale at the $z_d$ (denoted by $r_s$) is required as the data is given in terms of {$H(z) \, r_s/r_{s,fid}$, $D_M \, r_{s,fid}/r_s$, $D_V \, r_{s,fid}/r_s$}, where $r_{s,fid}$ is 147.78 Mpc in \cite{2017-Alam.others-Mon.Not.Roy.Astron.Soc.} and \cite{2018-Ata.others-Mon.Not.Roy.Astron.Soc.}, and 148.69 Mpc \cite{2015-Ross.etal-Mon.Not.Roy.Astron.Soc.} and the comoving angular diameter distance is given by 
\begin{equation}
D_M(z)=(1+z)D_A(z).
\end{equation}
The value of fiducial sound horizon, $r_{s,fid}$ which was calculated by assuming the $\Lambda$CDM model and the best fit values of parameters given by Planck-2018 \cite{2020-Aghanim.others-Astron.Astrophys.}, is model dependent, but not to a significant degree. The quantities $D_V \, r_{s,fid}/r_s$, $D_M \, r_{s,fid}/r_s$, $r_s$ and $r_{s,fid}$ is given in units of $Mpc$ while $H(z) \, r_s/r_{s,fid}$ is given in units of km s$^{-1}$Mpc$^{-1}$. We compute $r_s$ using the \emph{inverse distance ladder
method} given in Ref.~\cite{2015-Aubourg.others-Phys.Rev.D}.
In Sec. (IV), we have studied the perturbation evolution in the dark sector interacting model, and in the range $1500 < z < 20$, the evolution is nearly identical to the non-interacting case.  
Hence, we can use the inverse distance ladder to measure distances, and the Hubble parameter at the corresponding redshifts will be approximately the same for the interacting dark sector model.
The BAO data in terms of Acoustic parameter $A(z)$ is defined as~\cite{2005-Eisenstein.others-Astrophys.J.}:
\begin{equation}
 A(z)= \Big[\frac{100 D_V(z)\sqrt{(\Omega_{m}h^2)}}{cz}\Big]^{1/3}.
\end{equation} 
Thus, the BAO data consists of $A(z)$ and $D_V(z)$ (with associated errors) at different redshifts.  The measurement of these distances is a useful tool to constrain cosmological model parameters. The BAO data we use in the analysis lie in the redshift span of $0.106-2.36$ and contains 11 points reported in Refs.~\cite{2017-Alam.others-Mon.Not.Roy.Astron.Soc., 2011-Beutler.etal-Mon.Not.Roy.Astron.Soc.,2015-Ross.etal-Mon.Not.Roy.Astron.Soc.,2018-Ata.others-Mon.Not.Roy.Astron.Soc.,2017-Bautista.others-Astron.Astrophys.}. 
Among the data we use in the analysis, data points from BOSS DR12~\cite{2017-Alam.others-Mon.Not.Roy.Astron.Soc.} are correlated, and the rest of the data points are uncorrelated. In this work, we assume that the different data sets are independent of each other.

\noindent {\bf HIIG:} The third data set we use is the high redshift HII galaxy (HIIG) observations~\cite{2019-GonzalezMoran.etal-Mon.Not.Roy.Astron.Soc.,2014-Chavez.etal-Mon.Not.Roy.Astron.Soc.,2006-Erb.etal-Astrophys.J.,2014-Masters.others-Astrophys.J.,2014-Maseda.others-Astrophys.J.,2015-Terlevich.etal-Mon.Not.Roy.Astron.Soc.}.
These observations are new independent cosmological observations that use the correlation between the Balmer emission line velocity dispersion ($\sigma$) and luminosity ($L$) in HIIG to obtain the distance estimator. 
This $L$-$\sigma$ correlation is given by:
\begin{equation}
\label{eq::correlation_hiig}
    \log (L)=\beta \log(\sigma) + \gamma \, ,
\end{equation}
where, $\gamma$ and $\beta$ are the intercept and slope, respectively and $\log=\log_{10}$. The tight correlation between the Balmer line luminosity ($L$) and velocity dispersion ($\sigma$) of the emission lines can be used to constrain the cosmological model parameters.

{An extinction correction must be made to the observed fluxes to obtain the values of these parameters. We follow the method used in Ref. \cite{2019-GonzalezMoran.etal-Mon.Not.Roy.Astron.Soc.} and assume the extinction law given in Ref.~\cite{2003-Gordon.etal-TheAstrophysicalJournal}. The resulting value of the intercept and slope are:}
\begin{eqnarray}
    \label{eq:Gordon_beta}
    \beta &=& 5.022 \pm 0.058, \\
    \label{eq:Gordon_gamma}
    \gamma &=& 33.268 \pm 0.083,
\end{eqnarray}
respectively. In our analysis, we use these values of $\beta$ and $\gamma$. The $\beta$ and $\gamma$ values are obtained by fitting only the `local sample’, i.e. 36 Giant Extragalactic HII Regions for which the authors have distance estimates from Cepheids, and 107 HII galaxies with z $\leq$ 0.15. Together the two samples were used to calibrate the L-$\sigma$ relation and the value of H0 (cf. Ref.~\cite{2012-Chavez.etal-Mon.Not.Roy.Astron.Soc.,2018-FernandezArenas.etal-Mon.Not.Roy.Astron.Soc.}). Using these values in Eq.\eqref{eq::correlation_hiig}, we obtain the luminosity of a HII Galaxy. We then use the luminosity to obtain the distance modulus for that HII Galaxy:
\begin{equation}
\label{eq::mu_obs_hiig}
    \mu_{\rm obs} = 2.5\log L - 2.5\log f - 100.2,
\end{equation}
where $f$ denotes the measured flux of the HIIG, reported in the HIIG observational data along with the error associated with it. We can predict the distance modulus for a given cosmological model by using the theoretical definition:
\begin{equation}
\label{eq::mu_theo}
    \mu_{\rm th}\left( z\right) = 5\log D_{\rm L}\left( z\right) + 25,
\end{equation}
where the luminosity distance $D_L(z)$ (in the units of Mpc) is  related to the angular size distance $D_A(z)$ via distance duality relation and the transverse comoving distance $D_M(z)$ through $D_L(z)=(1+z)^2D_A(z)=(1+z)D_M(z)$. The HIIG data we use comprises 153 measurements that span the redshift range of 0.0088 to 2.429, covering a larger redshift range than the BAO data used in this analysis.

\noindent {\bf SN data (JLA):}
Type Ia supernovae, which are standardizable candles, is another useful tool to determine the expansion history of the Universe~\cite{2014-Betoule.others-Astron.Astrophys.,2018-Sako.others-Publ.Astron.Soc.Pac.,2011-Conley.others-Astrophys.J.Suppl.,2018-Balland.others-Astron.Astrophys.,2010-Guy.others-Astron.Astrophys.}.  The observable reported in the sample is the distance modulus, which is extracted from light curves by assuming that the intrinsic luminosity on average is the same for Type Ia supernovae with the identical color, shape, and environment, irrespective of the redshift measurement.  The standardized distance modulus, obtained by using the following linear empirical relation:
\begin{equation}
\label{mu_obs_sn}
\mu^{obs} = m_B^{*}+ \alpha x_1 -\beta C - M_B.
\end{equation}
Here, $m_B^*$ is the peak magnitude observed in the B-band rest frame, $\alpha$ and $\beta$ are nuisance parameters, C is the color of supernovae at peak brightness, and $x_1$ is `stretch' of the light curve. The values of the parameters $(m_B, x_1, C)$ are obtained by fitting supernovae spectral sequence to the photometric data.
The parameter $M_B$, which is the absolute B-band magnitude, depends on the host stellar mass.
The theoretical value of the distance modulus $\mu_{th}$ is given by  Eq.~\eqref{eq::mu_theo}, which depends on the cosmological model.

By measuring the apparent brightness and comparing it to other candles, one can estimate the distance the photons have traveled, and hence the rate of expansion of the Universe. Our analysis uses the full joint lightcurve analysis (JLA) sample comprising $740$ Type Ia Supernovae spanning a redshift range of z=0.01 to z= 1.4. We use the abbreviation `SN' {to denote} these 740 sample points.

\subsection{Data analysis technique}
\label{sec::technique}

For our analysis, we use the $\chi^{2}$ minimization technique. 
Any measurement data contains an observable quantity $X_{\rm obs}(z_{i})$ and its corresponding redshift $z_i$, along with the error associated with each point $\sigma_{i}$. Here, `$i$' takes the values up to N (number of data points in each observation).
We can also estimate these observable quantities theoretically [$X_{\rm th}(z_i)$] for the models considered in the analysis.

For $H(z)$ data,  the observable is the expansion rate, and we consider 31 points obtained using cosmic chronometer, and the $\chi^2$ is defined as:
\begin{equation}
\label{eq:chi2_Hz}
    \chi^2_{\rm H}(\textbf{p}) = \sum^{31}_{i = 1} \frac{[H_{\rm th}(\textbf{p}, z_i) - H_{\rm obs}(z_i)]^2}{\sigma_i^2},
\end{equation}
where, $\sigma_i$ is the uncertainty of $H_{\rm obs}(z_i)$. All these 31 points are independent of each other, and the expansion rate depends on the specific model chosen represented by 
`$\textbf{p}$' in the above expression.

For the BAO data points that are correlated (BOSS DR12), $\chi^2_{\rm BAO}$ is given by
\begin{equation}
\label{eq:chi2_BAO2}
    \chi^2_{\rm BAO}(\textbf{p}) = [X_{\rm th}(\textbf{p}) - X_{\rm obs}(z_i)]^T\textbf{C}^{-1}[X_{\rm th}(\textbf{p}) - X_{\rm obs}(z_i)],
\end{equation}
where superscripts $T$ and $-1$ denote the transpose and inverse of the matrices, respectively. For the data, we use the covariance matrix $\textbf{C}$ from Ref.~\cite{2017-Alam.others-Mon.Not.Roy.Astron.Soc.}.

For HIIG data consisting of 153 measurements, the $\chi^2$ is given by
\begin{equation}
\label{eq:chi2_HIIG}
    \chi^2_{\rm HIIG}(\textbf{p}) = \sum^{153}_{i = 1} \frac{[\mu_{\rm th}(\textbf{p}, z_i) - \mu_{\rm obs}(z_i)]^2}{\sigma_i^{\prime2}},
\end{equation}
where $\sigma_i^{\prime}$ is the uncertainty of the $i_{\rm th}$ measurement ({not to be confused} with the velocity dispersion ($\sigma$) term in HIIG measurements) and is given by
\begin{equation}
\label{eq:err_HIIG}
    \sigma^{\prime}=\sqrt{\sigma^{\prime2}_{\rm stat}+\sigma^{\prime2}_{\rm sys}} \, .
    \end{equation}
 $\sigma^{\prime}_{stat}$ is the statistical uncertainties and is given by:
\begin{equation}
\label{eq:stat_err_HIIG}
    \sigma^{{\prime}^2}_{\rm stat}=6.25\left[\sigma^{\prime2}_{\log f}+\beta^2\sigma^{\prime2}_{\log\sigma}+\sigma^{\prime2}_{\beta}(\log\sigma)^2+\sigma^{\prime2}_{\gamma}\right]+\left(\frac{\partial{\mu_{\rm th}}}{\partial{z}}\right)^2\sigma^{\prime2}_{z}.
\end{equation}
Due to the distance modulus term in the expression, the statistical uncertainty calculated this way is model-dependent. However,  when it comes to constraining the cosmological parameters, the model dependence is negligible~\cite{2020-Cao.etal-Mon.Not.Roy.Astron.Soc.}. In this analysis, we account for the reported systematic uncertainties i. e., $\sigma^2 = \sigma_{\rm stat}^2+\sigma_{sys}^2$ \footnote{In Ref.~\cite{2016-Chavez.etal-Mon.Not.Roy.Astron.Soc.}, the authors presented in greater detail a systematic error of $\sim$0.25, taking into account the uncertainties introduced from the size and age of the burst, abundances, and extinction [See also, Ref.~\cite{2014-Chavez.etal-Mon.Not.Roy.Astron.Soc.}].}. 

For the SN data with 740 joint light curves sample, the $\chi^2$ function is:
\begin{equation}
\label{eq:chi2_HIIGa}
    \chi^2_{\rm SN}(\textbf{p}) = \sum^{740}_{i,j = 1} [\mu_{\rm th}(\textbf{p}, z_i) - \mu_{\rm o}(z_i)] C^{-1}_{ij}[\mu_{\rm th}(\textbf{p}, z_j) - \mu_{\rm o}(z_j)]
\end{equation}
where $C_{ij}$ is the covariance matrix given in Ref.~\cite{2014-Betoule.others-Astron.Astrophys.}
.

For the joint analysis (Hz+BAO+HIIG+SN), we obtain the joint likelihood ($e^{-\chi^2}$) by multiplying individual likelihoods such that $\chi^2 = \chi^2_{H}+\chi^2_{BAO}+\chi^2_{HIIG}+\chi^2_{SN}$. Here, the maximum likelihood corresponds to the minimum value of $\chi^2$. 

\subsection{Parameter constraints} \label{sec::results}

Having discussed the data sets and the technique, we can now obtain parameter constraints for the interacting dark sector model discussed at the starting of this section.  More specifically,  we use the $\chi^2$ technique \footnote{For calculating $\chi^2$, the \blue{Metropolis-Hastings algorithm~\cite{1953-Metropolis.etal-J.Chem.Phys.,1970-Hastings-Biometrika,2005-Neal-arXivMathematicseprints}} is used to sample the parameter space, and we have modified the \blue{MCMC module by Benjamin Audren~\cite{2013-Audren.etal-JCAP}} to constrain the parameters. The convergence of Metropolis-Hastings runs depends on the value of the \blue{
 statistics ($R$)~\cite{1992-Gelman.Rubin-Statist.Sci.}}.} (described in Sec.~\ref{sec::technique}) to obtain the 1$\sigma$, 2$\sigma$, and 3$\sigma$ confidence regions corresponding to the four data sets for various cosmological parameters used in our dark energy- dark matter interaction model.  For a given value of $n$ in the scalar-field potential \eqref{eq:Potentialdef},  we obtain the constraints on the standard model parameters $H_0$, $\Omega_m$, $w_0$ and the parameter $C$, which describes the interaction strength in the dark sector. 

For the parameter fitting, we use priors that are consistent with the different constraints obtained from various observations. For the Hubble constant, we take the range to be $H_0 = 60-80~\rm{km}\, \rm{s}^{-1}~\rm{Mpc}^{-1}$. In Ref.~\cite{2021-Johnson.Shankaranarayanan-Phys.Rev.D}, two of the current authors studied the background evolution of the model for a range of initial conditions and showed that the accelerated attractor solution admitted by the model corresponds to $w = -1$. Thus, in the redshift range $1500 \geq z \geq 0$, $w \geq -1$ satisfy the dominant energy condition~\cite{2003-Carroll.etal-Phys.Rev.D}. It is important to note that $w<-1$ will lead to the destruction of all the structures ~\cite{2019-Bouali.etal-Phys.DarkUniv.}. Hence, we use the range $-1 < w_{\phi} < 1$ and the present-day value of the dark-energy equation of state parameter is set between $-1 \leq w_0 \leq 1$. The non-relativistic matter density is taken to be in the range  $0.01 \leq \Omega_m \leq 0.6$, and the interaction strength between dark matter-dark energy is taken to between $-1 \leq C \leq 1$. These priors are listed in Table~\ref{table::priors}.
\begin{table}[!htbp]
\centering
\begin{minipage}{80mm}
\begin{tabular}{|c|c|c|} \hline
Parameter & Lower Limit & Upper Limit\\ \hline
$H_0$&60.0 &80.0\\ \hline
$\Omega_{m}$&0.1 & 0.6\\ \hline
$w_0$& -1.0 & 1.0\\ \hline
$C$&-1.0& 1.0\\ \hline
\end{tabular}
\end{minipage}
\caption{Priors used in the analysis of parameter fitting.}
\label{table::priors}
\end{table}

\begin{figure}[!htbp]
\vspace*{-1.05cm}
\centering
\begin{tabular}{ccc}
\includegraphics[scale=0.29]{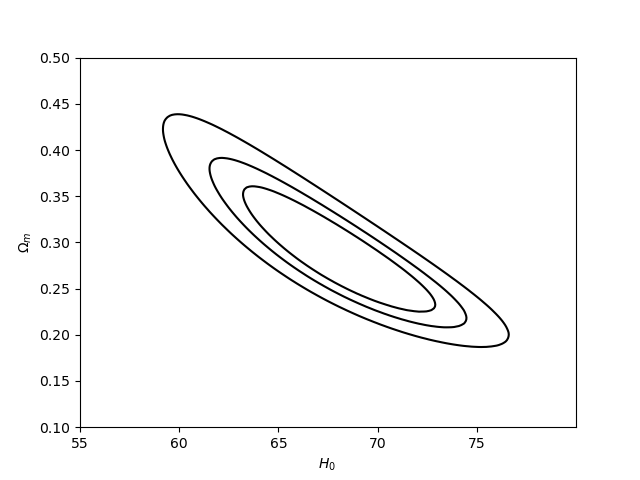} & \includegraphics[scale=0.29]{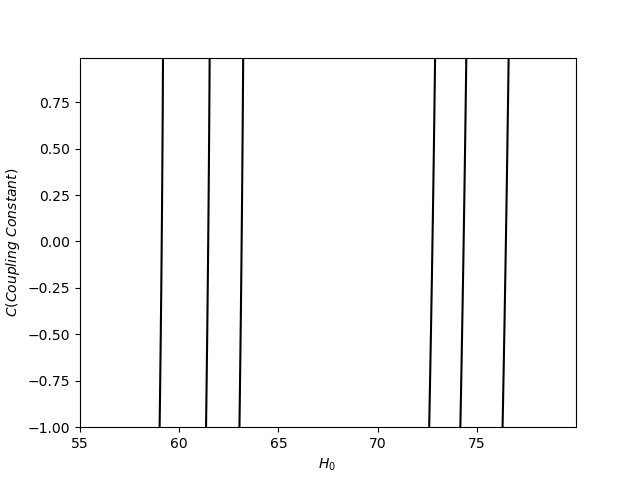} & \includegraphics[scale=0.29]{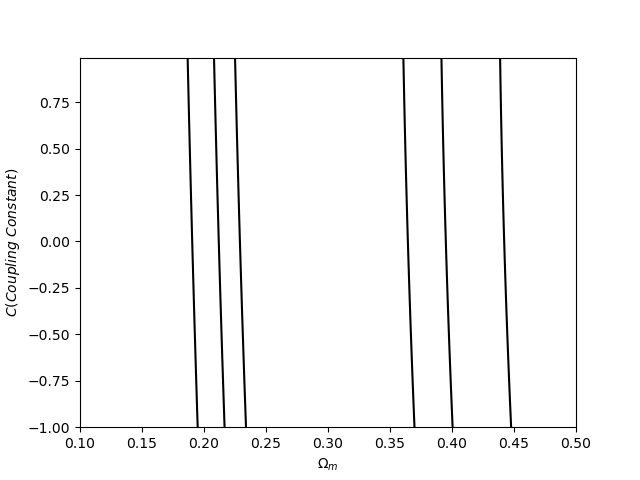} \\
\includegraphics[scale=0.29]{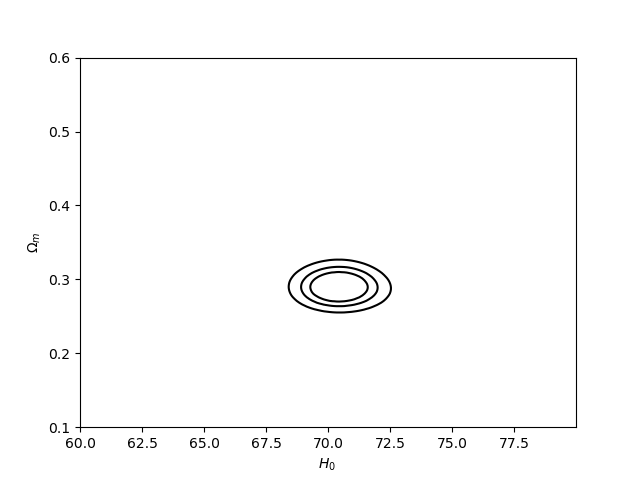} & \includegraphics[scale=0.29]{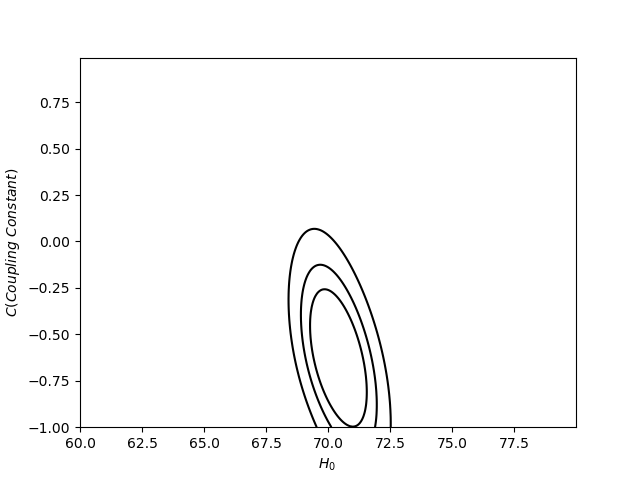} & \includegraphics[scale=0.29]{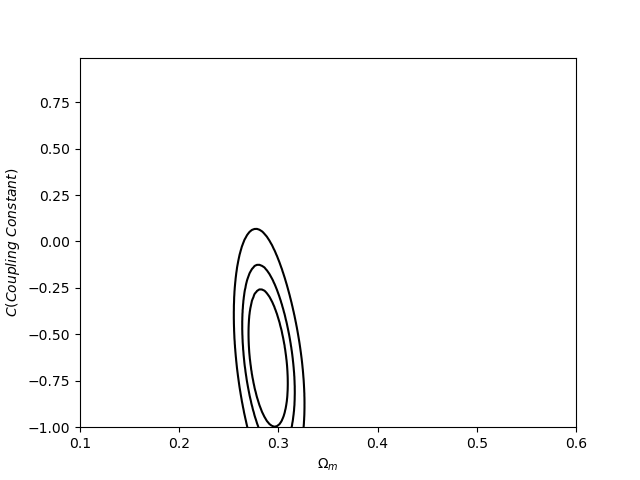} \\
\includegraphics[scale=0.29]{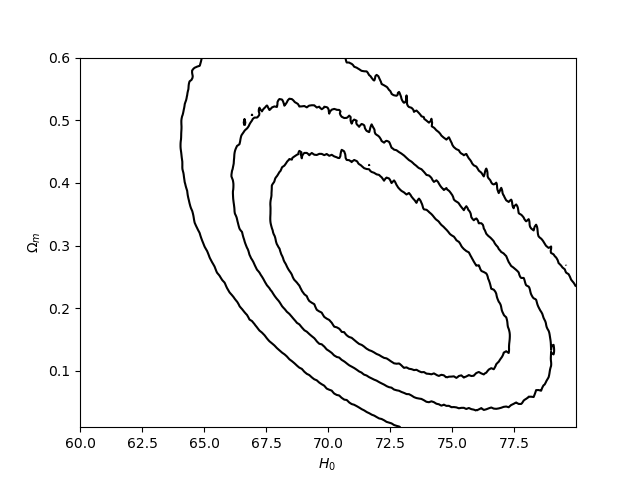} & \includegraphics[scale=0.29]{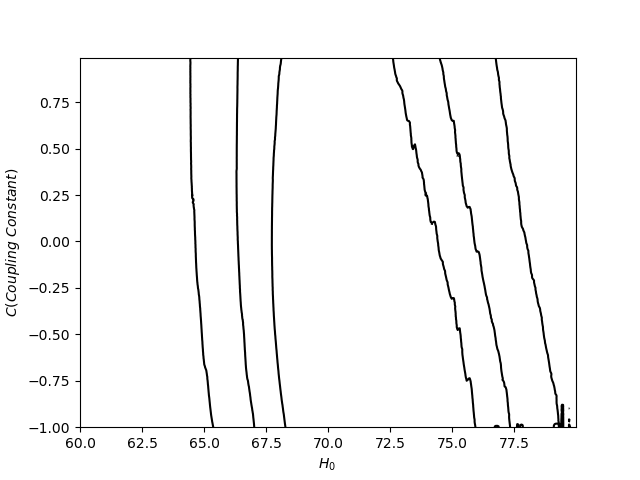} & \includegraphics[scale=0.29]{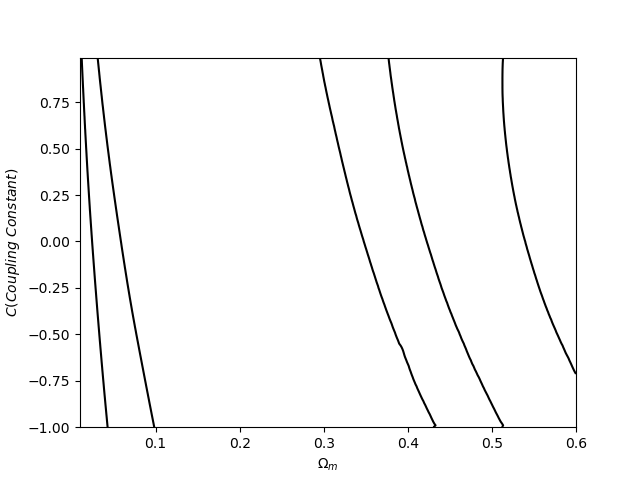} \\
\includegraphics[scale=0.29]{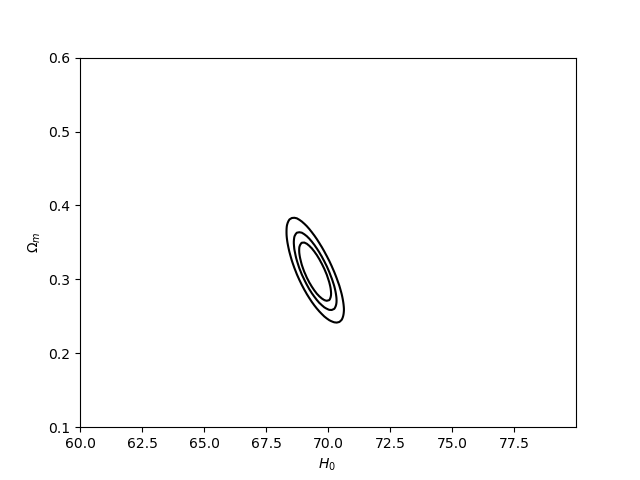} & \includegraphics[scale=0.29]{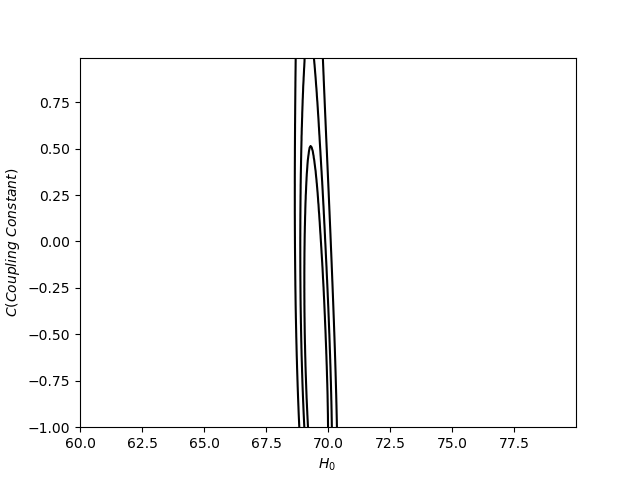} & \includegraphics[scale=0.29]{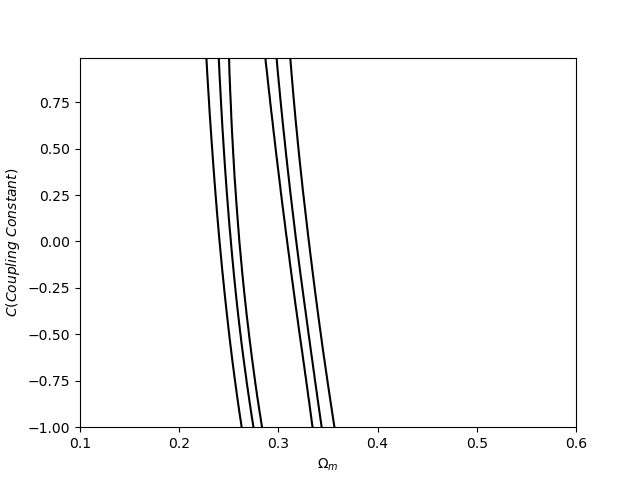} \\
\includegraphics[scale=0.29]{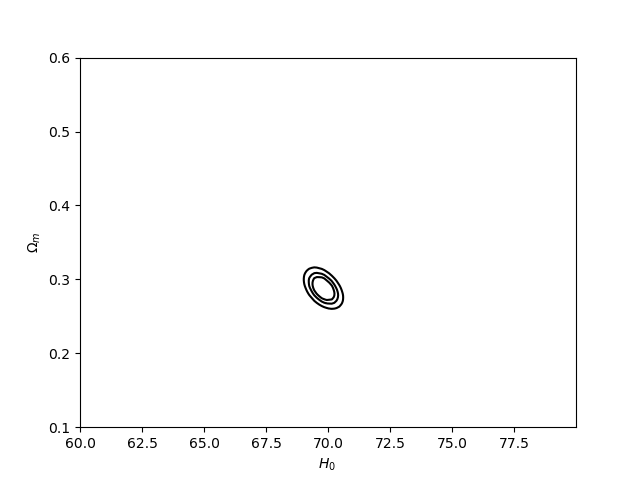} & \includegraphics[scale=0.29]{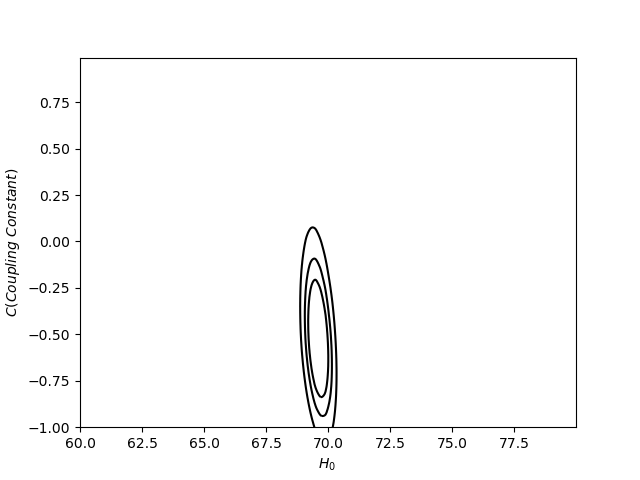} & \includegraphics[scale=0.29]{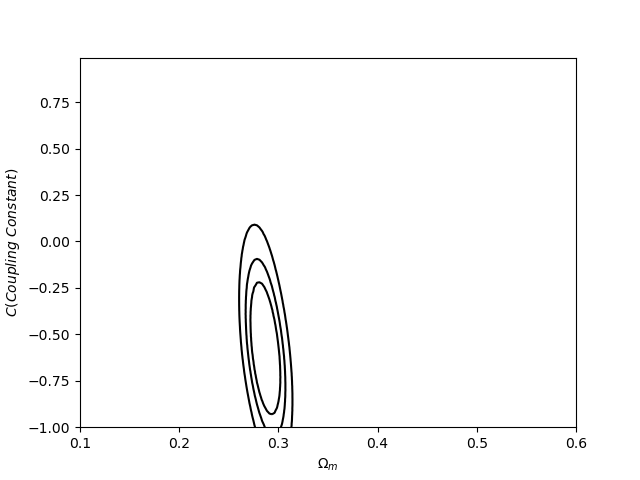}  
\end{tabular}
 \caption{1,2,3-$\sigma$ likelihood contours for Hz data (I row), BAO+Hz 
 data (II row),  HIIG data (III row), SN+Hz data (IV row) and all four data sets (V row). The two-dimensional contours are obtained by performing marginalization over other parameters.}
  \label{fig::h0_om_c}
\end{figure} 

\begin{figure}[!htbp]
\centering
\begin{tabular}{cc}
\includegraphics[scale=0.45]{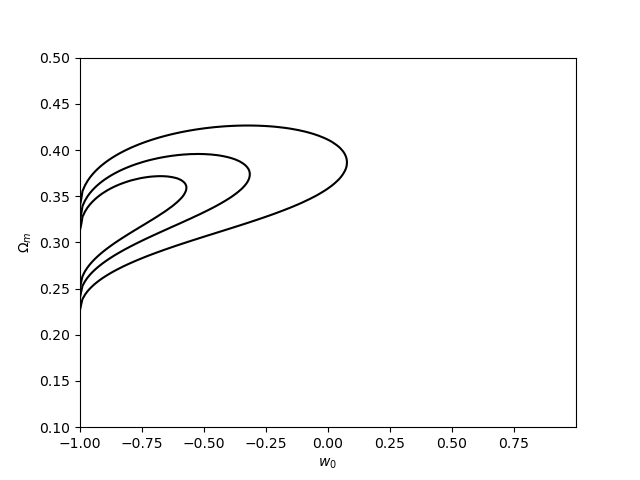}&\includegraphics[scale=0.45]{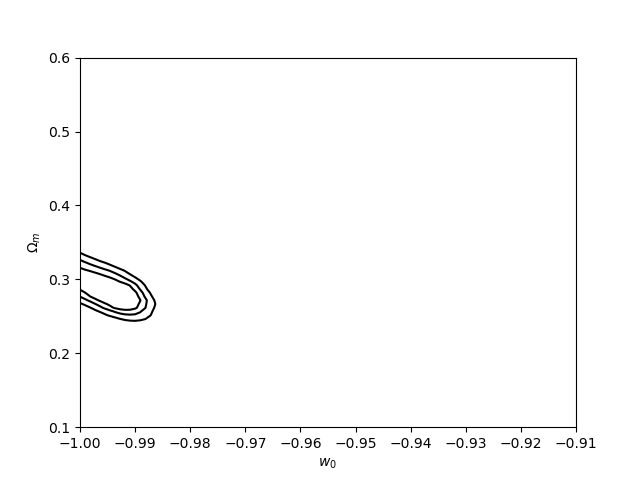}\\
\includegraphics[scale=0.45]{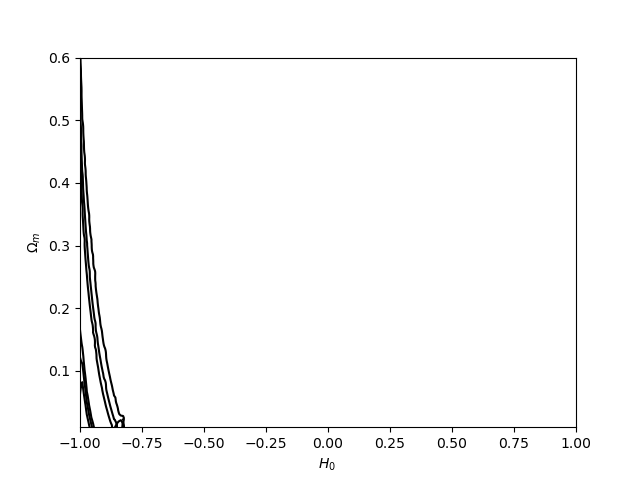}&\includegraphics[scale=0.45]{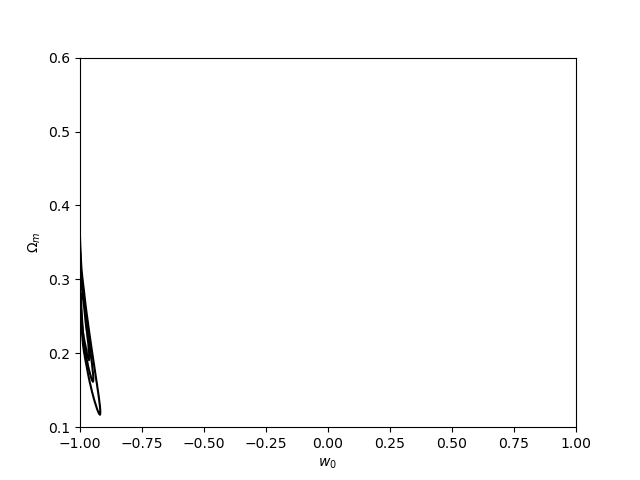}
\end{tabular}
\caption{1,2,3-$\sigma$ likelihood contours in `$w_0$-$\Omega_m$' plane. The top 
  row shows constraints from Hz data (left) and BAO+Hz observations (right). 
 The second row shows constraints from  HIIG measurements (left) and {SN+Hz} observations (right).} 
  \label{fig::w0_om} 
\end{figure}

Figure~\ref{fig::h0_om_c} contains the constraints on parameters $H_0$, interaction strength $C$, and $\Omega_m$ for the four observational data sets. The plots are for $n =1$ \eqref{eq:Potentialdef}. Analysis is also done for $n=2$ and $n = 3$; however, there is no significant change in the parameter constraints.  For completeness, in Appendix \ref{sec::n=2}, we have presented the results for $n=2$. The 1$\sigma$, 2$\sigma$, 3$\sigma$ contours corresponding to 67$\%$, 95$\%$ and 99$\%$ confidence regions respectively, are shown in two-dimensional planes in Figure~\ref{fig::h0_om_c}. The first, second, and third columns correspond to `$H_0 -\Omega_m$', `$H_0 - C$' and `$\Omega_m - C$' planes, respectively. To show these two-dimensional confidence regions, we have marginalized over the other parameters.
The two-dimensional confidence regions for standard parameters $w_0$ and $\Omega_m$ are shown in Figure~\ref{fig::w0_om}.
Table~\ref{table::range} contains the best fit values of the parameters. 
Table~\ref{table::range} contains the allowed range of parameters. 
In the first row, we show constraints from Hz measurements. In the second row, results from BAO+Hz observations are shown, and the third row represents the confidence contours from HIIG data, while results in the fourth row are from SN+Hz observations. The fifth row shows the constraints obtained from the combination of all the data sets mentioned in section~\ref{sec::observations}.

\begin{sidewaystable}
\renewcommand{\arraystretch}{1.5}
\hspace{-1.5cm}
\begin{tabular}{|c|c|c|c|c|c|c|c|c|c|} \hline
{\bf Data set}&{\bf $1\sigma$ Confidence} &{\bf $2\sigma$ confidence} &{\bf $3\sigma$ confidence}&Best-fit values&$\chi^2$&$dof,\nu$&$\chi_{red}^{2}$ &  $AIC$ & $BIC$ \\ \hline
\multirow{2}{*}{\bf Hz}& 64.19$\leq ~H_0 ~\leq$ 72.11 & 61.19$\leq ~H_0 ~\leq$74.12& 59.76$\leq ~H_0~\leq$75.91 & $H_0$=69.34  & 18.81 & 27 & 0.697 & 26.81 & 32.54\\
 & 0.24$\leq ~\Omega_m ~\leq$0.34 & 0.21$\leq ~\Omega_m~\leq$0.39 & 0.19$\leq~\Omega_m~\leq$0.43 & $\Omega_m$=0.29 &  &  & & &  \\
 & -1$\leq~w_0~\leq$-0.67 & -1$\leq~w_0~\leq$-0.24 & -1$\leq~w_0~\leq$0.04 & $w_0$=-0.989 &  &  &  &  &  \\
 & -1$\leq~C~\leq$1 & -1$\leq~C~\leq$1 & -1$\leq~C~\leq$1 & $C=0.98$ & & & & & \\ \hline
\multirow{2}{*}{\bf BAO+Hz}& 69.31$\leq ~H_0 ~\leq$71.61 & 68.95$\leq ~H_0 ~\leq$71.98 & 68.42$\leq ~H_0 ~\leq$72.57 & $H_0$=70.4 & 21.87 & 37 & 0.591 & 29.87 &36.72\\
 & 0.269$\leq ~\Omega_m ~\leq$0.309 & 0.264$\leq ~\Omega_m ~\leq$0.316 & 0.254$\leq ~\Omega_m ~\leq$0.32 & $\Omega_m$=0.29 & & & & & \\
 & -1$\leq~w_0~\leq$-0.989  & -1$\leq~w_0~\leq$-0.987 & -1$\leq~w_0~\leq$-0.985 & $w_0$=-0.997 & & & & &\\
 & -1$\leq~C~\leq$-0.261 & -1$\leq~C~\leq$-0.132 & -1$\leq~C~\leq$0.067 & $C$=-0.63 & & & & & \\ \hline
\multirow{2}{*}{\bf HIIG}& 67.78$\leq ~H_0 ~\leq$77.2 & 66.29$\leq ~H_0 ~\leq$78.9 & 64.07$\leq ~H_0 ~\leq$80& $H_0$=72.49 & 226.79 & 149 & 1.522 & 234.79&246.91\\
 & 0.091$\leq ~\Omega_m ~\leq$0.447 & 0.041 $\leq ~\Omega_m ~\leq$ 0.53 & 0.01$\leq ~\Omega_m ~\leq$0.6 & $\Omega_m$=0.25 & & & & &\\
 & -1$\leq~w_0~\leq$-0.87  & -1$\leq~w_0~\leq$-0.84 & -1$\leq~w_0~\leq$-0.81 & $w_0$=-0.92 & & & & &\\
 & -1$\leq~C~\leq$1  & -1$\leq~C~\leq$1 & -1$\leq~C~\leq$1 & $C$=-0.94 & & & & &\\ \hline
\multirow{2}{*}{\bf SN+Hz}& 69.18$\leq ~H_0 ~\leq$70.02 &69.06$\leq ~H_0 ~\leq$70.19 & 68.87$\leq ~H_0 ~\leq$70.36 & $H_0$=69.51 &737.21 &767 & 0.961& 745.21&763.8\\
 & 0.25$\leq ~\Omega_m ~\leq$0.33 & 0.24$\leq ~\Omega_m ~\leq$0.34 & 0.22$\leq ~\Omega_m ~\leq$0.35 & $\Omega_m$=0.31& & & &&\\
 & -1$\leq~w_0~\leq$-0.97  & -1$\leq~w_0~\leq$-0.93 & -1$\leq~w_0~\leq$-0.9 & $w_0$=-1 & & & & &\\
 & -1$\leq~C~\leq$-0.51 & -1$\leq~C~\leq$1 & -1$\leq~C~\leq$1 & $C$=-0.69 & & & & &\\ \hline
\multirow{2}{*}{\bf Hz+BAO}& 69.27$\leq ~H_0 ~\leq$70.08 & 69.07$\leq ~H_0 ~\leq$70.19 & 68.83$\leq ~H_0 ~\leq$70.59 & $H_0$=69.79 & 968.322 & 930& 1.041& 976.332&995.67\\
 \bf{+HIIG+SN}&0.27$\leq ~\Omega_m ~\leq$0.303 & 0.266$\leq ~\Omega_m ~\leq$0.308 & 0.26$\leq ~\Omega_m ~\leq$0.316 & $\Omega_m$=0.29 & (739.014)& (777)& (0.951)& (747.014)& (766.372)\\
 & -1$\leq~w_0~\leq$-0.99  & -1$\leq~w_0~\leq$-0.987 & -1$\leq~w_0~\leq$-0.985 & $w_0$=-0.99 & & & & & \\
 & -0.83$\leq~C~\leq$-0.21  & -0.93$\leq~C~\leq$-0.099 & -1$\leq~C~\leq$0.087 & $C$=-0.52 & & & & & \\\hline
\end{tabular}
\caption{Confidence limits, best fit values of parameters, AIC, BIC and $\chi^2$ values from various data sets for interacting dark sector cosmology. The values in the brackets in the last row correspond to the analysis of combined data set excluding the HIIG data.}
\label{table::range}
\end{sidewaystable}

\begin{figure}[!htbp]
\vspace*{-0.65cm}
\centering
\begin{tabular}{ccc}
\includegraphics[scale=0.29]{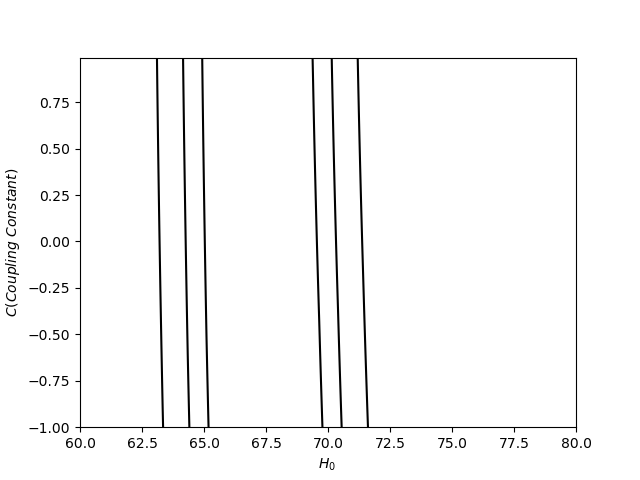}&\includegraphics[scale=0.29]{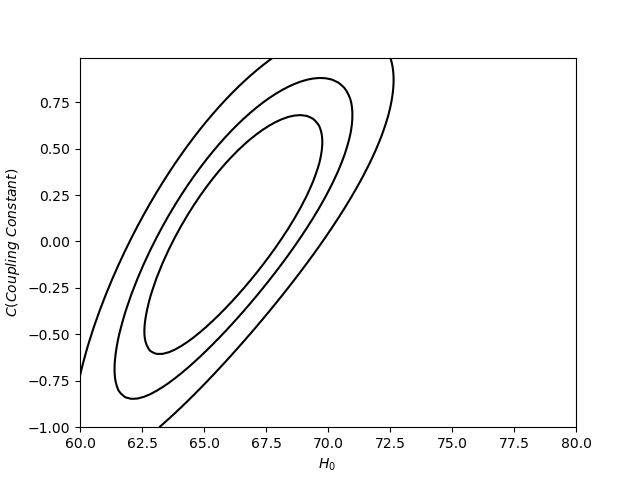}&\includegraphics[scale=0.29]{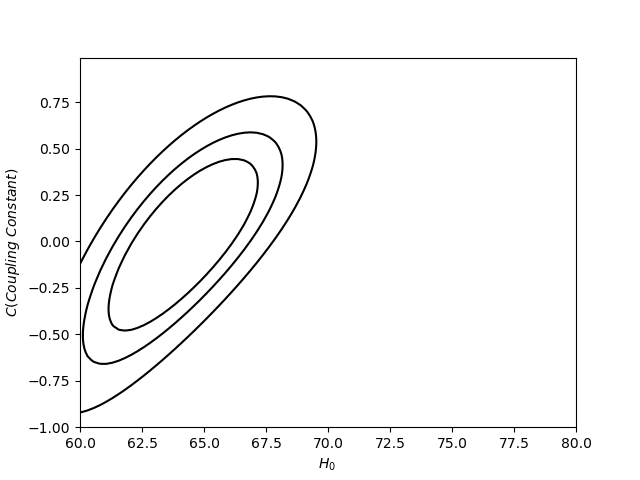}\\
\includegraphics[scale=0.29]{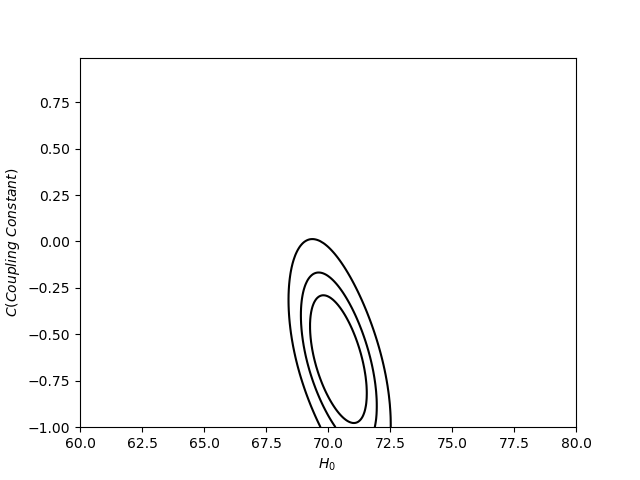}&\includegraphics[scale=0.29]{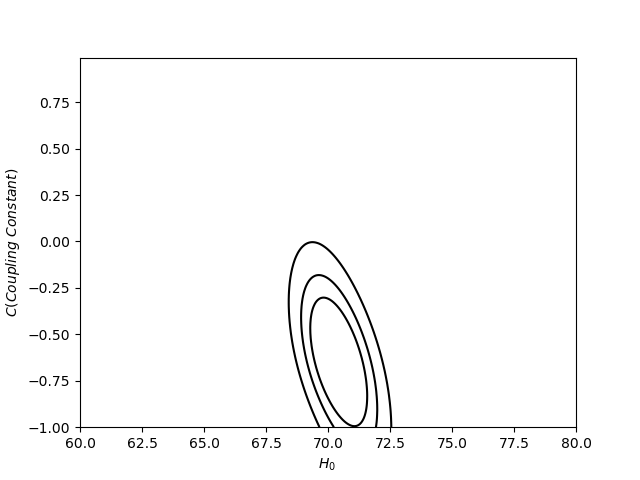}&\includegraphics[scale=0.29]{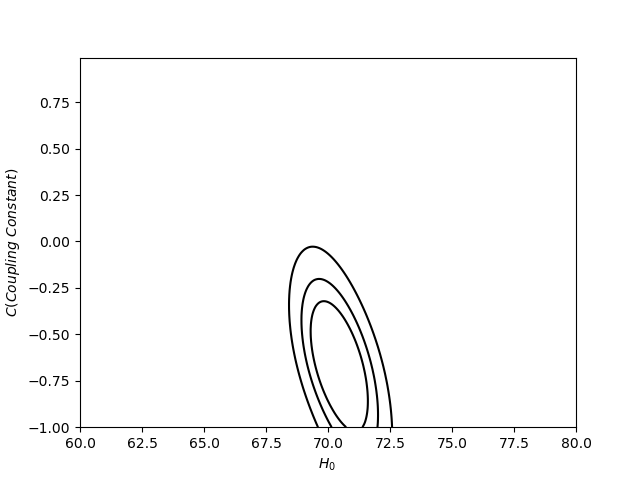}\\
\includegraphics[scale=0.29]{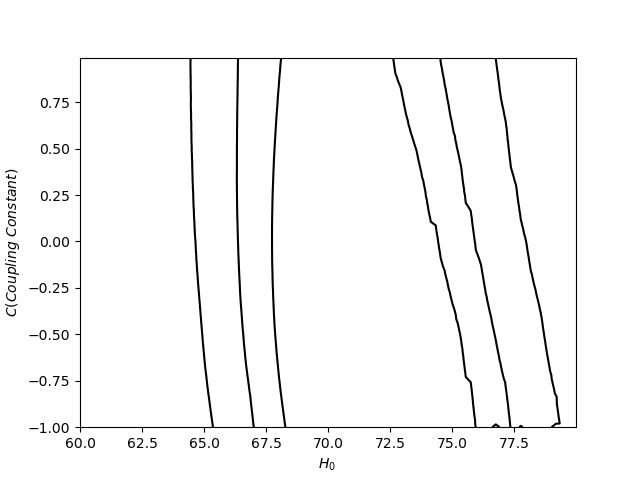}&\includegraphics[scale=0.29]{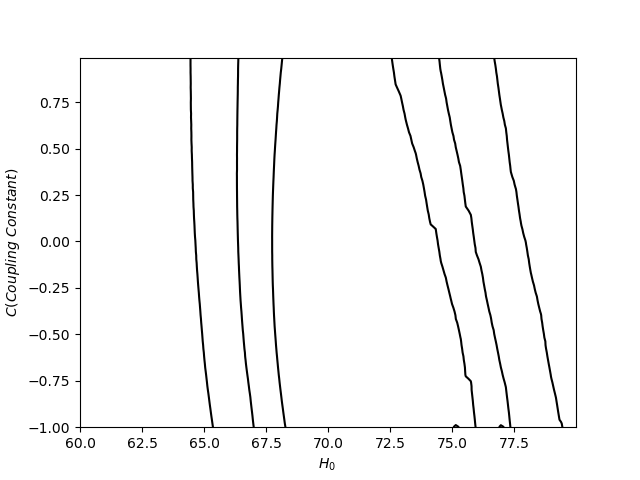}&\includegraphics[scale=0.29]{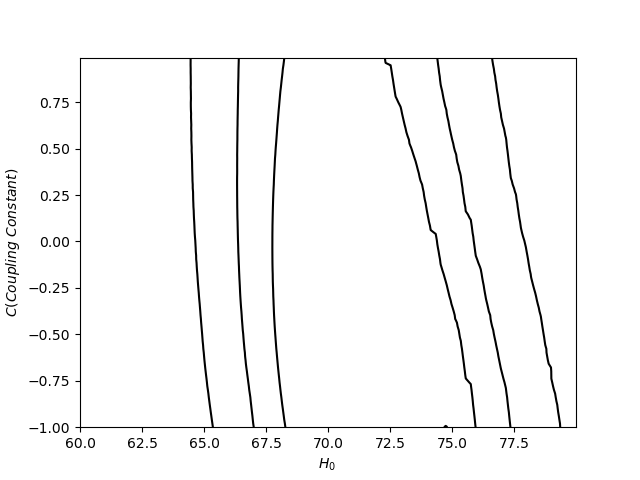}\\
\includegraphics[scale=0.29]{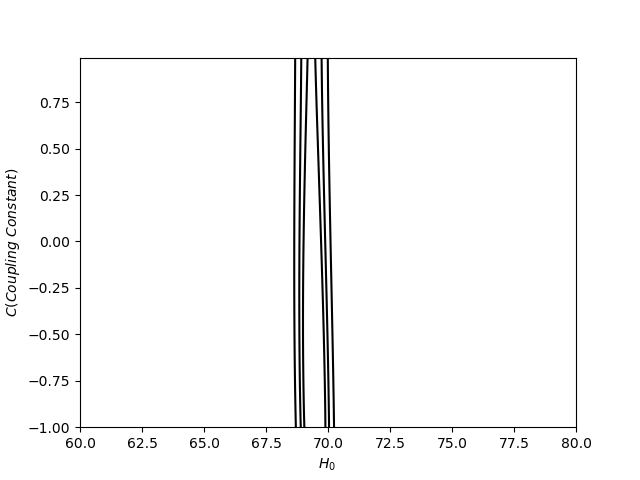}&\includegraphics[scale=0.29]{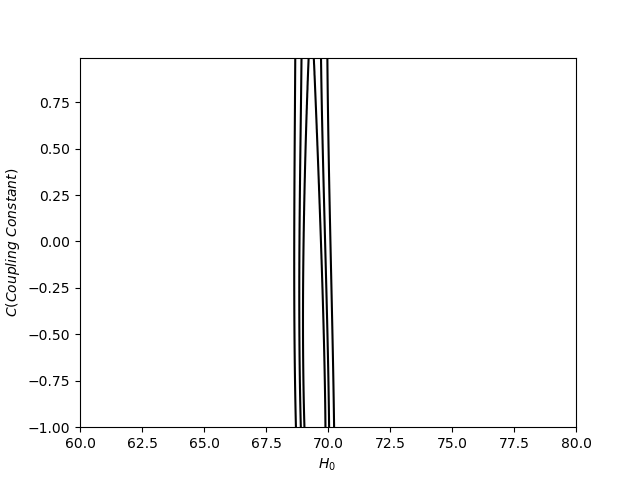}&\includegraphics[scale=0.29]{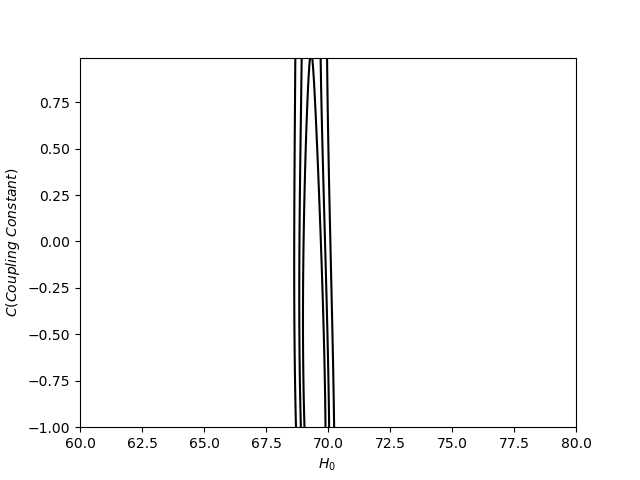}\\
\includegraphics[scale=0.29]{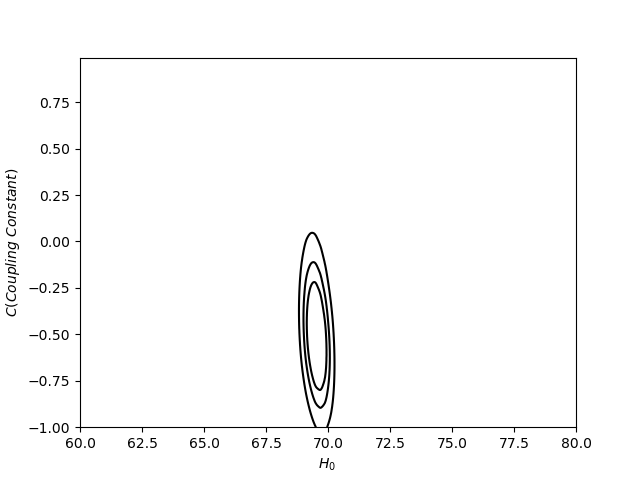}&\includegraphics[scale=0.29]{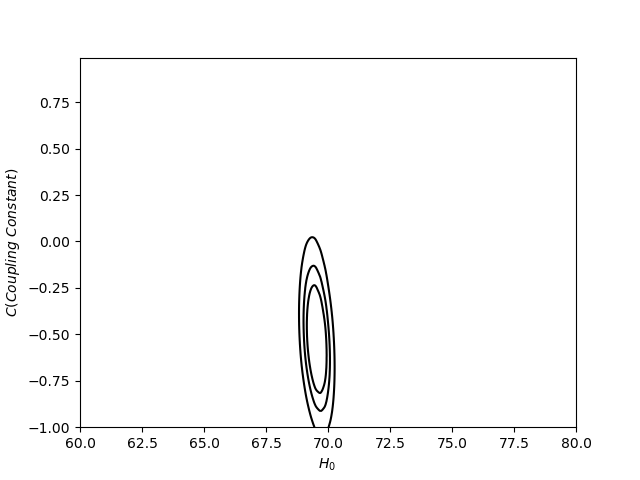}&\includegraphics[scale=0.29]{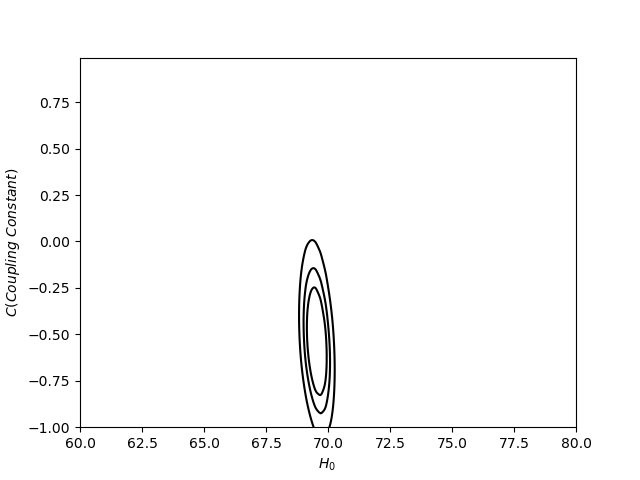}\\
\end{tabular}
\caption{1,2,3-{$\sigma$} likelihood contours in `$H_0-C$' plane for different values of $w_0$. Plots show constraints from Hz data (I row),  BAO+Hz (II row),
HIIG data (III row), {SN+Hz} (IV row) and all four data sets (V row). The left, middle and right plots  correspond to different $w_0$ values. For $H(z)$ data we choose $w_0 =[-1, -0.6, -0.1]$ and for the remaining data sets, we choose $w_0=[ -1, -0.99, -0.985]$.} 
\label{fig::varying-w0}
\end{figure}

The key inferences from the Hz data are as follows: First, the minimum value of $\chi^2$ is 18.81 which  corresponds to the best fit values of the parameters $H_0 = 69.34~\rm{km}\, \rm{s}^{-1}~\rm{Mpc}^{-1}$, $\Omega_m= 0.29, w_0=-0.98$ and the interaction strength $C=0.98$. Second, within the 2$\sigma$ region, the Hz data allows $H_0$ to take values between $ 61.19-74.12~\rm{km}\, \rm{s}^{-1}~\rm{Mpc}^{-1}$ which includes the values reported by Planck \cite{2020-Aghanim.others-Astron.Astrophys.} and the local measurements \cite{2019-Birrer.others-Mon.Not.Roy.Astron.Soc., 2016-Riess.others-Astrophys.J.}. Hence, with Hz observations, the interacting dark sector model is consistent with {both of these reported values}. Third, the best fit value, as well as the allowed range of non-relativistic density parameter, is also consistent with the constraints reported in the previous studies~\cite{2020-Aghanim.others-Astron.Astrophys.,2016-Riess.others-Astrophys.J.,2019-Birrer.others-Mon.Not.Roy.Astron.Soc.}. 
Fourth, after marginalizing over parameter $w_0$, the data allows the entire range of the coupling parameter ($C$) considered in the analysis within the $1\sigma$ region see Figure~\ref{fig::h0_om_c}. However, we also find that if we fix $w_0$ at a particular value, say $w_0=-1$, it does not constrain $C$ at all, but if we move away from $\Lambda$CDM like scenarios at present, and consider $w_0 \ge -1$ then we start getting a limit on $C$ as well. As the value of $w_0$ moves away from $-1$ towards $1$, the constraints on $C$ becomes tighter (cf. Figure \ref{fig::varying-w0}).
Fifth, from Figure~\ref{fig::w0_om}, we see that the Hz data does not provide a lower limit on $w_0$; however, an upper limit of -0.67 within $1\sigma$ and $w_0$=0.04 within $3\sigma$ region is allowed showing that this particular model does not allow for a non-accelerating universe within $1\sigma$ region. Also, Hz is the only observation that allows for a non-accelerating universe within the 3$\sigma$ region. The allowed range for $w_0$ is the widest compared to the other three observations considered in the analysis. The Hz measurements constrain $\Omega_m$ to take values within a range of $0.19 -0.43$ for 
$3\sigma$ confidence level, which is very wide compared to the ones obtained from {BAO+Hz} and {SN+Hz} data sets.
   

The key inferences from BAO+Hz data are as follows:  First, the minimum value of $\chi^2$ is {21.87} which corresponds to the best fit values of parameters giving $H_0 = 70.4$ km s$^{-1}$Mpc$^{-1}$, $\Omega_m$= 0.29, $w_0$=-0.997 and the interaction strength is $C$=-0.63. Second, within 1$\sigma$ region, 
{BAO+Hz} data allows $H_0$ to take a very small range given by  $69.31 - 71.61$ km s$^{-1}$Mpc$^{-1}$  which lies between the value of $H_0$ reported by Planck~\cite{2020-Aghanim.others-Astron.Astrophys.} and the local probes \cite{2016-Riess.others-Astrophys.J.,2019-Birrer.others-Mon.Not.Roy.Astron.Soc.}. Third, the best fit value of the non-relativistic matter density parameter is $\Omega_m=0.29$. The allowed range within the 3$\sigma$ region is very narrow and consistent with the constraints reported in the previous studies~\cite{2020-Aghanim.others-Astron.Astrophys.,2016-Riess.others-Astrophys.J.,2019-Birrer.others-Mon.Not.Roy.Astron.Soc.}. 
Fourth, within 1$\sigma$, {BAO+Hz} data also constrains the interaction strength $C$ within the range of -1.0 to -0.261 (cf. Figure~\ref{fig::h0_om_c}) and between -1 to 0.067 corresponding to 99$\%$ confidence region.  \emph{Thus, {BAO+Hz} data prefers negative values of $C$.} Here again, we find that if we fix $w_0$ at a particular value, say $w_0=-1$, the allowed range is narrower than when $w_0$ was a free parameter. And if we move away from $\Lambda$CDM-like scenarios at present, and consider $w_0 \ge -1$, then the upper limit on $C$ starts getting lower as the contours shift to the negative regions on $C$. As we change $w_0$ from $-1$ towards $1$, the constraints on $C$ become tighter as in Hz data, and we find that the {BAO+Hz} data prefers negative values of $C$.
Fifth, from Figure~\ref{fig::w0_om}, we see that the {BAO+Hz} data provide very small range on $w_0$ for 1$\sigma$, 2$\sigma$ region and within 3$\sigma$ region $\Lambda$CDM case is allowed. Therefore, the {BAO+Hz} observational data \emph{do not} allow for a non-accelerating universe and prefer a $\Lambda$CDM like scenario. It also provides the tightest constraints for the model parameters out of all the observational data-sets considered.


The key inferences from HIIG data are as follows: First, the minimum value of $\chi^2$ is 226.79 which corresponds to the best fit parameters $H_0 = 72.49$ km s$^{-1}$Mpc$^{-1}$, $\Omega_m= 0.25$, $w_0=-0.92$ and the interaction strength is $C=-0.94$. Second, HIIG data allows $H_0$ to take values in the range $67.78-77.2$ km s$^{-1}$Mpc$^{-1}$ within 1$\sigma$ region. {The best fit value for the model indicates the preference for the value of $H_0$ reported by local measurements \cite{2016-Riess.others-Astrophys.J.,2019-Birrer.others-Mon.Not.Roy.Astron.Soc.}. However, the interacting dark sector model is also consistent with the $H_0$ value reported in Ref.~\cite{2020-Aghanim.others-Astron.Astrophys.} within 3$\sigma$ region.} 
Third, the best fit value of the non-relativistic density parameter preferred by HIIG data is smaller than the value reported in the previous studies~\cite{2020-Aghanim.others-Astron.Astrophys.,2016-Riess.others-Astrophys.J.,2019-Birrer.others-Mon.Not.Roy.Astron.Soc.}. 
Fourth, similar to Hz data, HIIG data allows the entire range of coupling parameter ($C$)  within the $1\sigma$ region, see Figure~\ref{fig::h0_om_c}. Here again, we have marginalized over parameter $w_0$. We also found that if we fix $w_0$ at a particular value and consider $w_0 \ge -1$, then we see a slight shift in contour which is almost insignificant in changing $w=-1$ to $w\sim-0.985$. Fifth, from Figure~\ref{fig::w0_om}, we see that the HIIG data does not provide a lower limit on $w_0$. Still, an upper limit of $-0.81$ within the $3\sigma$  region is allowed, showing that similar to BAO+Hz data, this particular model does not allow for a non-accelerating universe within 3$\sigma$ region. The results are consistent with the $\Lambda$CDM model. These observations' constraints on $\Omega_m$ give the widest range amongst all data sets considered in the analysis. 


The key inferences from SN{+Hz} data are as follows: First, the minimum value of $\chi^2$ is {737.21} which  corresponds to the best fit values of the parameters $H_0 = 69.51$ km s$^{-1}$Mpc$^{-1}$, $\Omega_m$= 0.31, $w_0$=-1.0 and $C$=-1.
Second, the {SN+Hz} data allows $H_0$ to take values between $\sim$ $69.18-70.02$ km s$^{-1}$Mpc$^{-1}$ within 1$\sigma$ region, which lies between the values reported by Planck-2018~\cite{2020-Aghanim.others-Astron.Astrophys.} and the local $H_0$ measurements~\cite{2016-Riess.others-Astrophys.J.,2019-Birrer.others-Mon.Not.Roy.Astron.Soc.}. Interestingly, it provides a very narrow range for $H_0$ and, hence, \emph{the interacting dark sector model can potentially alleviate the $H_0$ tension.} 

Third, the best fit value, as well as the allowed range of non-relativistic density parameter, is also consistent with the constraints reported in previous studies~\cite{2020-Aghanim.others-Astron.Astrophys.}. Fourth, like Hz data, SN{+Hz} data also allows the entire range of the interaction strength ($C$) within the 3$\sigma$ region. However, within 1$\sigma$ region, it constrains $C$ to be less than 0.5, (cf. Figure~\ref{fig::h0_om_c}). Here again, we have marginalized over parameter $w_0$. We find that like other data sets, {SN+Hz} also prefers negative values of interaction strength.

Fifth, from Figure~\ref{fig::w0_om}, we see that the SN+Hz data does not provide a lower limit on $w_0$. However, within the $1\sigma$, there is an upper limit of $-0.97$, and $w_0=-0.9$ within the $3\sigma$ region. Thus, the analysis shows that a non-accelerating universe is not allowed.  The allowed values of $\Omega_m$ are very narrow and consistent with previous studies. This model is also consistent with the $\Lambda$CDM model. We have done the analysis with Pantheon compilation of SN for some specific values of model parameters $H_0$, $\Omega_m$, and $C$, and we get similar results (cf. appendix \ref{sec::SN_comparison}).

The key inferences from the combined data are as follows: First, the minimum value of $\chi^2$ is 968.332, which corresponds to the best fit values of the parameters are $H_0 = 69.79$ km s$^{-1}$Mpc$^{-1}$, $\Omega_m$= 0.29, $w_0$=-0.99 and the interaction strength is $C$=-0.52.
Second, the Hz+BAO+HIIG+SN data allows $H_0$ to take values between $\sim$ $69.27-70.08$ km s$^{-1}$Mpc$^{-1}$ within 1$\sigma$ region, which lies between the values reported by Planck-2018~\cite{2020-Aghanim.others-Astron.Astrophys.} and the local probes \cite{2016-Riess.others-Astrophys.J.,2019-Birrer.others-Mon.Not.Roy.Astron.Soc.}. It provides a very narrow range for $H_0$ within the 3$\sigma$ confidence region. Therefore, 
this dark sector interaction model puts very narrow constraints on model parameters with the joint analysis. Third, the best fit value, as well as the allowed range of non-relativistic density parameter, is also consistent with a narrow range of allowed values with $0.316\ge\Omega_m\ge0.26$ within 3$\sigma$ region, and these constraints are consistent with the ones reported in previous studies~\cite{2016-Riess.others-Astrophys.J.,2020-Aghanim.others-Astron.Astrophys.}. Fourth, we get a very narrow range for the coupling parameter for the joint analysis, $C$, which restricts it to take values only within -0.83 to -0.21 for 1$\sigma$ and from -1 to 0.087 for 3$\sigma$ confidence regions, see Figure~\ref{fig::h0_om_c}. In the joint analysis, the constraints are driven by the BAO observation, which has the most constraining capacity, followed by SN, Hz, and HIIG observations. Like the individual cases, if we fix $w_0$ at a particular value, say $w_0=-1$, the combination data gives a slightly narrower range. If we move away from $\Lambda$CDM-like scenarios at present,
for $w_0 \ge -1$, there is a preference for negative values of $C$.

Fifth, from Figure~\ref{fig::w0_om}, we see that the combined data does not provide a lower limit on $w_0$. However, within 1$\sigma$, we get
the upper limit of $-0.993$ and $w_0=-0.99$ within 3$\sigma$ region. This again shows that the model does not allow for a non-accelerating universe and constrains $w_0$ to a value close to -1, and is consistent with the $\Lambda$CDM model.

In Figure \ref{fig::varying-w0}, instead of marginalizing $w_0$, we assume a value of $w_0$ within 3$\sigma$ allowed range reported in this work and see the change in the $H_0-C$ plane. The first row is obtained for Hz data, and the value of $w_0$ considered are -1, -0.6, and -0.1 (left, middle, and right plots, respectively). In the second, third, fourth, and fifth rows, the results correspond to BAO+Hz, HIIG, SN+Hz, and combined analysis, respectively. For the left, middle and right plots, respectively, we fix $w_0$ at -1, -0.99, and -0.985. For Hz measurements, we see a significant change in the constraints as $w_0$ changes from $-1$ to $-0.1$, and we start getting constraints on $C$. But for BAO+Hz, HIIG, SN+Hz, and combined case, there is a slight shift in contours in contours when $w_0$ is varied from $w_0$=-1 to -0.985 (within 3$\sigma$ range).

Thus, from these analyses, we make the following conclusions: 
\begin{enumerate}
\item All the observational data sets considered constrain $H_0$ to be close to $70$ km s$^{-1}$Mpc$^{-1}$. The dark sector interaction model puts very narrow constraints on the model parameters.

\item The constraints on $\Omega_m$ obtained from various data sets are consistent with each other. 

\item The constraints on $w_0$ are consistent with the $\Lambda$CDM model, and only Hz data allows for a non-accelerating universe.

\item  All data sets, except Hz, prefer negative value for the interaction strength ($C$). 

\item  We have also done the combined analysis excluding the HIIG data and found that the best fit values and constraints are approximately the same. The results corresponding to this analysis are given in the brackets in the last row of Table~\ref{table::range}.

\item Our analysis shows no significant difference in the best-fit values for different values of $n$ (specifically, for $n=1$ and $n=2$ (See Appendix \ref{sec::n=2})).  However, we notice that the allowed range of cosmological parameters ($H_0$, $\Omega_m$, $w_0$) increases for HIIG as we go from $n=1$ to $n=2$. In the case of  interaction parameter $C$, Hz shows no significant change, HIIG data allows a wider range, whereas for BAO+Hz and SN+Hz case, the contour shifts lower, thereby giving a smaller value for the upper limit of $C$.
\end{enumerate}

The reduced $\chi^2$ values for the interacting dark sector model we have considered is closer to one (except for HIIG and combined data sets) compared to $\Lambda$CDM and $w$CDM models. (see Appendix~\ref{sec:wlambda}). Thus, our analysis points to the fact that there is a strong degeneracy between the interacting and non-interacting dark sector models with respect to these low-redshift background observations. In the next section, we explicitly show that the first-order perturbations can break the degeneracy between these two scenarios.

\section{Evolution of the scalar perturbations and predictions of the model}
\label{sec:pertevo}

In the previous sections, we obtained the constraints on the various model parameters based on the observational data related to the background evolution of the Universe. From these constraints, it is apparent that we need to go beyond the background observations to distinguish between non-interacting and interacting dark sector models. This section looks at the evolution of first-order perturbations for negative value for the interaction strength ($C$).

The perturbed perturbed FRW metric in the Newtonian gauge given by~\cite{2005-Mukhanov-PhysicalFoundationsCosmology}:
\begin{equation}
\label{eq:newtmetric}
g_{00}=-\left(1+2\Phi \right),\quad g_{0i}=0, \quad g_{ij} = a^2 (1-2\Psi)\delta_{ij} \, ,
\end{equation}
where $\Phi \equiv \Phi(t,x,y,z)$ and $\Psi \equiv \Psi(t,x,y,z)$ are the Bardeen Potentials.

We obtain the evolution of three perturbed quantities, which are relevant to three different cosmological observations:
\begin{enumerate}
\item Structure formation: $\delta_m(t,x,y,z) \equiv \frac{\delta \rho_m(t,x,y,z)}{ \overline{\rho_m}(t)}$
\item Weak lensing : $\Phi+\Psi$
\item Integrated Sachs-Wolfe (ISW) effect: $\Phi'+\Psi'$
\end{enumerate}
where $\delta_m$ is the density perturbation of dark matter fluid. We study the evolution of these perturbed quantities for various length scales specified by the wavenumber $k$. 

To analyze the difference in the evolution of the scalar perturbations in dark sector interactions compared to standard cosmology, we study the following quantities:
\begin{subequations}
\begin{eqnarray}
\label{def:Deltam}
\Delta \delta_m = \delta_{m_i} - \delta_{m_{ni}}, & \qquad& \Delta \delta_{m_{rel}} = \dfrac{\delta_{m_i} - \delta_{m_{ni}}}{\delta_{m_{ni}}} = \dfrac{\Delta \delta_m}{\delta_{m_{ni}}} \\
\label{def:DeltaPhi}
\Delta \Phi = \Phi_i - \Phi_{ni}, & \qquad &  \Delta \Phi_{rel} = \dfrac{\Phi_i - \Phi_{ni}}{\Phi_{ni}} = \dfrac{\Delta \Phi}{\Phi_{ni}} \\
\label{def:DeltaPhiPrime}
\Delta \Phi' = \Phi'_i - \Phi'_{ni}, &  \qquad & \Delta \Phi'_{rel} = \dfrac{\Phi'_i - \Phi'_{ni}}{\Phi'_{ni}} = \dfrac{\Delta \Phi'}{\Phi'_{ni}}
\end{eqnarray}
\end{subequations}
where the subscripts \textit{i} and \textit{ni} denote the interacting and non-interacting scenarios, respectively.

The perturbed interaction term in the fluid description is given by
\begin{equation}
 \delta Q^{\rm (F)} =-(\delta \rho_m - 3 \delta p_m)\alpha_{,\phi}(\overline{\phi})\dot{\overline{\phi}}-(\overline{\rho}_m-3\overline{p}_m)\left[\alpha_{,\phi \phi}(\overline{\phi}) \dot{\overline{\phi}} \delta \phi + \alpha_{,\phi}(\overline{\phi}) \dot{\delta \phi}\right] 
\end{equation}
In appendix B of Ref.~\cite{2021-Johnson.Shankaranarayanan-Phys.Rev.D}, the authors obtained the scalar perturbations equations for the interacting dark sector model. We rewrite these equations in terms of the following dimensionless variables:

\begin{eqnarray} \nonumber
x = \sqrt{\dfrac{8 \pi}{6}}\dfrac{\dot{\phi}}{H m_{Pl}}, \quad y = \sqrt{\dfrac{8 \pi}{3}}\dfrac{\sqrt{U}}{H m_{Pl}}, &&
\lambda = - \dfrac{m_{Pl}}{\sqrt{8 \pi}}\dfrac{U_{,\phi}}{U}, \quad \Gamma = \dfrac{U U_{,\phi \phi}}{U_{,\phi}^2}
\\
\alpha = \alpha(\phi), \quad  \beta =  -\dfrac{m_{Pl}}{\sqrt{8 \pi}}\dfrac{\alpha_{,\phi}}{\alpha}, && \gamma = \dfrac{\alpha \alpha_{,\phi \phi}}{\alpha_{,\phi}^2}
\end{eqnarray}

In terms of these dimensionless variables, the scalar perturbation equations are~\cite{2021-Johnson.Shankaranarayanan-Phys.Rev.D}:
\begin{eqnarray}
\label{eq:perturbed}
\nonumber
\delta \phi '' + \left[ \dfrac{3}{2}\left(y^2 - x^2  - \omega \Omega_m + 1 \right) - 3\sqrt{6} \alpha \beta x \left( c_s^2 - \dfrac{1}{3} \right)\right] \delta \phi' \\  \nonumber
+ \left[ -9 \beta \left(\Omega_m \gamma \left( \omega - \dfrac{1}{3} \right) \beta -y^2 \left(c_s^2 - \dfrac{1}{3} \right) \lambda\right)\alpha + 3 \Gamma \lambda^2 y^2 + \dfrac{k^2}{a^2 H^2}\right] \delta \phi \\ \nonumber
 + \left[ -18 \sqrt{2}\alpha \beta \left(c_s^2 - \dfrac{1}{3} \right) - 8 \sqrt{3} x \right] \Phi' \\ 
 - 18 \sqrt{2} \left[ \alpha \beta \left( \left( c_s^2 - \dfrac{1}{3}\right) \left( y^2 + \dfrac{k^2}{3 a^2 H^2} \right)  + (c_s^2 - \omega) \right) + \dfrac{ \lambda y^2}{3} \right] \Phi  = && 0
\\
\nonumber
\Phi'' + \dfrac{3}{2}\left[y^2-x^2-\Omega_m \omega + 2 c_s^2
+\dfrac{5}{3} \right] \Phi' + 3 \left[ c_s^2 \left(\dfrac{k^2}{3 a^2 H^2} - x^2 +1 \right) - \Omega_m \omega +y^2 \right] \Phi \\ + \dfrac{\sqrt{3} x}{2}(c_s^2 -1)\delta\phi'
-\dfrac{3 \sqrt{2} \lambda y^2}{4} (c_s^2+1) \delta \phi  = && 0
\\
\nonumber
\delta' + 3 (\omega -c_s^2)(\sqrt{6}\alpha\beta x -1) \delta + \dfrac{2}{3} \dfrac{k^2}{a^2 H^2 \Omega_m} \Phi + \left(-3 \omega - 3 + \dfrac{k^2}{a^2 H^2 \Omega_m} \right) \Phi' \\
- \dfrac{1}{\sqrt{2}} \alpha \beta (3\omega -1) \delta \phi' + \sqrt{3} \left[ \alpha \beta^2 \gamma (3\omega -1) - \dfrac{1}{9} \dfrac{k^2}{a^2 H^2 \Omega_m} \right] x \delta\phi  = &&0 \, ,
\end{eqnarray}
where $\omega$ and $c_s$ denote the equation of state and sound speed of the dark matter fluid, respectively. We solve these equations for the redshift range $0 \leq z \lesssim 1500$. Hence for the analysis in this section, we include the contributions of radiation, dark matter, and dark energy to the total energy density of the Universe.
The calculations are done in the rest frame of the pressureless dark matter fluid, for which $\omega = c_s^2 = 0$ (cf. Appendix \ref{app:Sound}). As mentioned in Sec~\ref{sec:IDSmodel}, this analysis is done for $U(\phi) \sim 1/\phi$ and $\alpha(\phi) \sim \phi$. Analysis is also done for $n=2$, however, the results are not sensitive to $n$.  For completeness, in Appendix \ref{app:phi2}, we have presented the results for $n=2$.

To understand the effect of the interaction between dark energy and dark matter on the perturbed quantities, we define scaled interaction function $\delta q$:
\begin{equation}
\label{def:scaledq}
\delta q = \dfrac{\delta Q}{H^3 M_{Pl}^2}.
\end{equation}
\begin{figure}[!htb]
\begin{minipage}[b]{.45\textwidth}
\includegraphics[scale=0.4]{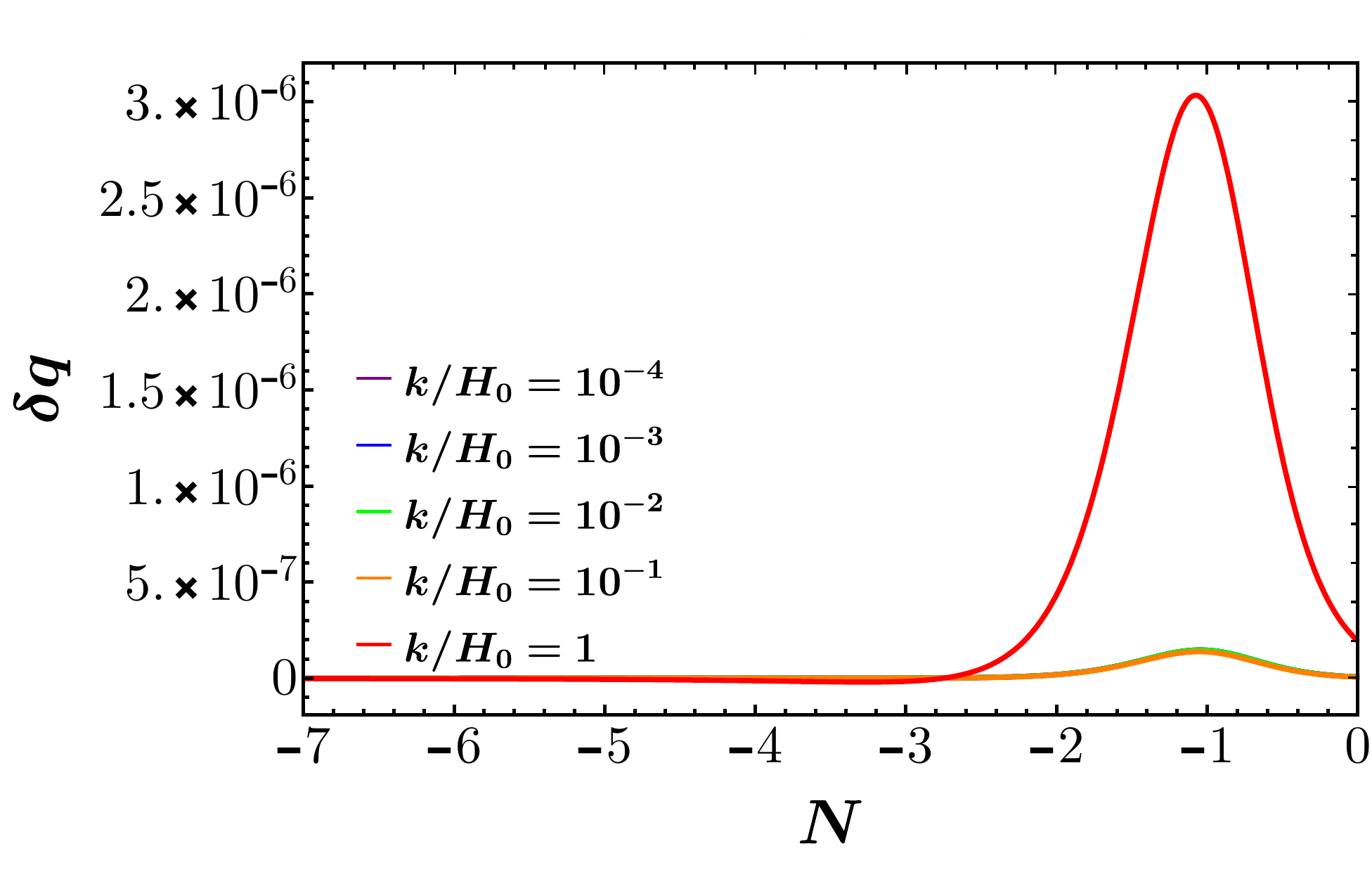}
\end{minipage}\hfill
\begin{minipage}[b]{.45\textwidth}
\includegraphics[scale=0.4]{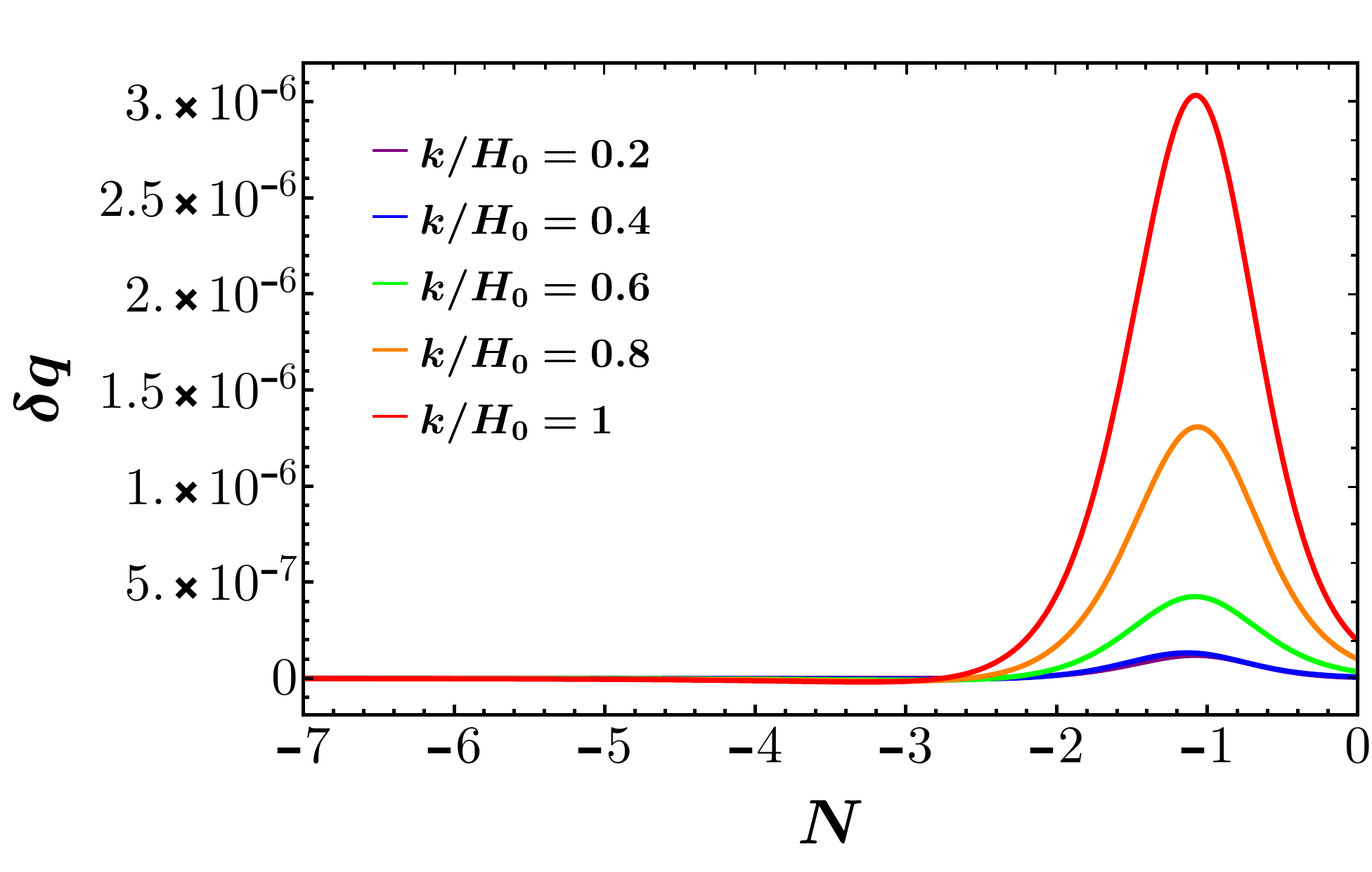}
\end{minipage}
\caption{Evolution of $\delta q$ as a function of $N$ for different values of $k$, with $C=-0.6$.}
\label{fig:dqevophi1}
\end{figure}
Figure \ref{fig:dqevophi1} is the plot of $\delta q$ as a function of number of e-foldings ($N$) for different $k$ values. Since this forms the basis of the rest of the analysis, we would like to stress the following points: First,  we see that the interaction function peaks around $N\sim -1$ ($z\sim 1.5 - 2.5$), and the interaction increases with increasing values of $k$. Second, 
since the interaction in the dark sector is a local interaction, the effect of the interaction should be least at the largest length scales (smallest $k$), and this is what we see from the plots. In other words, the interaction strength introduces a new length scale in the dynamics and leads to a preference for the growth of perturbations in certain length scales. We will see this feature for all the three quantities $\delta_m$, $\Phi + \Psi$ and $\Phi' + \Psi'$.

In the following subsections, we obtain the evolution of the perturbed quantities relevant to the upcoming cosmological observations and determine the constraints to distinguish the interacting dark sector model from standard cosmology.

\subsection{Structure formation}

Over the last few decades, the three-dimensional distribution of galaxies is available due to many surveys. With the redshift measurement of millions of galaxies, there are two key conclusions: First, if we smoothen the distribution on the largest scales, it approaches a homogeneous distribution consistent with the FRW model. Second, in the smaller scales, there are overdense regions (clusters) and underdense regions (voids); around $10~\rm{Mpc}$, the RMS density-fluctuation amplitude is of the order unity. Since the interaction function, $\delta q$
increases with increasing values of $k$, we can expect that the cold matter density 
perturbations in our model may have a different profile compared to standard cosmology. 

Hence, first we look at the evolution of the matter density perturbation $\delta_m$. More specifically, $\delta_m$, $\Delta \delta_m$ and $\Delta \delta_{m_{rel}}$ defined in Eq. \eqref{def:Deltam}. 

\begin{figure}[!htb]
\begin{minipage}[b]{.45\textwidth}
\includegraphics[scale=0.4]{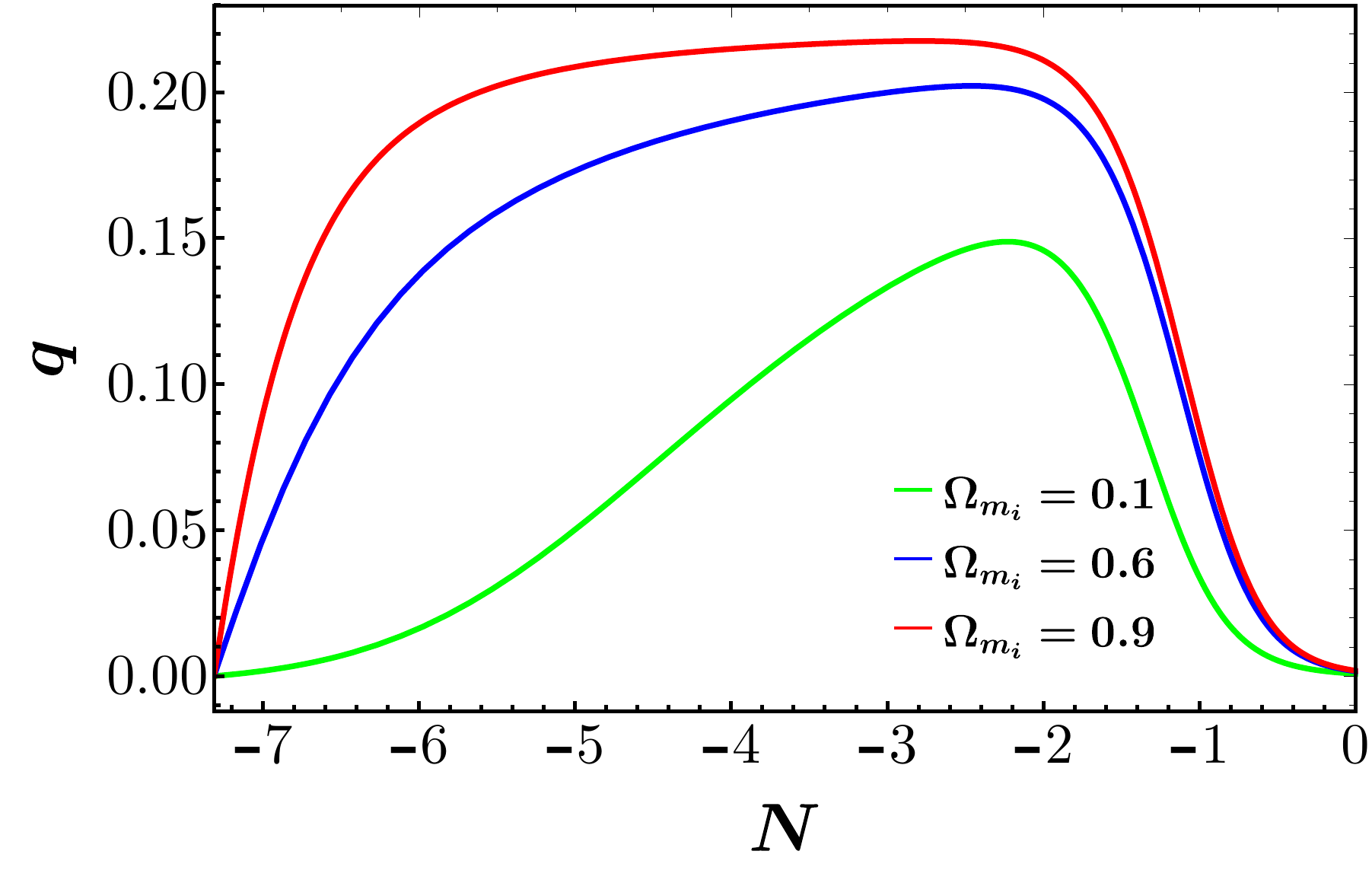}
\end{minipage}\hfill
\begin{minipage}[b]{.45\textwidth}
\includegraphics[scale=0.4]{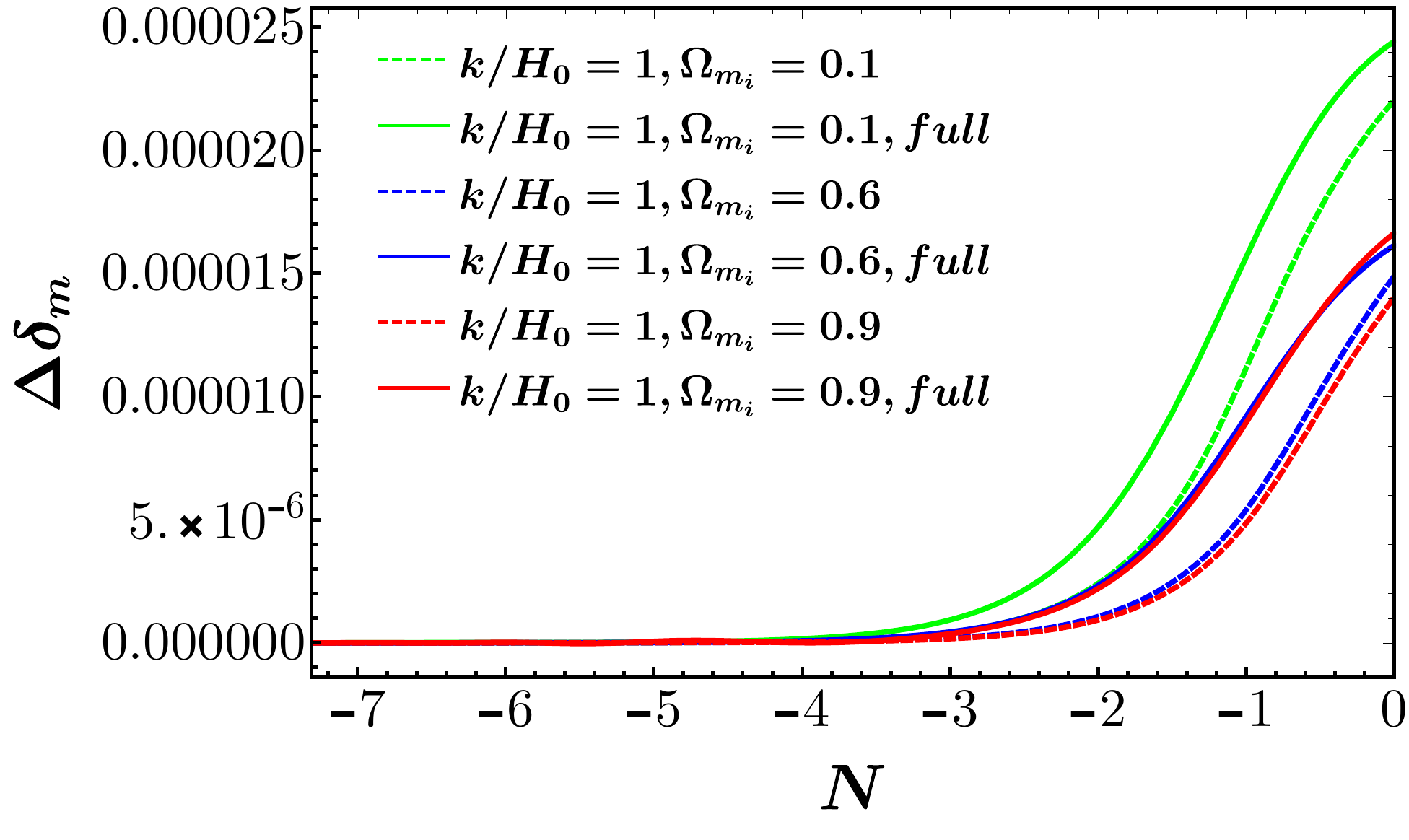}
\end{minipage}
\caption{Evolution of $q$ (left panel) and $\Delta \delta_m$ (right panel) as a function of $N$ with $C=-0.6$.}
\label{fig:partialint}
\end{figure}

To gain a physical understanding of the effect of the interaction term on the evolution of the perturbed quantities, we consider an approximation in which the perturbed interaction term is switched off ($\delta q$ = 0). The right panel in Figure \ref{fig:partialint} contains the evolution of $\Delta \delta_m$ with and without this approximation. Dashed lines refer to the evolution with the approximation, and the solid lines refer to the full evolution (without any approximation).

The background interaction term $q \equiv \alpha \beta x \Omega_m$ determines the rate of growth of $\delta_m$. Larger the initial value of $\Omega_m$, the value of $q$ increases at an earlier epoch and stays at a higher value till $N \sim -2 (z\sim 7.4)$. This behaviour can be seen in the left panel of Figure \ref{fig:partialint}.  The right panel of Figure \ref{fig:partialint} contains the evolution of $ \Delta \delta_m$ .Here, we see that the growth of $\delta_m$ is suppressed for the larger initial values of $\Omega_m$. A better analytical understanding will shine a light on the role of $Q^{F}$ in the evolution of the perturbed quantities.

In the rest of this section, we numerically evolve the Eqs.~\ref{eq:perturbed} and obtain the evolution of the perturbed quantities.

\begin{figure}[!htb]
\begin{minipage}[b]{.45\textwidth}
\includegraphics[scale=0.4]{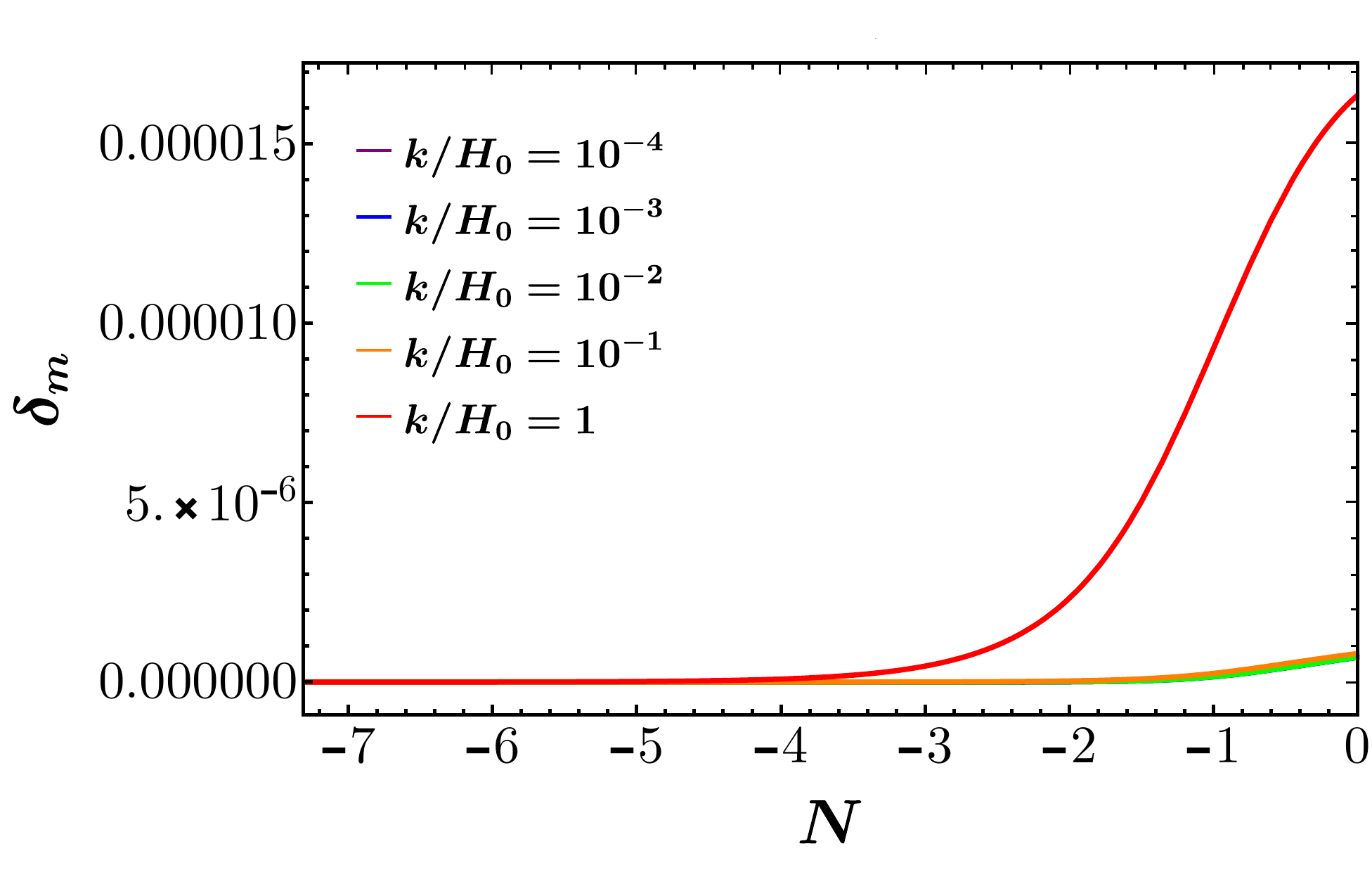}
\end{minipage}\hfill
\begin{minipage}[b]{.45\textwidth}
\includegraphics[scale=0.4]{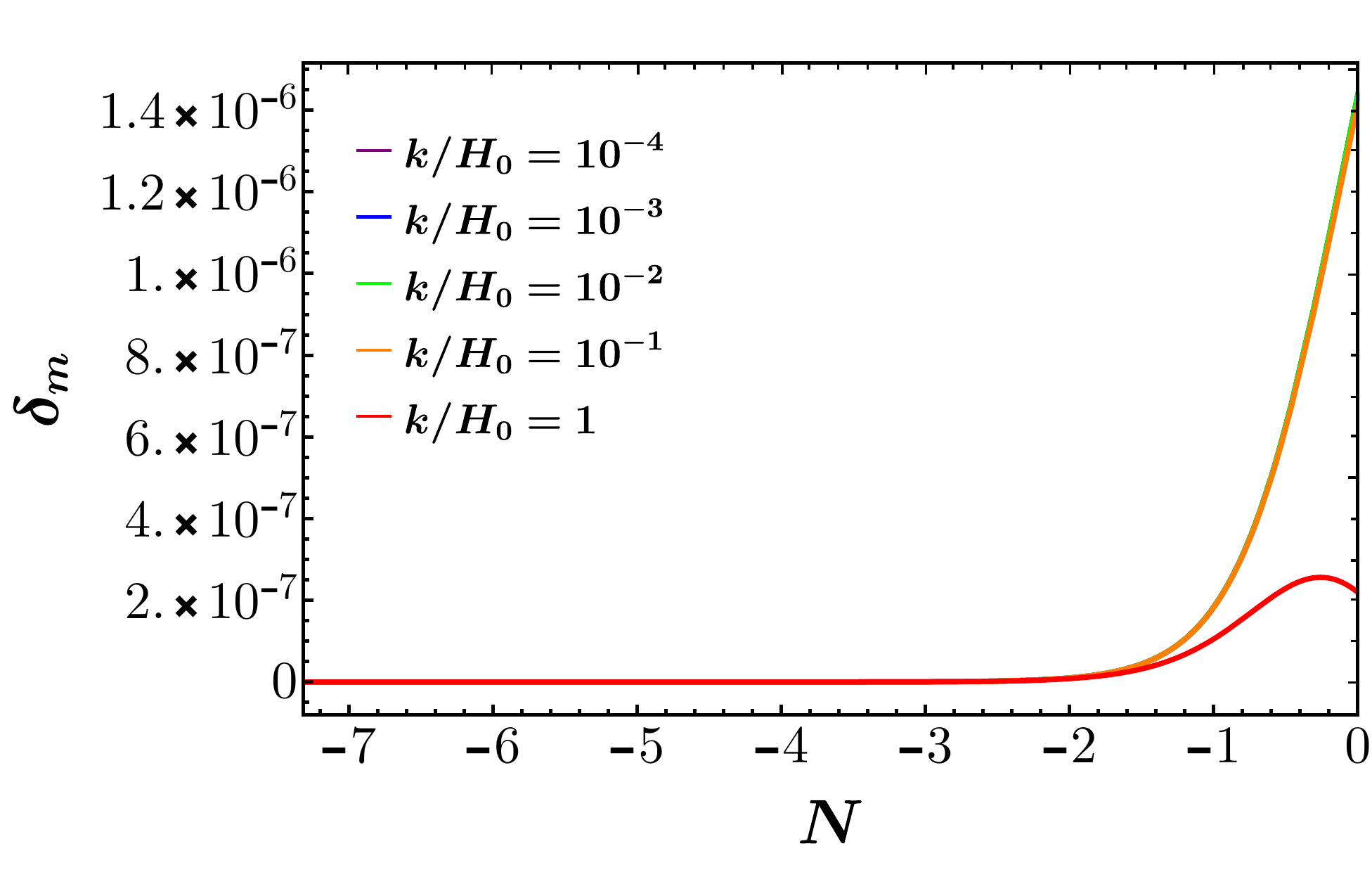}
\end{minipage}
\caption{Evolution of $\delta_m$ as a function of $N$. Left: $C=-0.6$, Right: $C=0$.}

\label{fig:deltaevophi1}
\end{figure}

\begin{figure}[!htb]
\begin{minipage}[b]{.45\textwidth}
\includegraphics[scale=0.4]{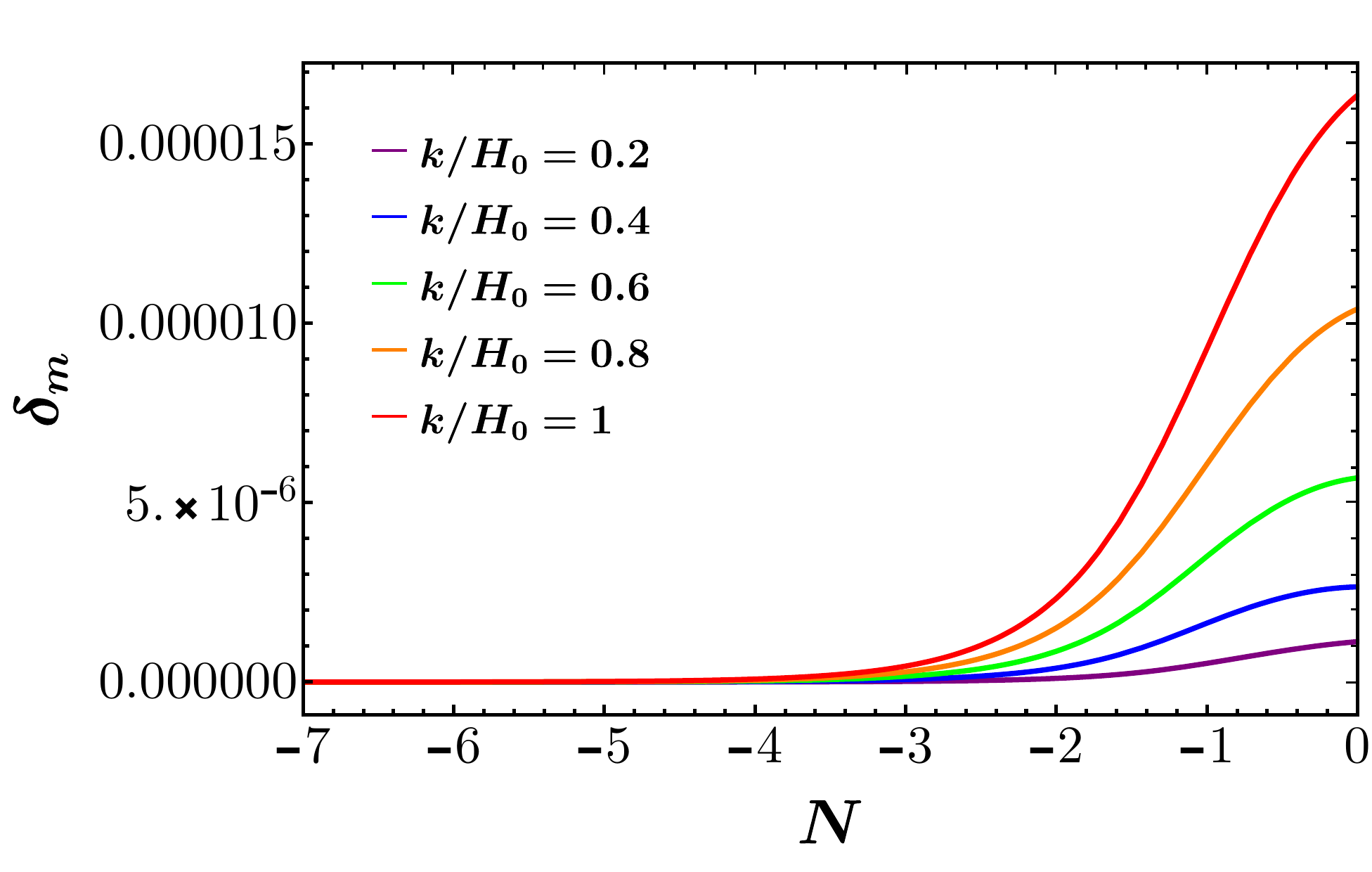}
\end{minipage}\hfill
\begin{minipage}[b]{.45\textwidth}
\includegraphics[scale=0.4]{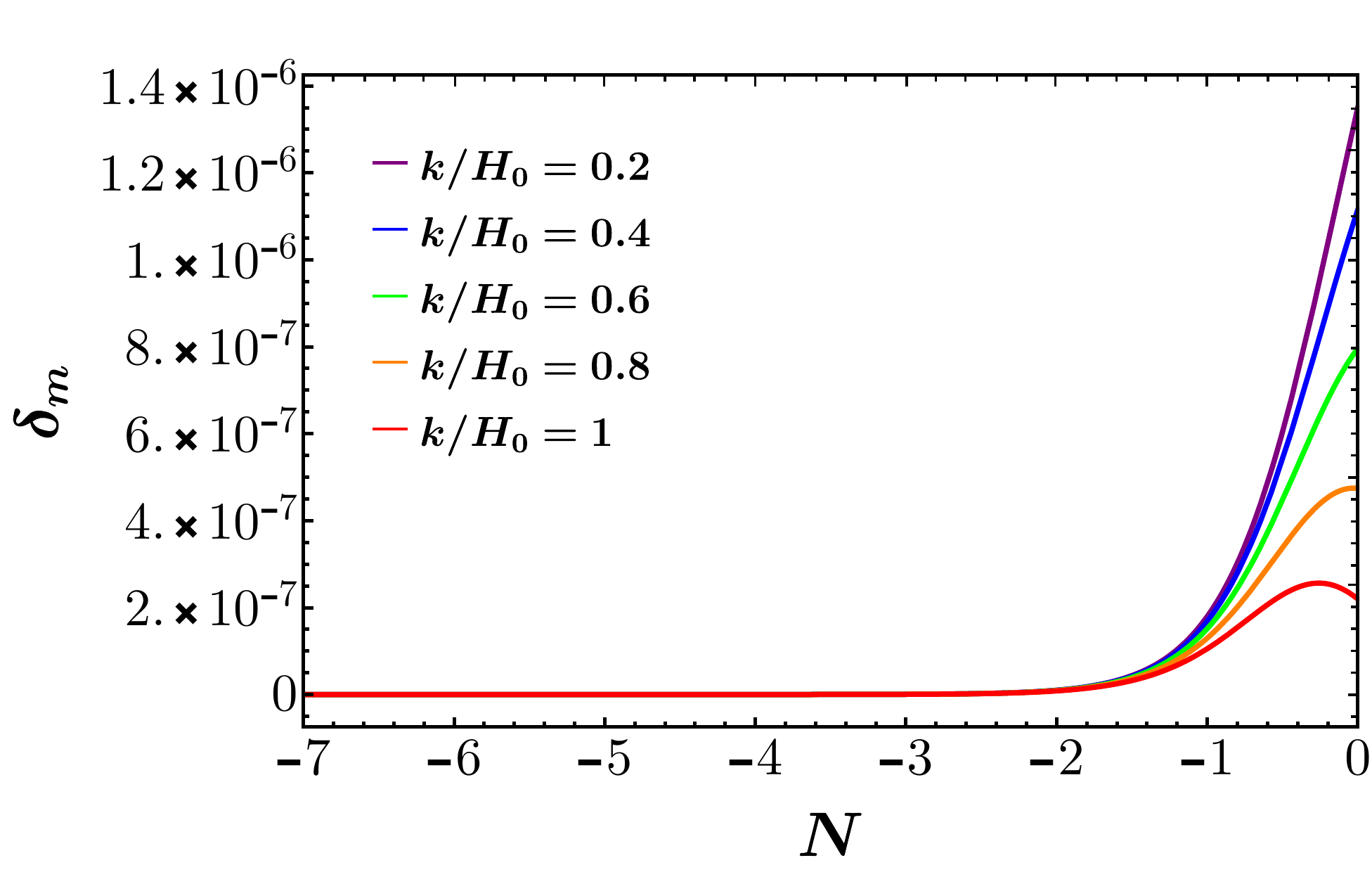}
\end{minipage}
\caption{Evolution of $\delta_m$ as a function of $N$. Left: $C=-0.6$, Right: $C=0$.}
\label{fig:deltaevozoomphi1}
\end{figure}

\begin{figure}[!htb]
\begin{minipage}[b]{.45\textwidth}
\includegraphics[scale=0.4]{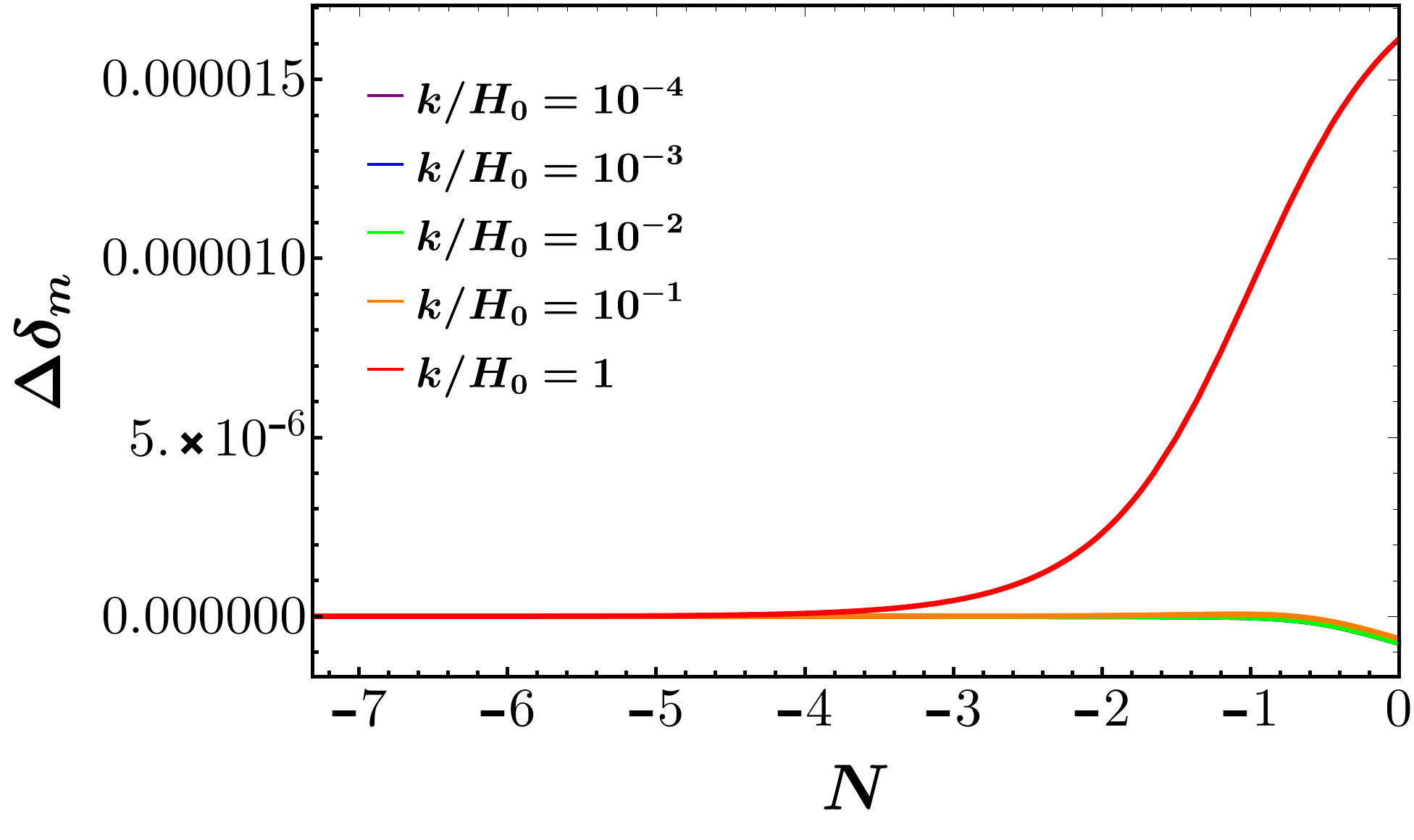}
\end{minipage}\hfill
\begin{minipage}[b]{.45\textwidth}
\includegraphics[scale=0.4]{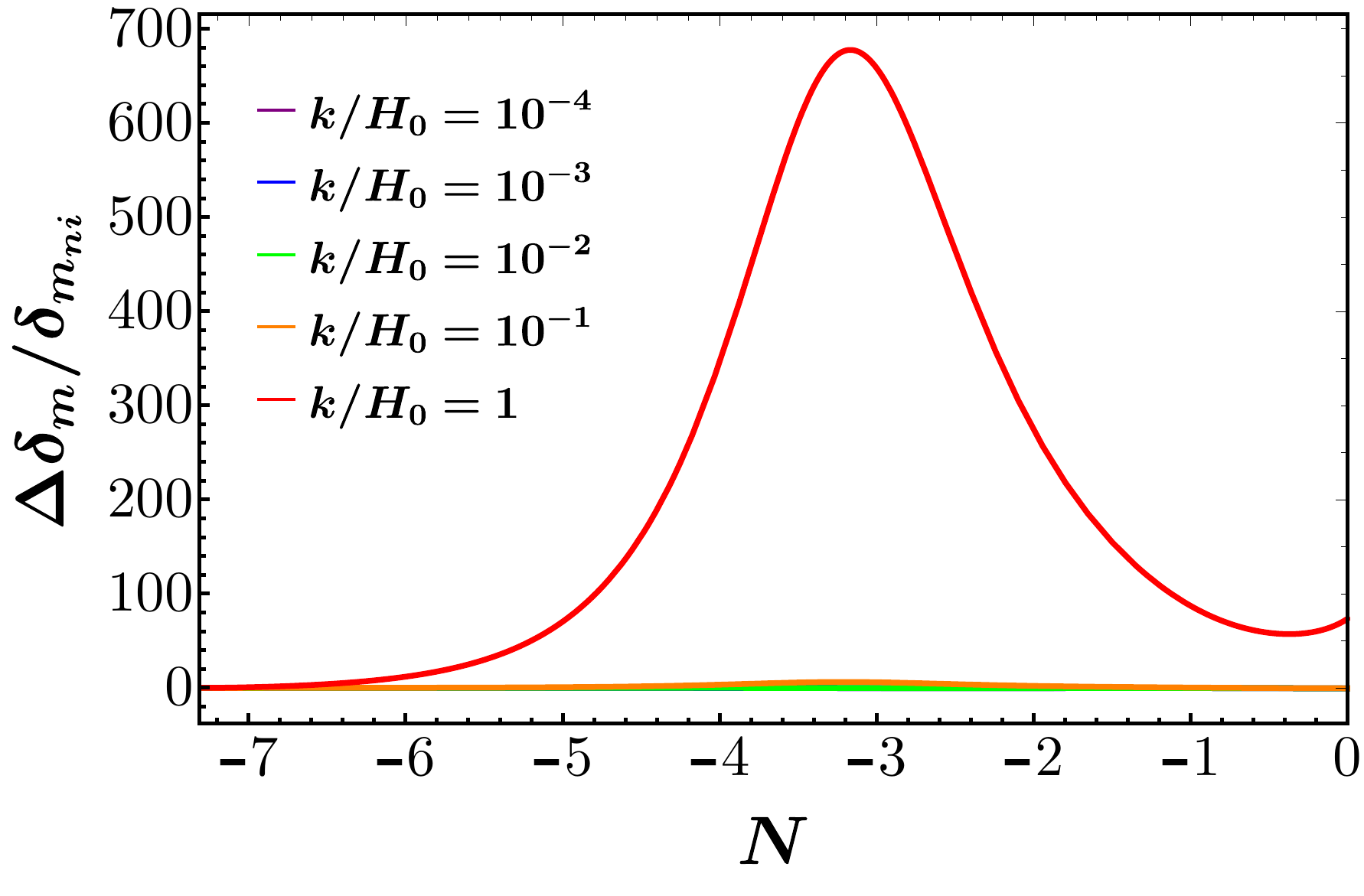}
\end{minipage}
\caption{Evolution of $\Delta \delta_m$ (left), $\Delta \delta_m/ \delta_{m_{ni}}$ (right)  as a function of $N$.}
\label{fig:ddeltaphi1}
\end{figure}

\begin{figure}[!htb]
\begin{minipage}[b]{.45\textwidth}
\includegraphics[scale=0.4]{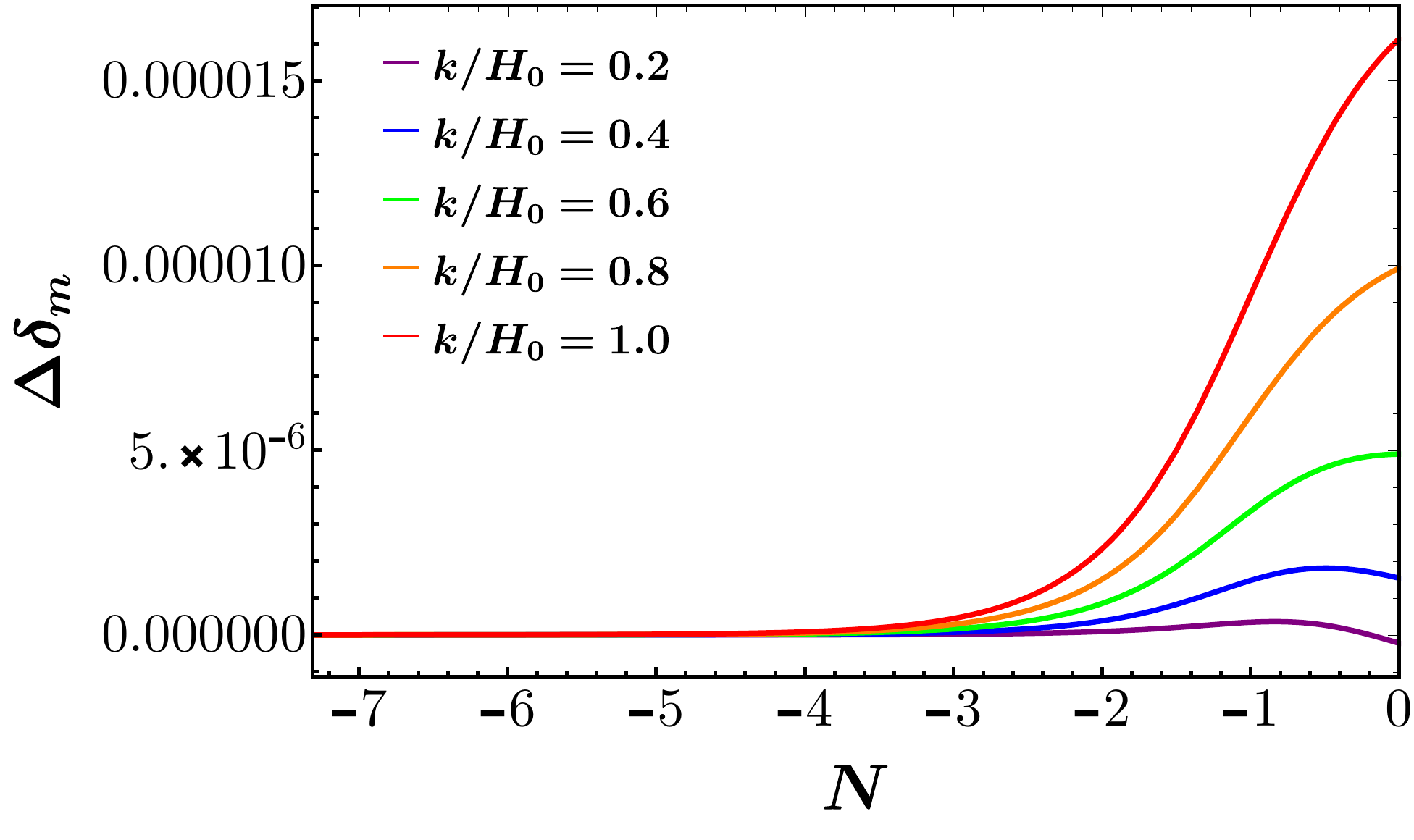}
\end{minipage}\hfill
\begin{minipage}[b]{.45\textwidth}
\includegraphics[scale=0.4]{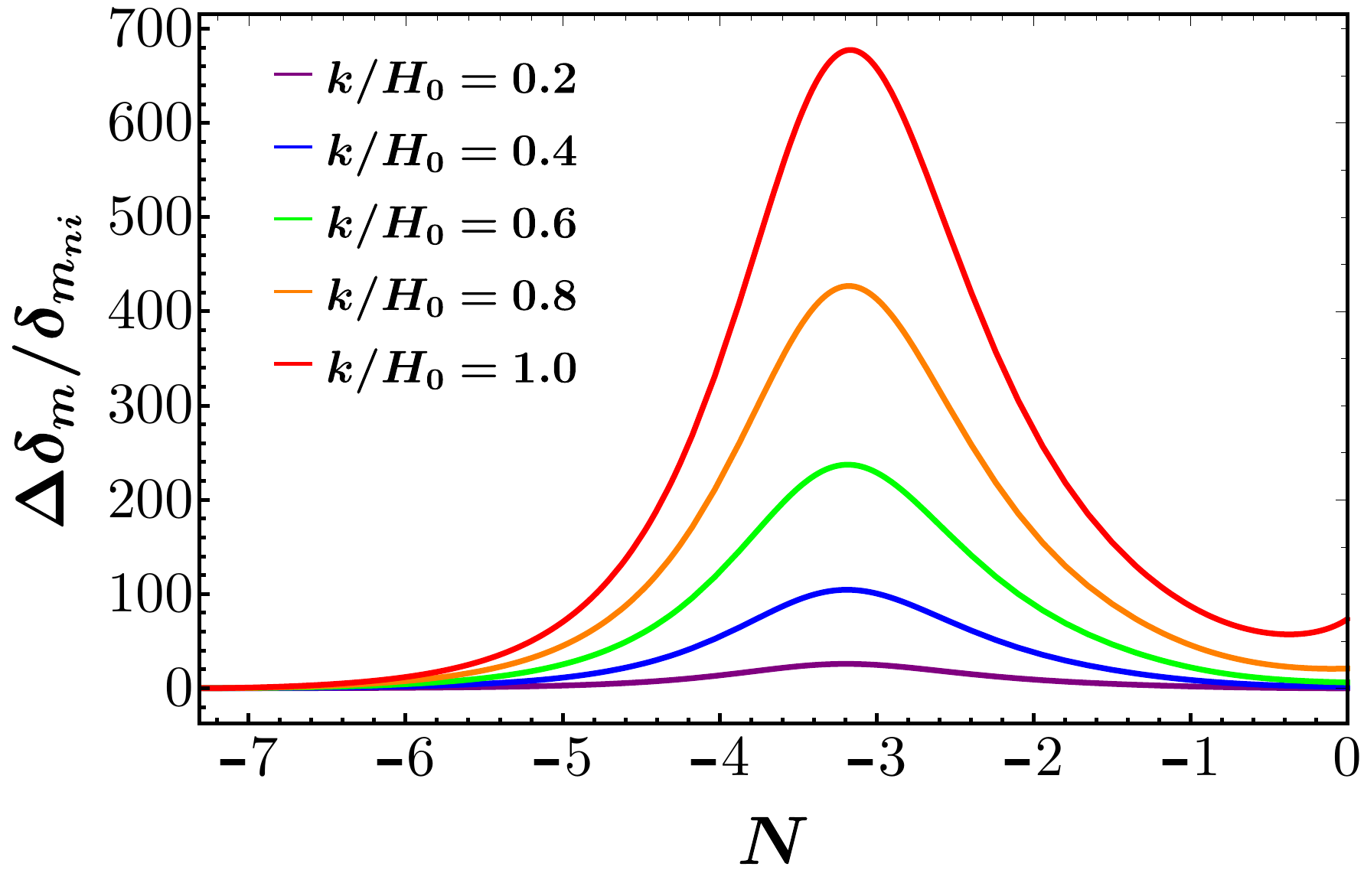}
\end{minipage}
\caption{Evolution of $\Delta \delta_m$ (left), $\Delta \delta_m/ \delta_{m_{ni}}$ (right)  as a function of $N$.}
\label{fig:ddeltazoomphi1}
\end{figure}

Figures \ref{fig:deltaevophi1} and  \ref{fig:deltaevozoomphi1} [\ref{fig:ddeltaphi1}  and \ref{fig:ddeltazoomphi1}]
contain plots of $\delta_m$ [$\Delta \delta_m$, $\Delta \delta_{m_{rel}}$] as a function of $N$ for different length scales in interacting and non-interacting scenarios.

From these plots, we infer the following: First, the difference in the evolution of $\delta_m$ between the interacting and non-interacting scenarios is significant after $N \sim -3$. Second, this difference increases with the increase in the value of the wavenumber $k$. This means that the interaction has a more significant effect on the evolution of the scalar perturbations in the smaller length scales (large values of $k$) than the larger length scales (smaller values of $k$). Third, these deviations become significant for $z \sim 10-20$ and lie in the epoch of reionization. During this epoch, a predominantly neutral intergalactic medium was ionized by the emergence of the first luminous sources. Before the reionization epoch, the formation and evolution of structure were dominated by dark matter alone. However, the interacting dark sector leads to the exchange of density perturbations at smaller length scales. This indicates that it will be possible to detect the signatures of dark energy - dark matter interaction in large-scale structure observations. This provides a possible way to detect the signatures of dark sector interaction in the existing and upcoming cosmological observations like Euclid satellite~\cite{2018-Amendola.Others-LivingRev.Rel.}, GMRT~\cite{2017-Gupta.etalCurrent-Science}, SKA~\cite{2015-Maartens.etal-PoS} and LOFAR \cite{2013-Haarlem.others-Astron.Astrophys.}.

\subsection{Weak gravitational lensing}

The matter content of the Universe is dominated by dark matter. Most of the cosmological observations to study the matter distribution in the Universe depend on the observations of the luminous matter, which gives us little information regarding the total mass distribution in the Universe. Gravitational lensing provides important information regarding the total mass distribution in the Universe, as it is independent of the nature of the matter and its interaction with electromagnetic radiation. Hence, weak gravitational lensing holds enormous promise as it can reveal the distribution of dark matter independently of any assumptions about its nature. 
The quantity $\Phi+\Psi$ determines the geodesic of a photon, which affects the weak gravitational lensing~\cite{2005-Mukhanov-PhysicalFoundationsCosmology}. 
Like the standard cosmology, for the dark-sector interacting model considered here, 
$\Phi(t,x,y,z) = \Psi(t,x,y,z)$. Hence, it is sufficient to study the evolution of $\Phi$ to distinguish the dark sector model from standard cosmology.

\begin{figure}[!htb]
\begin{minipage}[b]{.45\textwidth}
\includegraphics[scale=0.4]{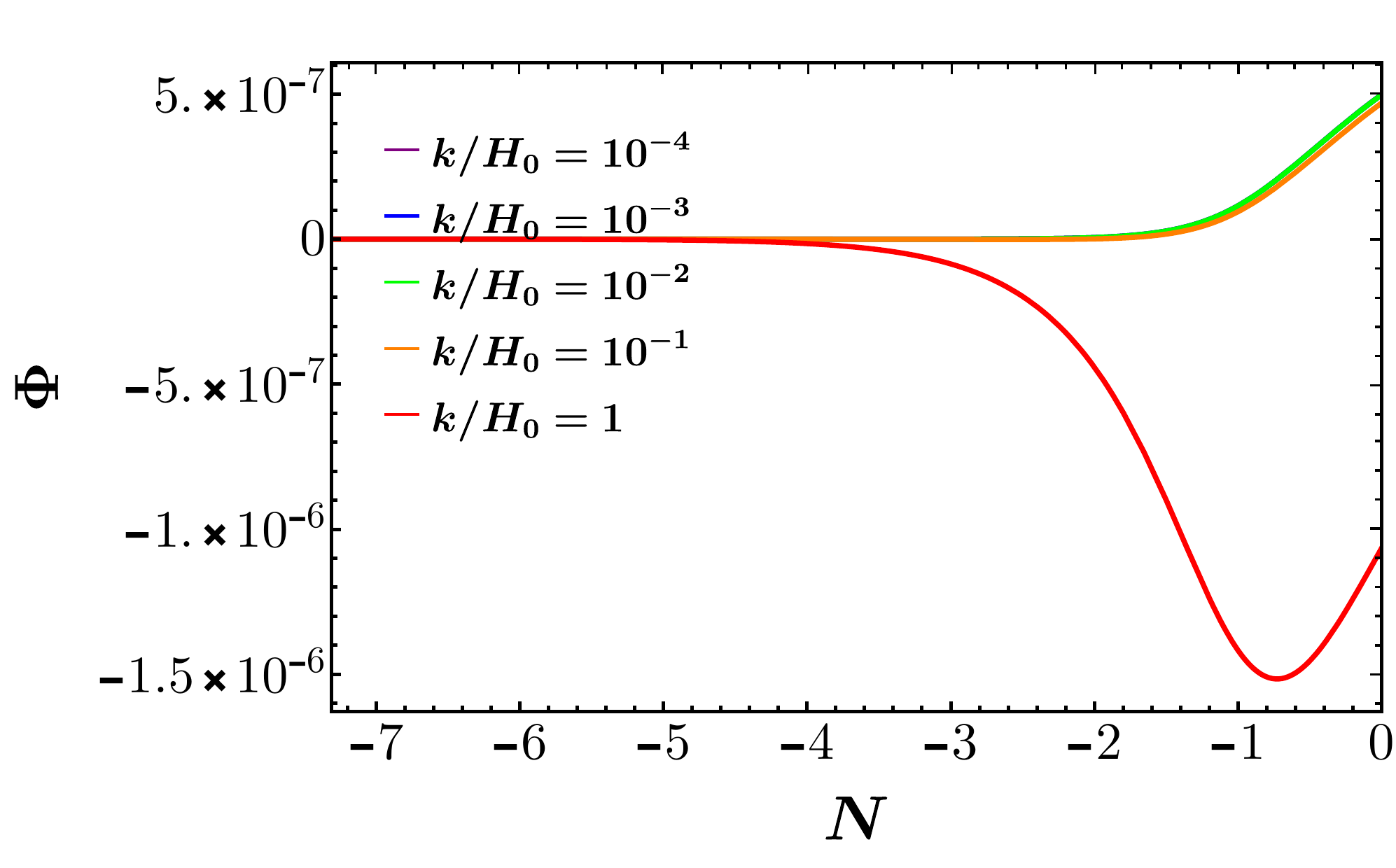}
\end{minipage}\hfill
\begin{minipage}[b]{.45\textwidth}
\includegraphics[scale=0.4]{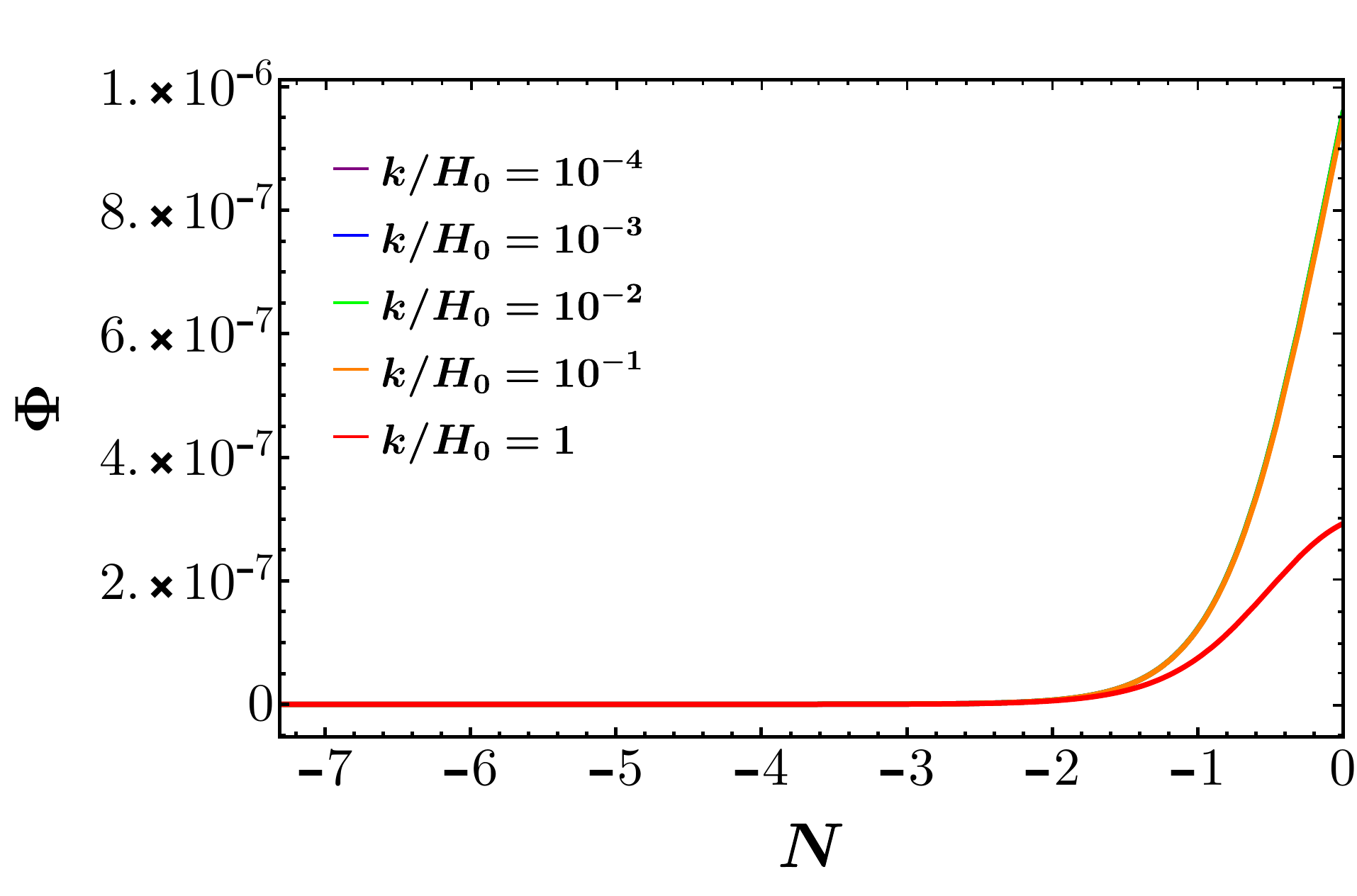}
\end{minipage}
\caption{Evolution of $\Phi$ as a function of $N$. Left: $C=-0.6$, Right: $C=0$.}
\label{fig:Phievophi1}
\end{figure}
\begin{figure}[!htb]
\begin{minipage}[b]{.45\textwidth}
\includegraphics[scale=0.4]{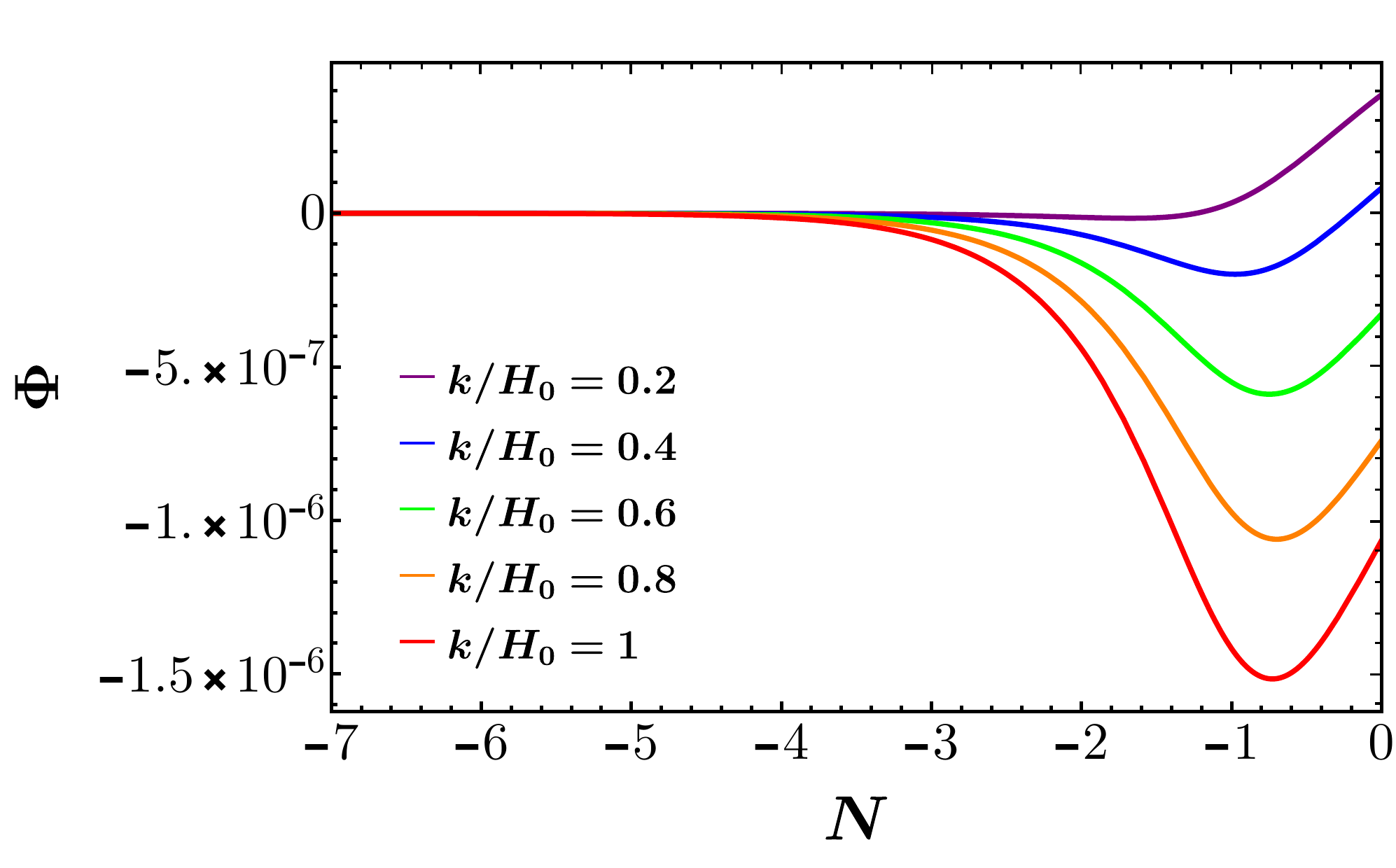}
\end{minipage}\hfill
\begin{minipage}[b]{.45\textwidth}
\includegraphics[scale=0.4]{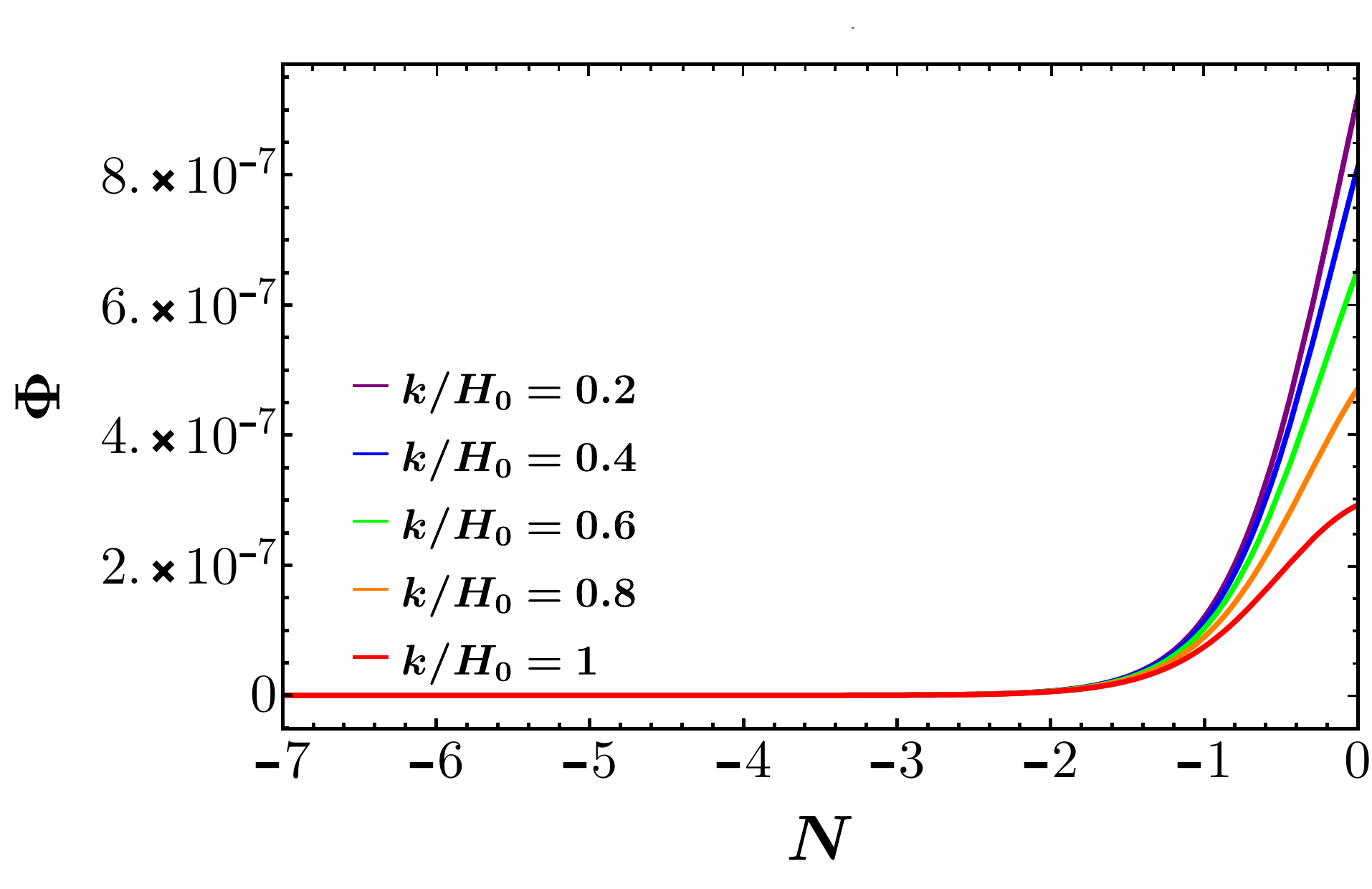}
\end{minipage}
\caption{Evolution of $\Phi$ as a function of $N$. Left: $C=-0.6$, Right: $C=0$.}
\label{fig:Phievozoomphi1}
\end{figure}

\begin{figure}[!htb]
\begin{minipage}[b]{.45\textwidth}
\includegraphics[scale=0.4]{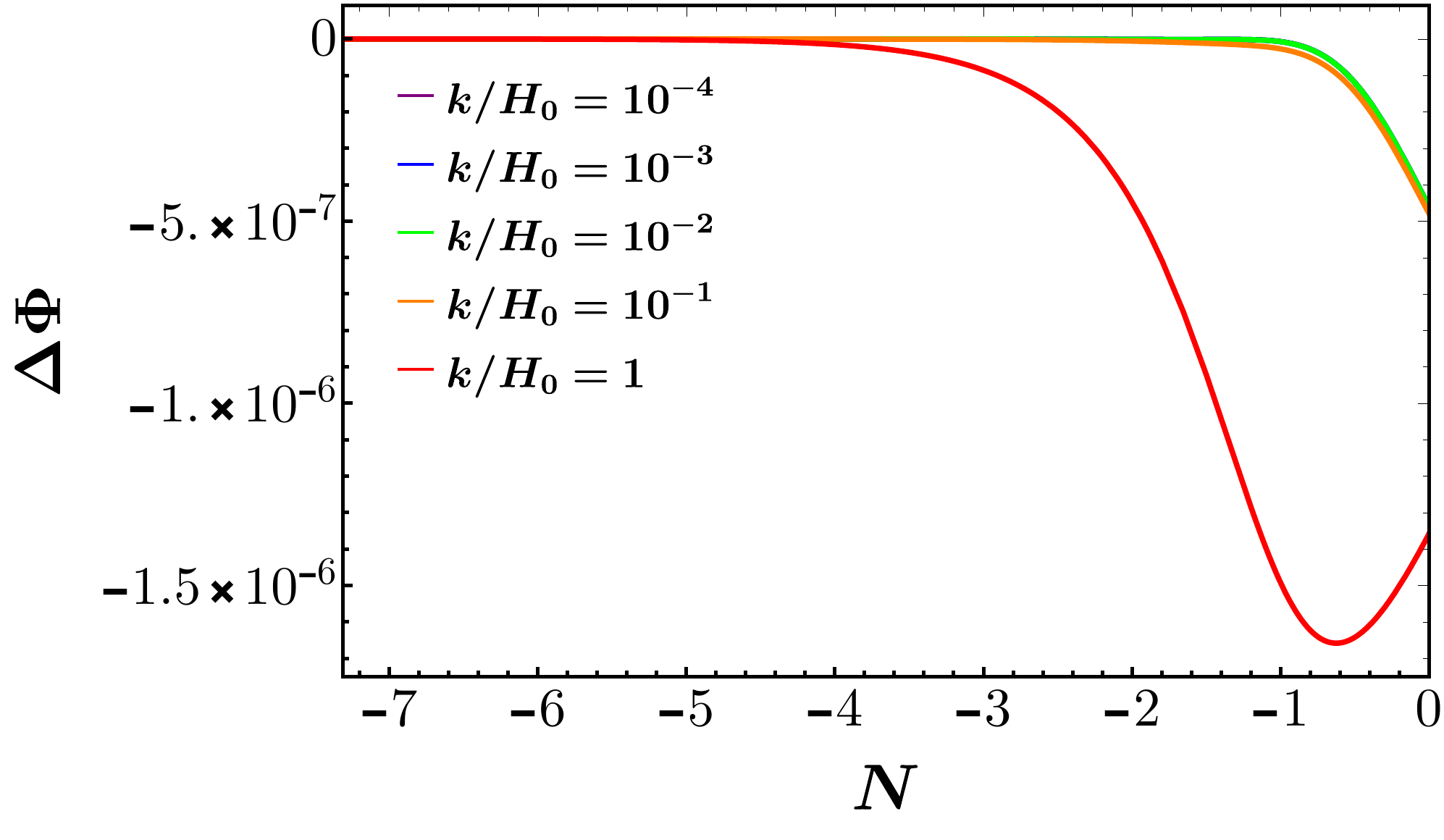}
\end{minipage}\hfill
\begin{minipage}[b]{.45\textwidth}
\includegraphics[scale=0.4]{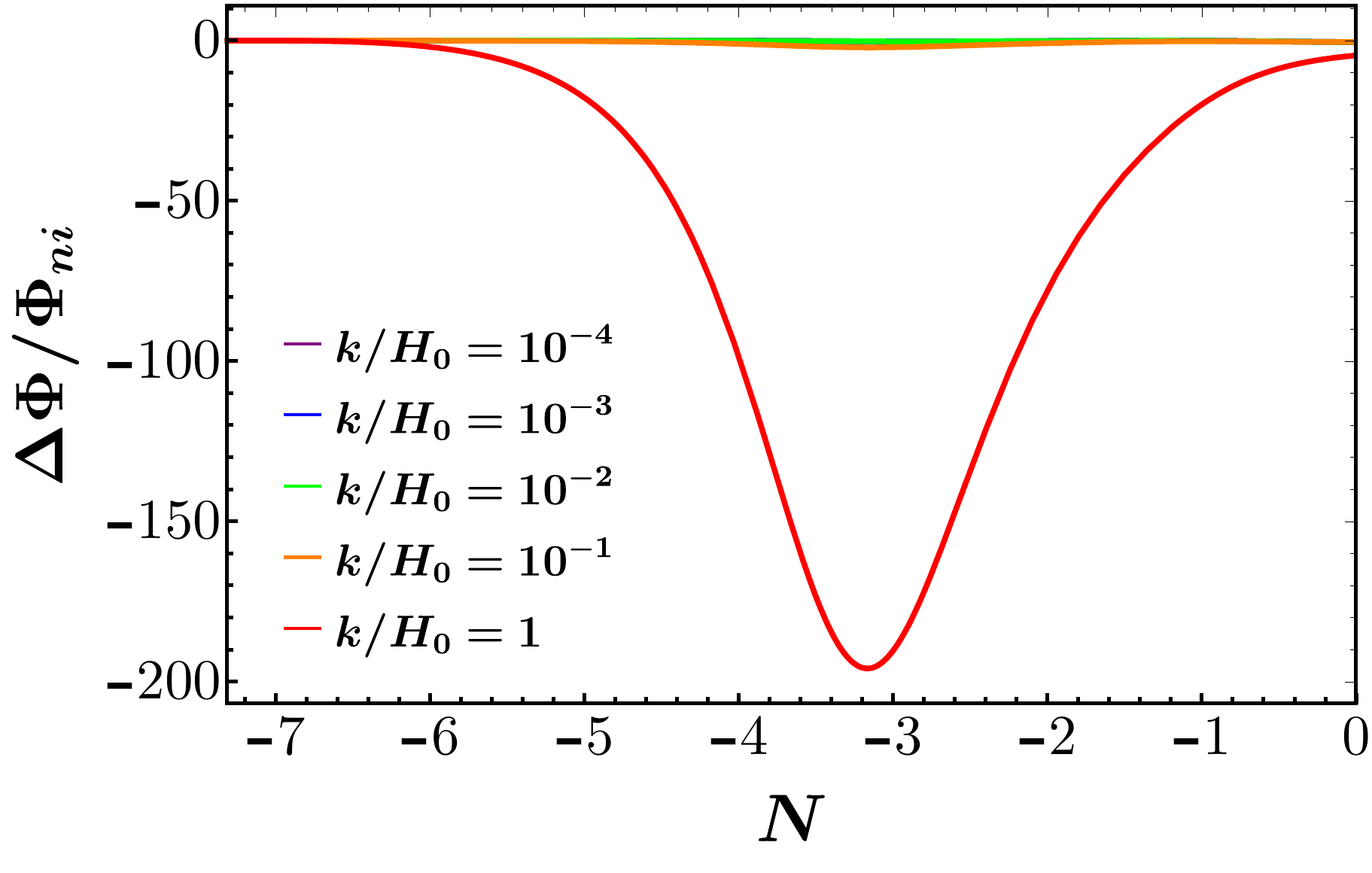}
\end{minipage}
\caption{Evolution of $\Delta \Phi$ (left), $\Delta \Phi/ \Phi_{{ni}}$ (right)  as a function of $N$.}
\label{fig:dlensphi1}
\end{figure}

\begin{figure}[!htb]
\begin{minipage}[b]{.45\textwidth}
\includegraphics[scale=0.4]{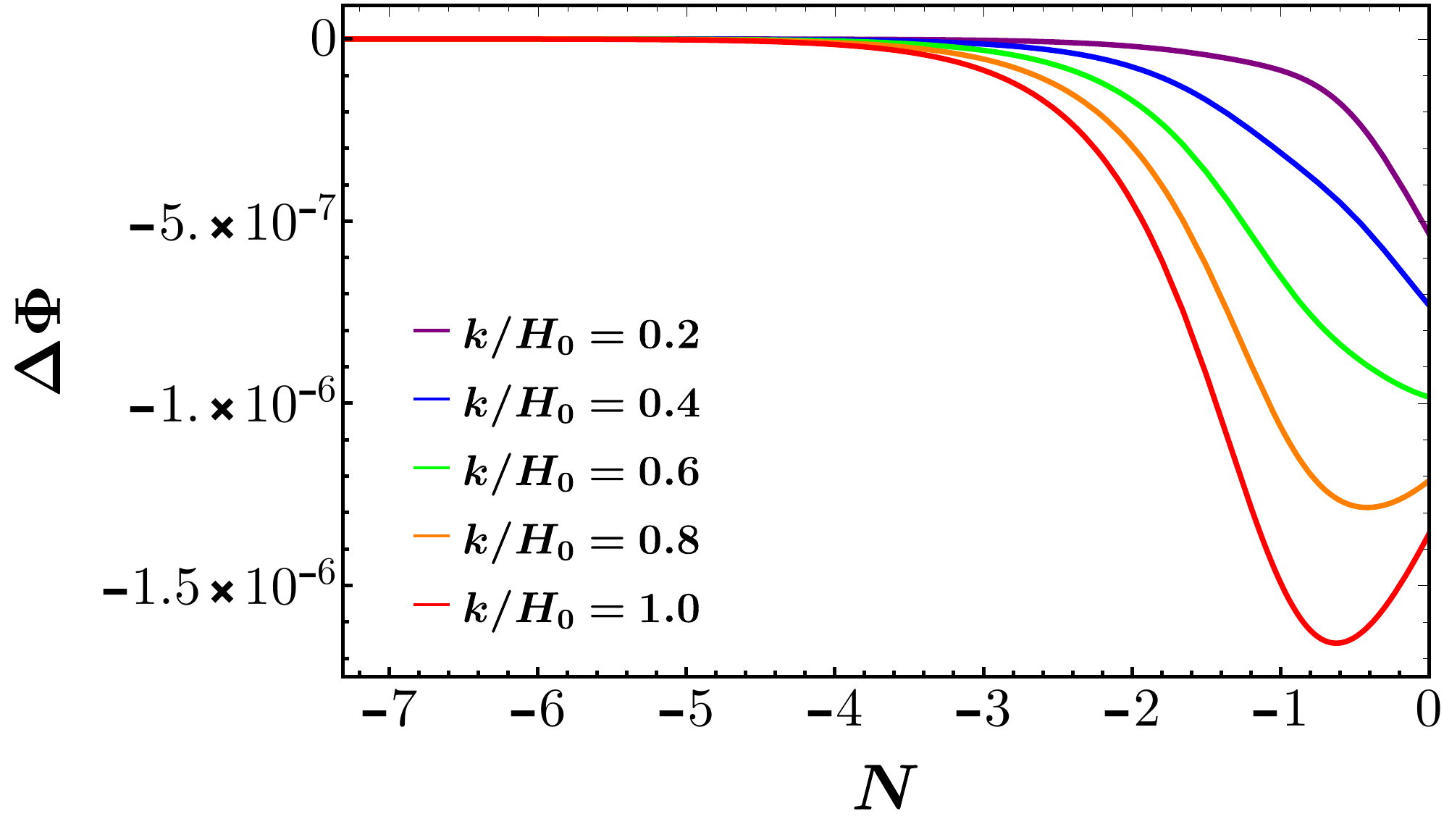}
\end{minipage}\hfill
\begin{minipage}[b]{.45\textwidth}
\includegraphics[scale=0.4]{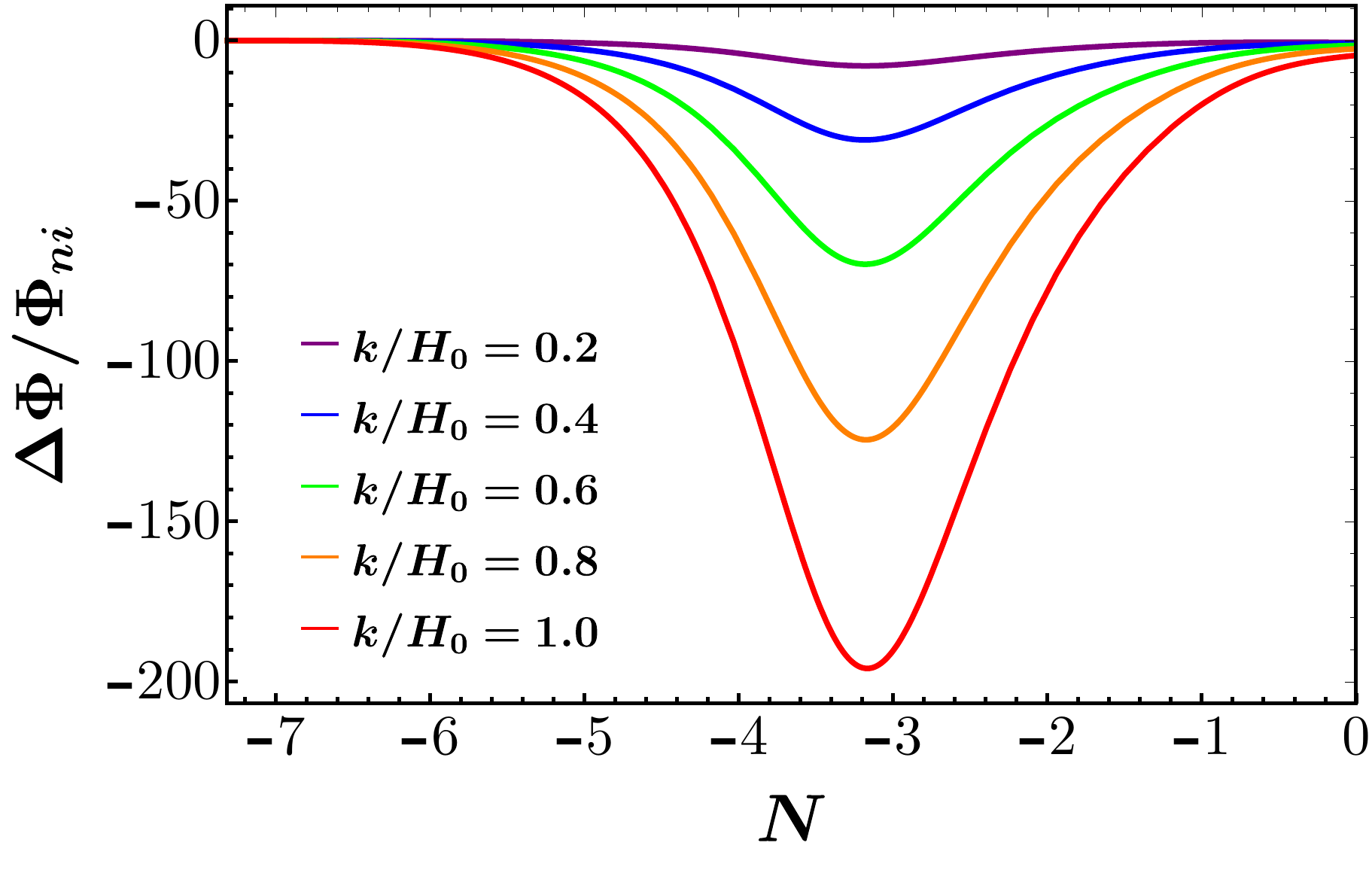}
\end{minipage}
\caption{Evolution of $\Delta \Phi$ (left), $\Delta \Phi/ \Phi_{{ni}}$ (right)  as a function of $N$.}
\label{fig:dlensrelphi1}
\end{figure}

To study the signatures of the interacting dark sector, we look at the evolution of scalar metric perturbation $\Phi$ for different length scales starting from $z \sim 1500$. To analyze the difference in the evolution of $\Phi$ in the two scenarios, we also look at $\Delta \Phi$ and $\Delta \Phi_{rel}$. Figures \ref{fig:Phievophi1} and \ref{fig:Phievozoomphi1} contain plots of $\Phi$ as a function of $N$ for different length scales in interacting and non-interacting scenarios. Evolution of $\Delta \Phi$ and $\Delta \Phi_{rel}$ as a function of $N$ 
are plotted in Figures \ref{fig:dlensphi1} and \ref{fig:dlensrelphi1}.

From the evolution of these quantities, we see that starting from the same initial conditions at $z \sim 1500$, the evolution of $\Phi$ begins to show the effect of dark energy - dark matter interaction at about $N \sim -3$. This effect becomes even more prominent towards the lower redshifts $z < 5$. By looking at the $k-$dependence of the evolution, this effect is enhanced at lower length scales. 
This means that the interaction has a larger effect on the evolution of the scalar perturbations in the smaller length scales (large values of $k$) than the larger length scales (smaller values of $k$). Thus, this indicates that observations of weak lensing can help us potentially distinguish between interacting and non-interacting scenarios and potentially provide a way to resolve the tension between Planck-2018 and KiDS-450, KiDS-1000~\cite{2017-Hildebrandt.others-Mon.Not.Roy.Astron.Soc.,2021-Heymans.others-Astron.Astrophys.} in the $\sigma_8-\Omega_m$ plane.

\begin{figure}[!htb]
\begin{minipage}[b]{.45\textwidth}
\includegraphics[scale=0.4]{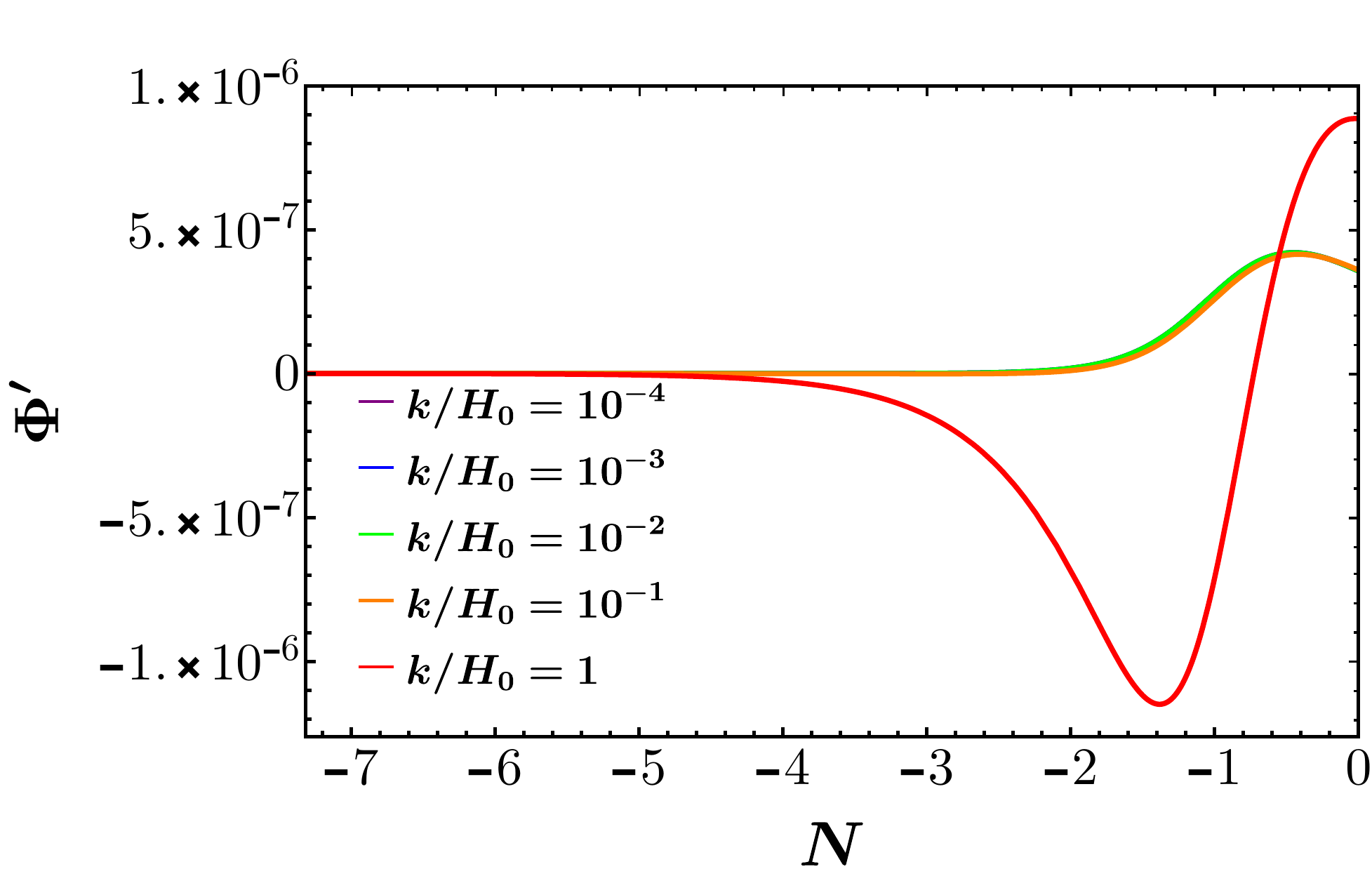}
\end{minipage}\hfill
\begin{minipage}[b]{.45\textwidth}
\includegraphics[scale=0.4]{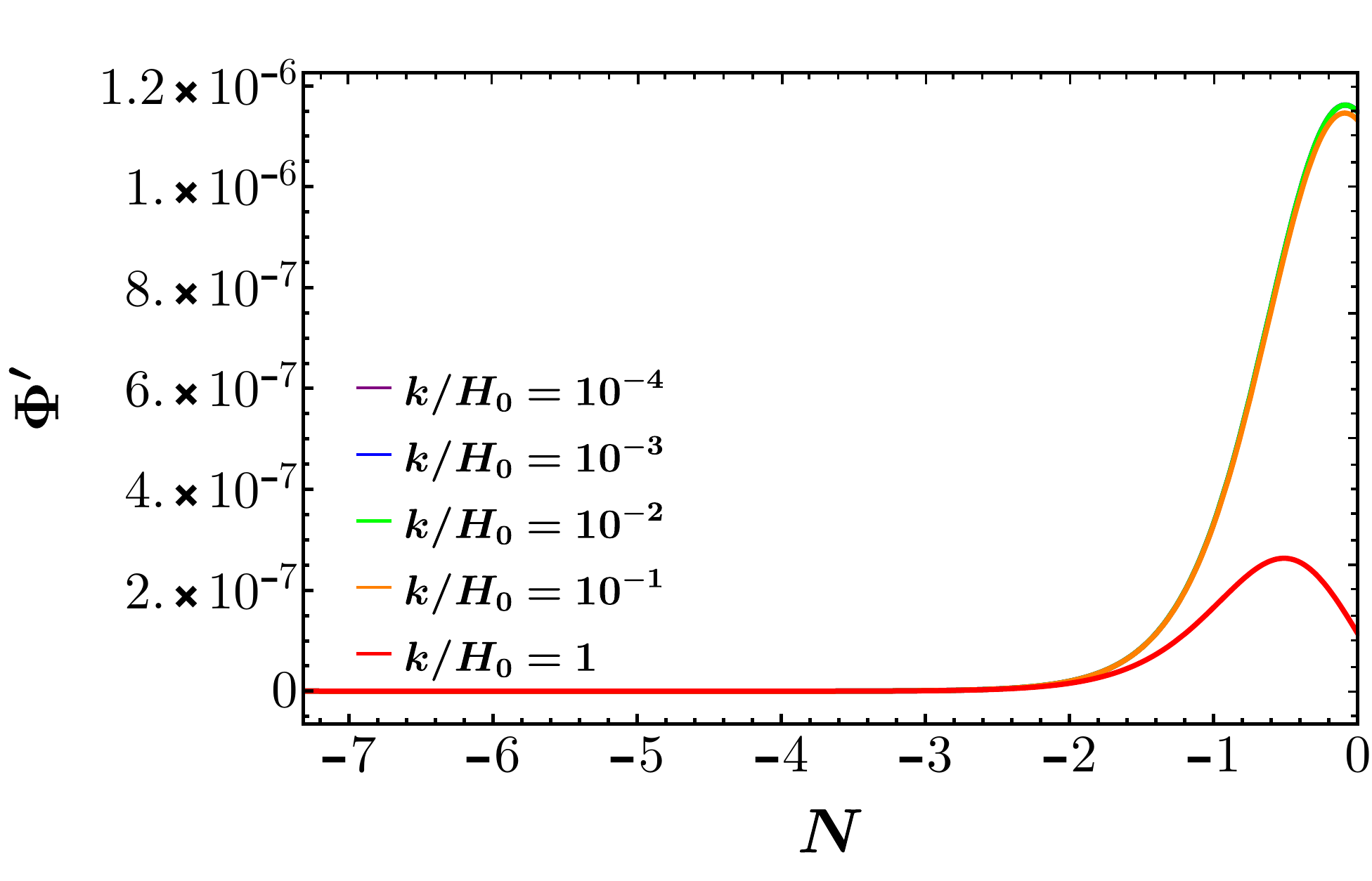}
\end{minipage}
\caption{Evolution of $\Phi'$ as a function of $N$. Left: $C=-0.6$, Right: $C=0$.}
\label{fig:iswevophi1}
\end{figure}
\begin{figure}[!htb]
\begin{minipage}[b]{.45\textwidth}
\includegraphics[scale=0.4]{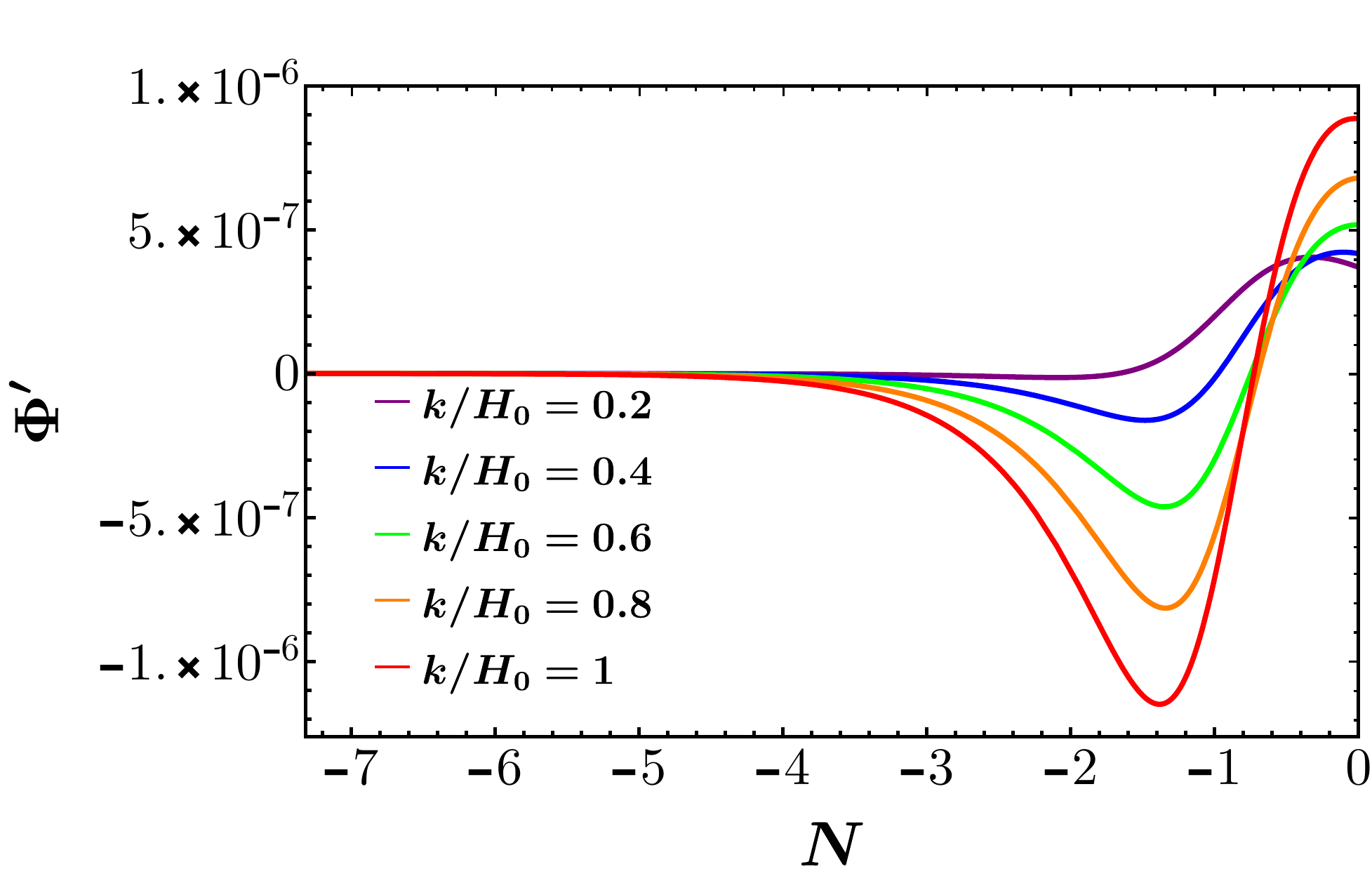}
\end{minipage}\hfill
\begin{minipage}[b]{.45\textwidth}
\includegraphics[scale=0.4]{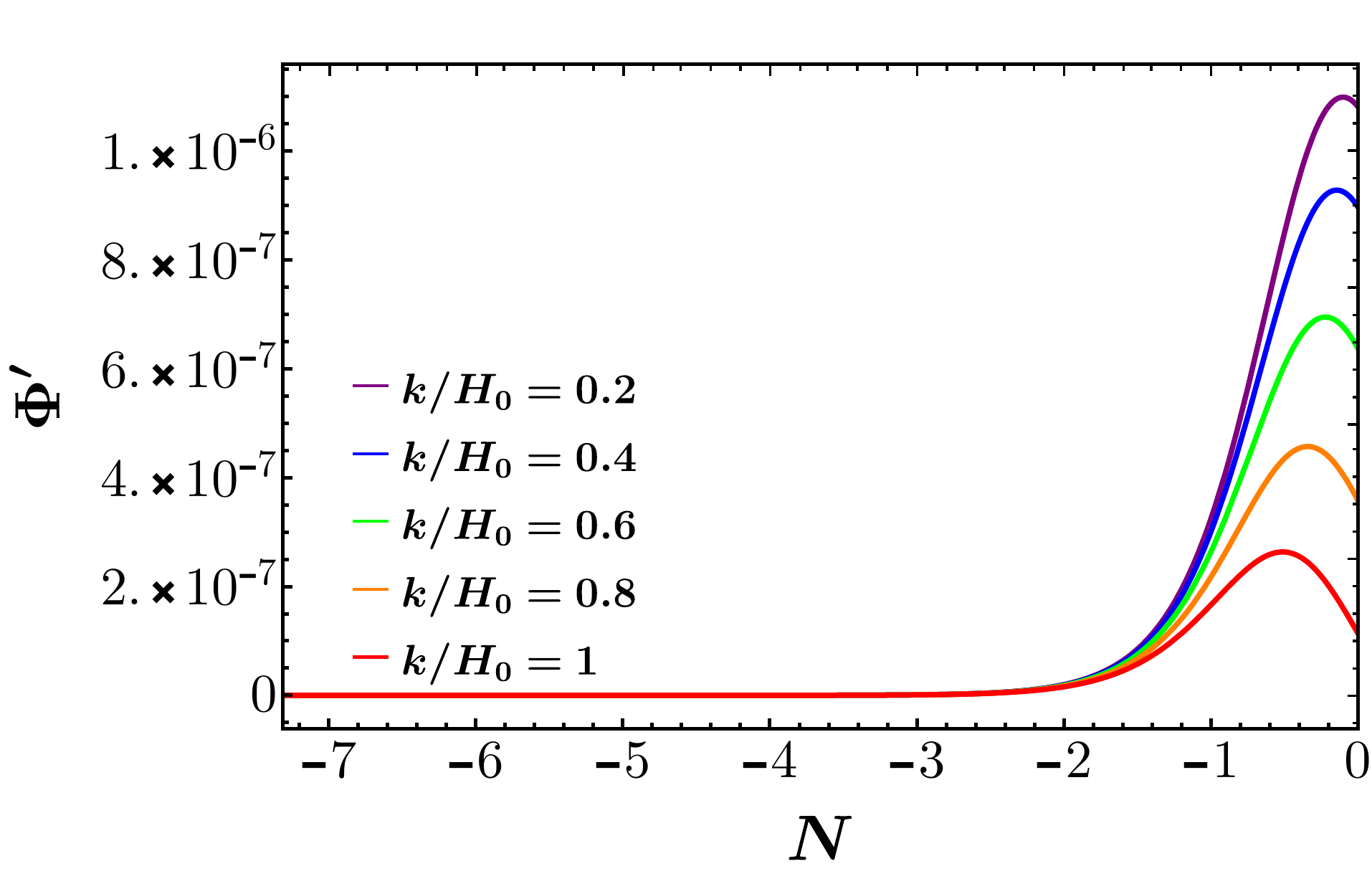}
\end{minipage}
\caption{Evolution of $\Phi'$ as a function of $N$. Left: $C=-0.6$, Right: $C=0$.}
\label{fig:iswevozoomphi1}
\end{figure}

\subsection{Integrated Sachs-Wolfe effect}

The integrated Sachs-Wolfe (ISW)  effect is a secondary anisotropy of the cosmic microwave background (CMB), which arises because of the variation in the cosmic gravitational potential between local observers and the surface of the last scattering~ \cite{2014-Nishizawa-PTEP}.
The ISW effect is related to the rate of change of ($\Phi + \Psi$) w.r.t. conformal time ($\eta$)~\cite{2011-Gorbunov.Rubakov-IntroductionTheoryEarly}. While weak gravitational lensing is determined by the spatial dependence of the metric scalar perturbation $\Phi$, the ISW effect provides valuable information about the time evolution of the same, especially in the late accelerating Universe. Even though its detectability is weaker than weak lensing, it is a powerful tool to study the underlying cosmology. It can be detected using the cross-correlation between the observational data on CMB and large-scale structures. In the flat $\Lambda$CDM model, detection of the ISW signal provides direct detection of dark energy \cite{1996-Crittenden.Turok-Phys.Rev.Lett.}. 
\begin{figure}[!htb]
\begin{minipage}[b]{.45\textwidth}
\includegraphics[scale=0.4]{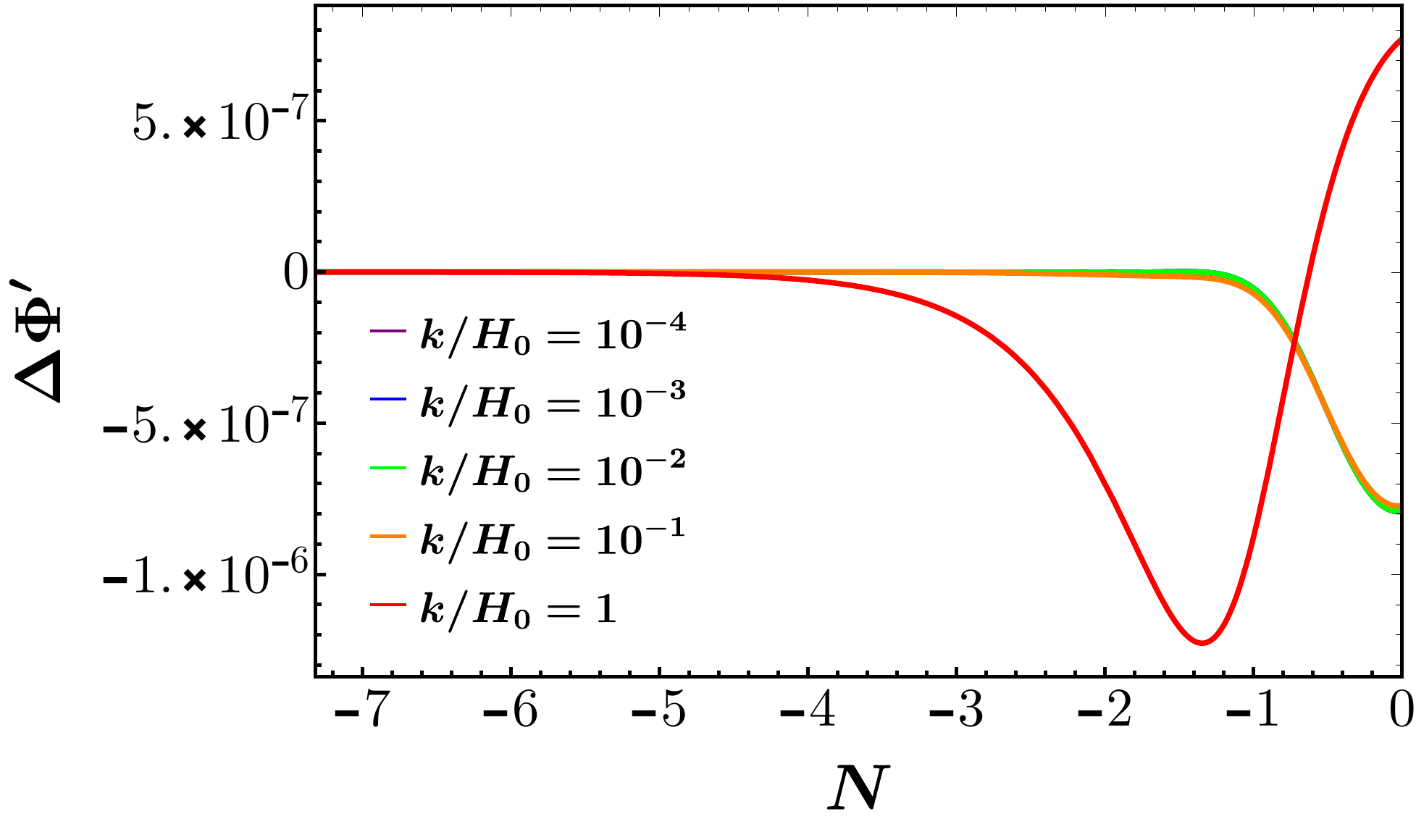}
\end{minipage}\hfill
\begin{minipage}[b]{.45\textwidth}
\includegraphics[scale=0.4]{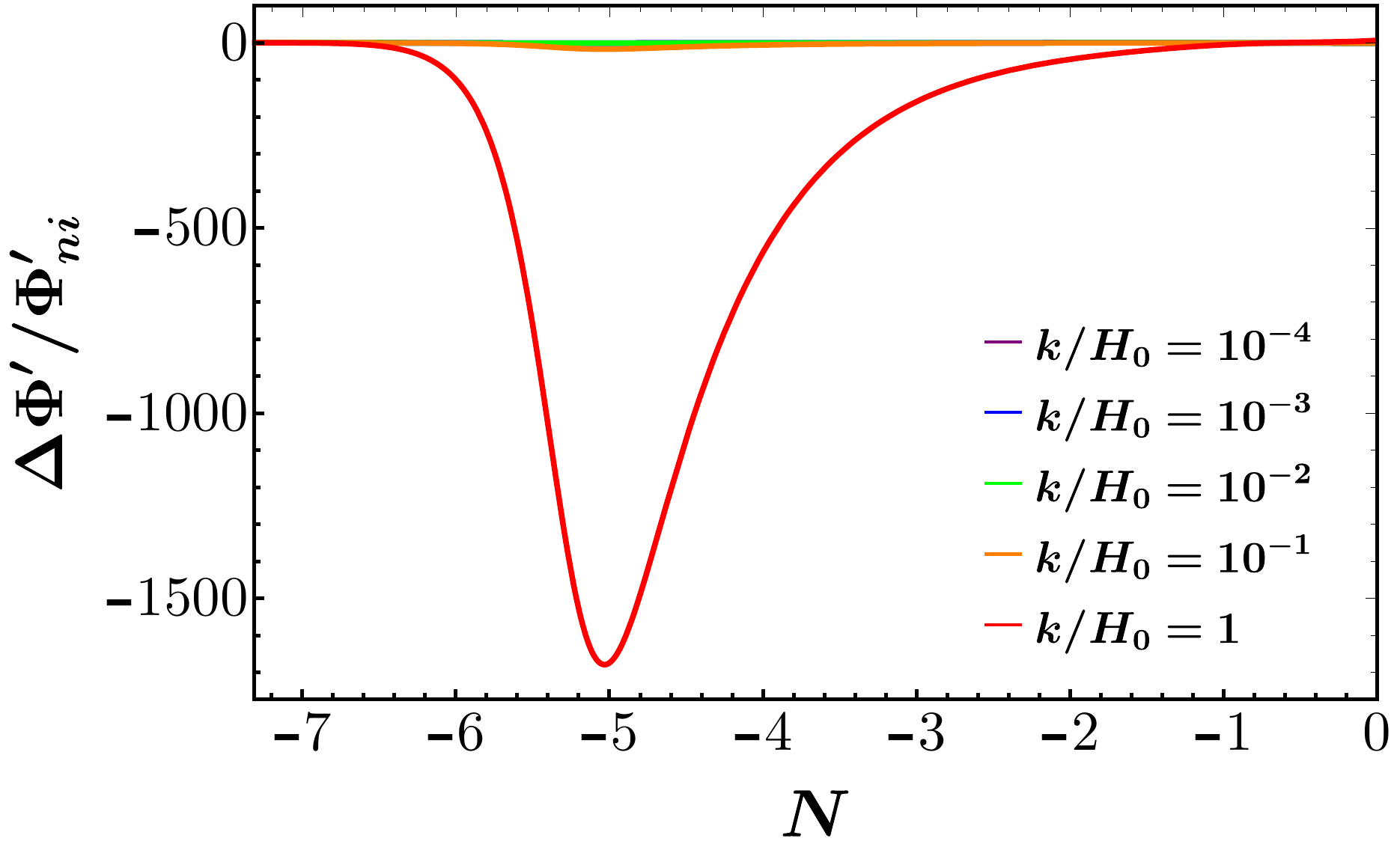}
\end{minipage}
\caption{Evolution of $\Delta \Phi'$ (left), $\Delta \Phi'/ \Phi'_{{ni}}$ (right)  as a function of $N$.}
\label{fig:diswphi1}
\end{figure}
\begin{figure}[!htb]
\begin{minipage}[b]{.45\textwidth}
\includegraphics[scale=0.4]{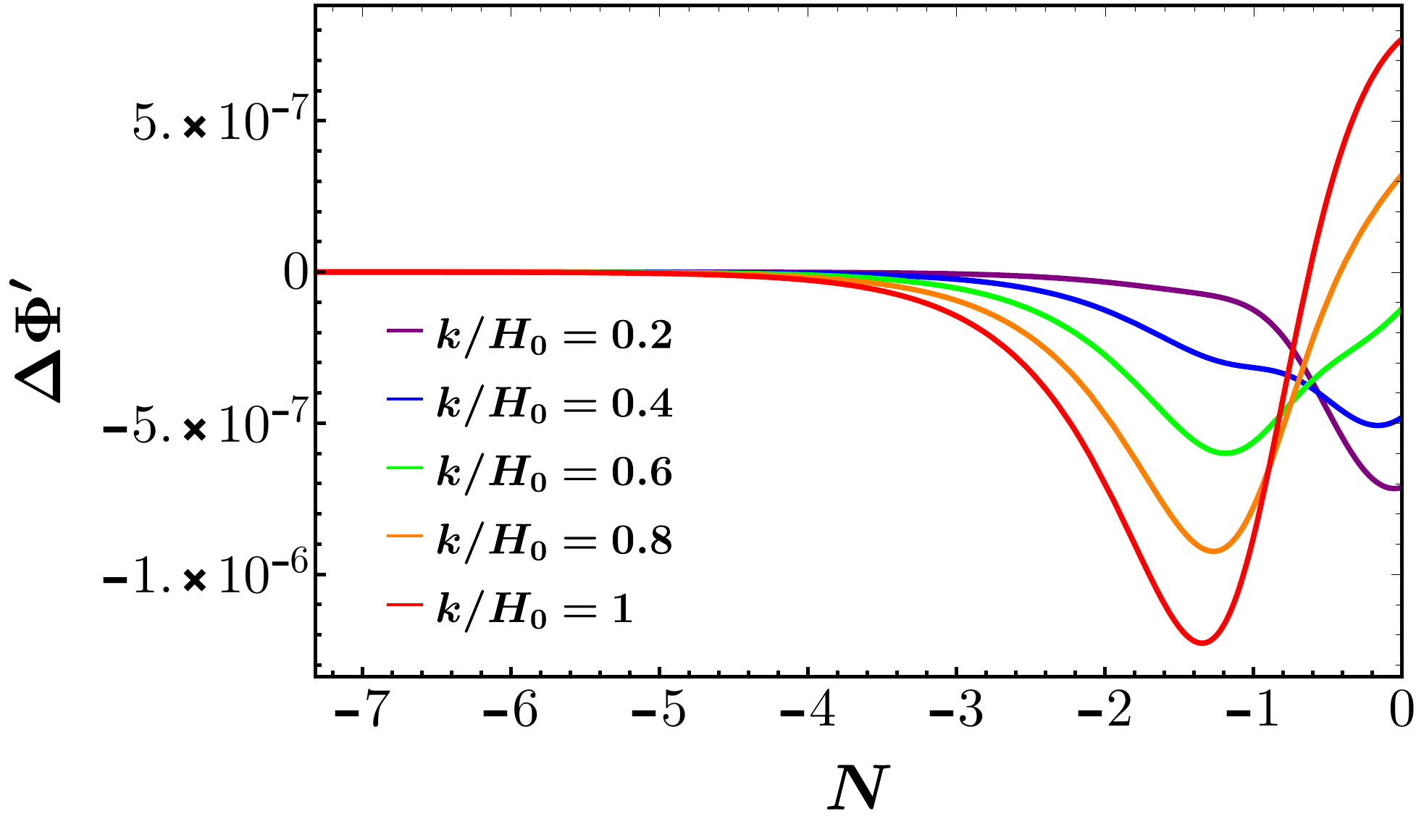}
\end{minipage}\hfill
\begin{minipage}[b]{.45\textwidth}
\includegraphics[scale=0.4]{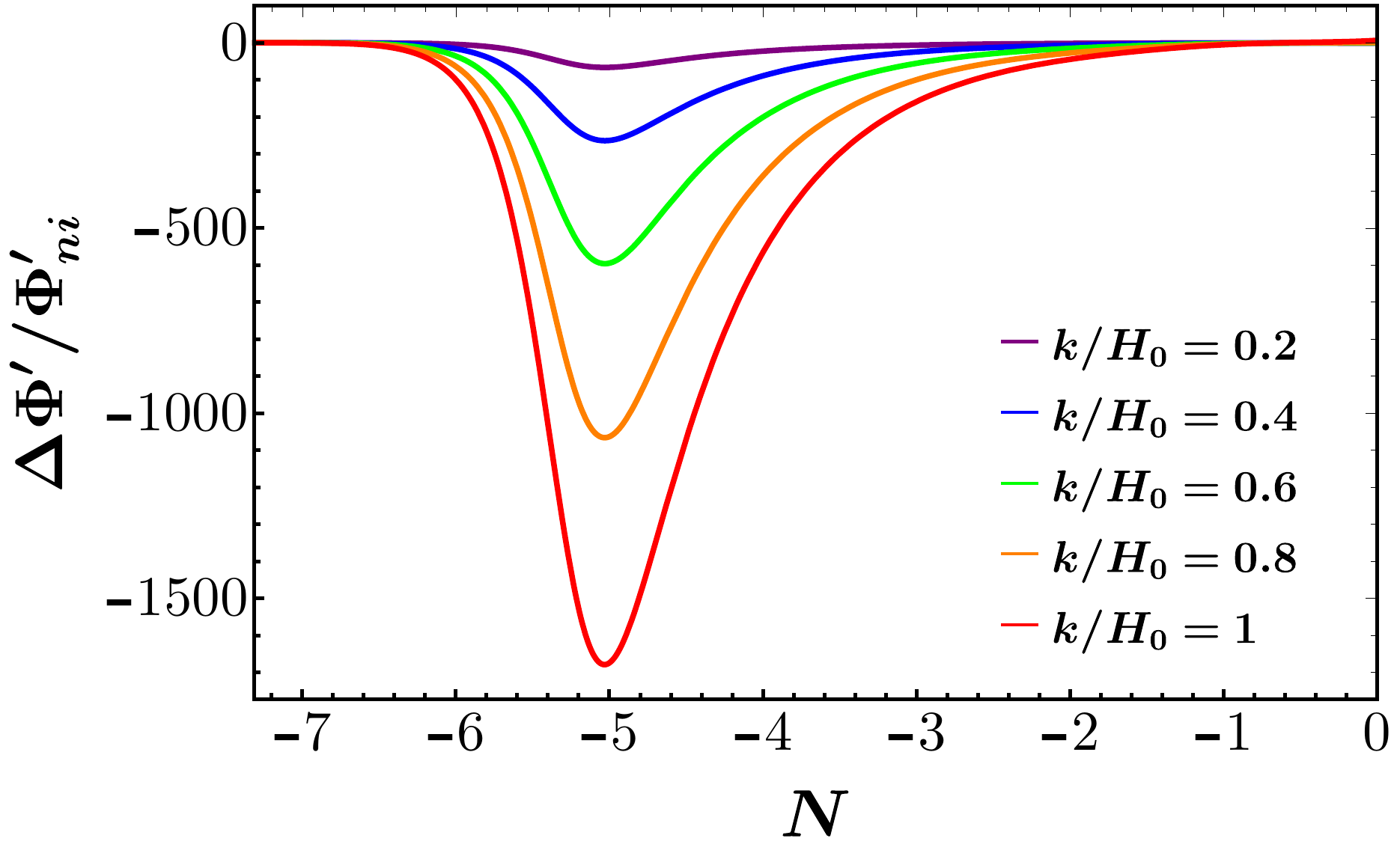}
\end{minipage}
\caption{Evolution of $\Delta \Phi'$ (left), $\Delta \Phi'/ \Phi'_{{ni}}$ (right)  as a function of $N$.}
\label{fig:diswzoomphi1}
\end{figure}

Since the Bardeen potential $\Phi$ evolve differently in the interacting and non-interacting scenarios, this change should potentially change the temperature fluctuations of the CMB photons. 
Figures \ref{fig:iswevophi1} and \ref{fig:iswevozoomphi1} contain plots of $\Phi'$ as a function of $N$ for different length scales in interacting and non-interacting scenarios. Evolution of $\Delta \Phi'$ and $\Delta \Phi'_{rel}$ as a function of $N$ 
are plotted in Figures \ref{fig:diswphi1} and \ref{fig:diswzoomphi1}.

Like $\delta_m$ and $\Phi$, we see that the difference in the evolution of $\Phi'$ in these two scenarios becomes significant at $N \sim -3$. Consistent with the fact that the first-order interaction term is larger at the smaller length scales, the difference in the evolution of $\Phi'$ in the interacting and non-interacting scenarios is enhanced for larger values of $k$. This indicates that observations on the ISW effect can detect or constrain dark energy and dark matter interaction.

\section{Conclusions}
\label{sec:conclusion}

In Ref.~\cite{2021-Johnson.Shankaranarayanan-Phys.Rev.D} two of the current
authors found a mapping between phenomenological models of the dark-energy dark matter coupling functions $Q$ from a consistent classical field theory.  We showed that the mapping holds both at the background and first-order perturbations level.  In this work we used this interacting field theory framework for a specific scalar field  potential $U(\phi) \sim 1/ \phi^n$ and  linear interaction function $\alpha(\phi) \sim \phi$. We analyzed the background cosmological evolution in this model and obtained the model parameters from cosmological observations. We evolved the perturbed equations in the redshift range $1500 \lesssim z \leq 0$ and obtained testable predictions of the model with future cosmological observations.

\noindent {\bf Constraints from observations:}~We obtained constraints for the model parameters from four observational data sets --- Hubble parameter measurements, baryon acoustic oscillation observation,  high-$z$ HII galaxy measurements, and Type Ia supernovae observations. For numerical analysis, we rewrote the evolution equations in terms of dimensionless variables. Using the $\chi^2$ minimization technique, we obtained the constraints on $H_0$, $\Omega_m$, $w_0$, and the interaction strength $C$. 

The key conclusions of the analysis for $n = 1$ ($U(\phi) \sim \phi^{-1} $)  case
are: (i) All the four data sets 
constrain the value of $H_0$ to be close to $70$ km s$^{-1}$Mpc$^{-1}$. 
BAO+Hz and SN+Hz observations provide the tightest constraints, followed by HIIG and Hz measurements.
(ii) When a combined analysis of all four data sets is performed, the {constraints} are impacted by BAO and SN observations the most, and the allowed range for $H_0$ becomes even narrower. 
(iii) {The constraints on $\Omega_m$ obtained from various data sets are consistent with each other, and BAO+Hz provides the smallest allowed range, which drives the limit for combined analysis, followed by SN + Hz, Hz, and then HIIG data}. 
(iv) When it comes to constraining $w_0$, all the observations are consistent with the $\Lambda$CDM model, and only Hz data allows for a non-accelerating universe. 
(v) As for the constraints on $C$, we find that only {BAO+Hz} data constrains $C$ within 3$\sigma$ confidence region, and hence, when analysis with a combination of the data sets is performed, the allowed values of $C$ is influenced by {BAO+Hz data} the most. We also find that, except for Hz measurements, all the three data sets show a preference for a negative value $C$ ({cf. Table~\ref{table::range}}). {The Hz data is nearly insensitive to the sign and value of $C$ within the considered range}. 
(vi) Our analysis points to the fact that there is a strong degeneracy between the interacting and non-interacting dark sector models with respect to these low-redshift background observations.

The key conclusions of the analysis for $n = 2$ ($U(\phi) \sim \phi^{-2} $)  case
are: 
(i) Constraints from the Hz data set do not change significantly. For other data sets, there is a slight shift in the contours. 
(ii) The observations prefer slightly higher values of $\Omega_m$, the contours from Hz-data shift towards higher values of $\Omega_m$. HIIG data allows a significantly larger range of $\Omega_m$ compared to $n=1$. For SN+Hz observations, there is no significant change in the lower range but the upper limit on $\Omega_m$ shifts slightly higher. For BAO+Hz data, the change in the allowed range of $\Omega_m$ is insignificant. 
(iii) For $H_0$, the change is not noticeable when we go from $U(\phi) \sim \phi^{-1}$ to $\phi^{-2}$. For $w_0$, there is no noticeable change from Hz data, but the allowed ranges increase when the $n=1$ is changed to $n=2$ for BAO+Hz, HIIG, and SN+Hz observations. 
(iv) The constraints on coupling parameter $C$ change significantly when $n$ changes. For $n=2$, constraints on $C$ from Hz do not show much change. Still, for BAO+Hz and SN+Hz data, we get upper limits on $C$, and the contours shift towards negative values of $C$, showing their preference for a negative value of interaction strength. We can also see this in constraints obtained from the combination of data sets.  All the observations are consistent with $C=-1$, but BAO+Hz and SN+Hz observations do not agree with $C=1$ within 1$\sigma$ confidence regions for $n=2$ case.

Our analysis shows no significant difference in the best-fit values for different values of $n$ (specifically, for $n=1$ $n=2$ (See Appendix \ref{sec::n=2})).  However, we notice that the allowed range of cosmological parameters ($H_0$, $\Omega_m$, $w_0$) increases for HIIG as we go from $n=1$ to $n=2$. In the case of  interaction parameter $C$, Hz shows no significant change, HIIG data allows a wider range, whereas for BAO+Hz and SN+Hz case, the contour shifts lower, thereby giving a smaller value for the upper limit of $C$.

\noindent {\bf Distinguishing dark sector interacting model from standard cosmology:} As we have shown, there is a strong degeneracy between the interacting and non-interacting dark sector models with respect to these background observations. To distinguish between the two scenarios, we looked at the evolution of the scalar perturbations in the interacting dark sector model. We considered a inverse potential $U(\phi) \sim 1/\phi$ and a linear interaction function $\alpha(\phi) \sim \phi$ with negative values of interaction strength $C$. We evolved three perturbed quantities ($\delta_m$, $\Phi$, $\Phi'$) from last scattering surface to present epoch ($1500 \lesssim z \leq 0$).  These three perturbed quantities are related to structure formation, weak gravitational lensing, and the ISW effect, respectively.

The density perturbation $\delta_m$ grows faster in the interacting scenarios, especially at the lower length scales. The difference in the evolution becomes significant for $z < 20$, for all length scales, and the difference peaks at smaller redshift values $z <  5$. This means that cosmological observations related to the formation of large-scale structures can potentially detect the signatures of dark matter - dark energy interaction. We see a similar trend in the evolution of $\Phi$ and $\Phi'$. This indicates an interaction between dark energy and dark matter will be reflected on the observational data on weak gravitational lensing and ISW effect. We get a similar behaviour for inverse-square potential $U(\phi) \propto 1 / \phi^2$.  The evolution of the perturbations in the interacting dark sector also differs from the ones in modified gravity models like $f(R)$ gravity, which describes the late-time acceleration of the Universe~\cite{2019-Johnson.Shankaranarayanan-Phys.Rev.D}. It was shown that, in the case of $f(R)$ models, the identity $\Phi = \Psi$ does not hold, and 
the evolution of perturbations are monotonic. As we have shown in this work, for a class of interacting dark sector models, the evolution is more complicated due to the interaction between dark energy and dark matter. Hence these models can potentially be distinguished using future observations.

It is interesting to note that all the perturbed quantities are significant for $z \sim 10-20$ and lie in the epoch of reionization. During this epoch, a predominantly neutral intergalactic medium was ionized by the emergence of the first luminous sources. Before the reionization epoch, the formation and evolution of structure were dominated by dark matter alone. However, the interacting dark sector leads to the exchange of density perturbations at smaller length scales. This indicates that it is 
possible to distinguish these models from the observations at the epoch of reionization.  

We have shown that the interacting dark sector model is consistent with the low-redshift background  observations and obtained the parameter constraints. The constraints on the dark energy-dark matter interaction model parameters can be used as priors in future studies. We have not addressed the issue of the tension in the $\sigma_8-\Omega_m$ plane between Planck and cosmic shear experiments~\cite{2017-Hildebrandt.others-Mon.Not.Roy.Astron.Soc.}. We plan to address this in future work. 

Currently, we are looking to obtain the constraints on the model from the evolution of the perturbations using the relevant observational data sets. It will also be interesting to look at the observational consequences of the difference in the evolution of the density perturbation. Since interaction is higher for the smaller length scales, it can significantly affect the evolution of the mass distribution of the binary black holes detected by the gravitational wave observations~\cite{2019-Farr.etal-Astrophys.J.Lett.}.


\section{Acknowledgements}
We thank T. Padmanabhan for fruitful discussions. We thank Ana Luisa Gonzalez-Moran for providing Gordon extinction corrected HIIG measurements and useful information related to the measurements (Ref.~\cite{2012-Chavez.etal-Mon.Not.Roy.Astron.Soc.,2018-FernandezArenas.etal-Mon.Not.Roy.Astron.Soc.}). JPJ is supported by CSIR Senior Research Fellowship, India. The work is partially supported by the ISRO-Respond grant.

\appendix

\section{Best fit values: $\Lambda$CDM, $w$CDM, and interacting dark sector models}
\label{sec:wlambda}

 In the tables below, we present the \textit{reduced chi-square} values ($\chi^2_{red}$) and the best fit values of the parameters for the simple $\Lambda$CDM model and $w$CDM model compared to the interacting dark sector model.
	
\begin{table}[!htb]
	\centering
	\makebox[0pt][c]{\parbox{1.2\textwidth}{%
			\begin{minipage}[b]{0.45\linewidth}\centering
			\begin{tabular}{|l|l|l|l|} \hline
{\bf Observations} &
$H_0$ & $\Omega_m$ &$\chi^2_{red}$\\ \hline
{\bf Hz} &68.19 & 0.29 &0.652\\ \hline
{\bf BAO+Hz} &68.52 & 0.28 &0.567\\ \hline
{\bf  HIIG} & 72.41 & 0.259&1.509\\ \hline
{\bf  SN+Hz} &69.8& 0.29 &0.959\\ \hline
{\bf  All combined} &69.69& 0.29 &1.039\\ \hline
\end{tabular}
\label{table::best_fit_lcdm}
			\end{minipage}
			\begin{minipage}[b]{0.45\linewidth}\centering
				\begin{tabular}{|l|l|l|l|l|} \hline
{\bf Observations} &
$H_0$ & $\Omega_m$ & $w_0$ &$\chi^2_{red}$\\ \hline
{\bf Hz} &70.09 & 0.28 & -1.13 &0.673\\ \hline
{\bf BAO+Hz} &67.34 & 0.28 &-1.03 &0.578\\ \hline
{\bf  HIIG} & 71.43 & 0.25&-0.89 &1.514\\ \hline
{\bf  SN+Hz} &69.89& 0.29 &  -1.01 &0.9599\\ \hline
{\bf  All combined} &70.01& 0.28 &  -1.03 &1.040\\ \hline
\end{tabular}
\label{table::best_fit_wcdm}
			\end{minipage}
}}
\caption{The best fit values of the parameters obtained for $\Lambda$CDM model (left panel) and $w$CDM model (right panel).}
\end{table}
\begin{table}[!htb]
	\centering
	\begin{tabular}{|l|l|l|l|l|l|} \hline
{\bf Observations} &
$H_0$ & $\Omega_m$ & $C$ & $w_0$ &$\chi^2_{red}$\\ \hline
{\bf Hz} &69.34 & 0.29 & 0.98 &-0.989 &0.697\\ \hline
{\bf BAO+Hz} &70.4 & 0.29 & -0.63 &-0.997 &0.591\\ \hline
{\bf  HIIG} & 72.49 & 0.25& -0.94 &-0.92 &1.522\\ \hline
{\bf  SN+Hz} &69.51& 0.31 & -0.69 & -1.0 &0.961\\ \hline
{\bf  All combined} &69.79& 0.29 & -0.52 & -0.99 &1.041\\ \hline
\end{tabular}
\caption{The best fit values of the parameters obtained for the dark-energy dark-matter interaction model.}
\label{table::best_fit01}
\end{table}

As we can see, $\chi^2_{red}$ values for the interacting dark sector model we have considered is closer to one (except for HIIG and combined data sets) compared to $\Lambda$CDM and $w$CDM models. In the analysis for HIIG observations, we have used $\sigma$=$\sigma_{stat}$; therefore, we get a higher $\chi^2_{red}$. See Ref.~\cite{2016-Chavez.etal-Mon.Not.Roy.Astron.Soc.}, where the authors present in greater detail that there is a systematic error of $\sim$0.25. If we include it in the analysis, we will get $\chi^2_{red}$ $\sim$ 1 \cite{2019-GonzalezMoran.etal-Mon.Not.Roy.Astron.Soc.}.
Looking at the $\chi^2_{red}$ values, we see that there is a strong degeneracy between the interacting and non-interacting dark sector models with respect to these low-redshift background observations.

%


\begin{table}[h]
\centering
\begin{tabular}{|l|l|l|l|l|l|l|}
   \hline
& \multicolumn{3}{|c|}{AIC}&\multicolumn{3}{|c|}{BIC}\\
\hline
{\bf Observations} & 
$\Lambda$CDM & $w$CDM &  DEDM & $\Lambda$CDM & $w$CDM &  DEDM \\ \hline
{\bf Hz} &22.93 & 24.87 & 26.81 &25.798 & 29.192& 32.545\\ \hline
{\bf BAO+Hz}  &26.12 & 27.99 & 29.87 &29.547& 33.131& 36.724\\ \hline
{\bf  HIIG} & 231.87 & 233.12& 234.79 & 237.93 & 242.211& 246.911\\ \hline
{\bf  SN+Hz}  &741.82& 743.21 & 745.21 &751.03& 757.03& 763.63\\ \hline
{\bf  All combined} &972.57& 974.449 &976.322 &982.248& 988.967 &995.656\\ \hline
\end{tabular}
\caption{AIC and BIC for different models.}
\label{table::AIC_BIC}
\end{table}

The introduction of new parameters often gives a better fit for the data, irrespective of the relevance of the parameter. Therefore, to select which model is better regardless of the number of free parameters, some information criteria are used to penalize additional parameters in the analysis. We use the \blue{Akaike information criterion (AIC)~\cite{1974-Akaike-IEEETAC} and the Bayesian information criterion (BIC)~\cite{1978-Schwarz-AnnalsStatist.}} which are defined as $AIC = -2\ln\mathcal{L} + 2k= \chi_{min}^2 + 2k$ and $BIC = -2\ln\mathcal{L}+ k\ln N=\chi_{min}^2+k\ln N$. Here, $\mathcal{L}$ and $\chi_{min}$ denote the maximum likelihood and minimum $\chi$ value for a model, respectively, $k$ denotes the number of parameters of the model, and $N$ denotes the number of data points in the observations.

\blue{From the above table, we infer that the interacting dark sector model can provide a good fit for the observations while not being favoured over the $\Lambda$CDM model w.r.t. AIC. However, our model is strongly disfavoured w.r.t. BIC for the supernovae data.
}

\section{Comparing the parameter constraints from JLA and Pantheon data sets}
\label{sec::SN_comparison}
In this appendix we obtain the constraints on our model
parameter for the Pantheon compilation of type Ia Supernovae for fixing values of $\Omega_m, H_0$ and $C$. We also compare these results with the constraints from JLA in Sec.~\ref{sec::results}. 

\begin{figure*}[!htbp]
\centering
\begin{tabular}{cc}
\includegraphics[scale=0.42]{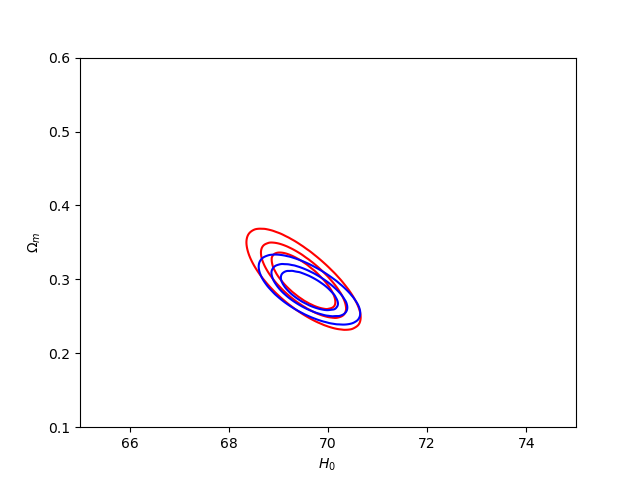}&\includegraphics[scale=0.42]{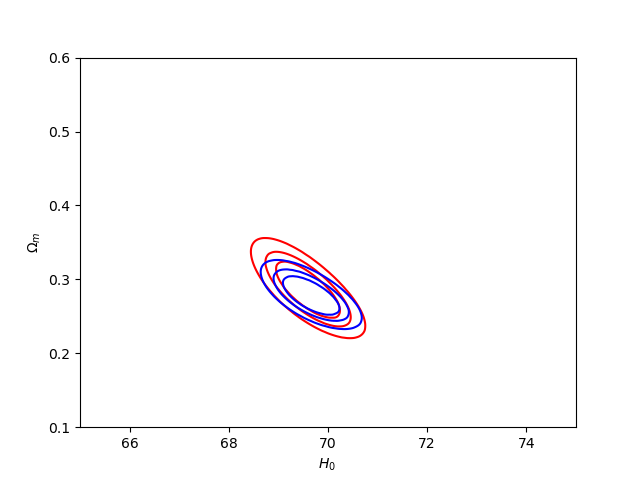}\\
\includegraphics[scale=0.42]{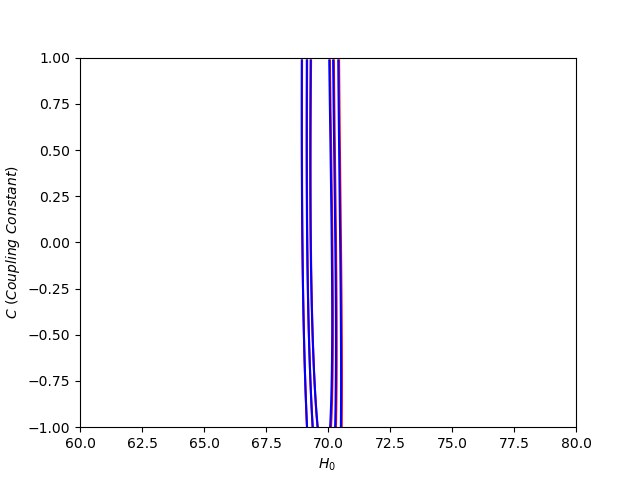}&\includegraphics[scale=0.42]{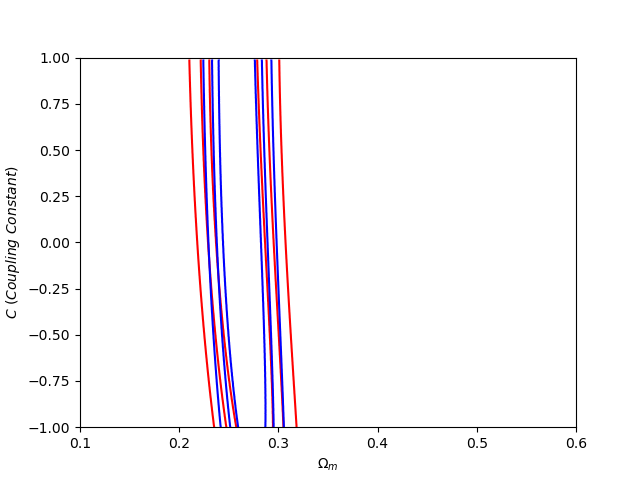}\\
\end{tabular}
\caption{The figure represents $1\sigma$, $2\sigma$ and $3\sigma$ contours obtained from JLA compilation (red contours) of Type Ia supernovae and from Pantheon data (blue contours). The left figure in the first row shows constraints obtained in $H_0-\Omega_m$ plane for $C=-1$ and $w_0=-1$, the right figure in the first row shows results in $H_0-\Omega_m$ plane for $C=-0.5$ and $w_0=-1$ from the two SN compilation data sets. The left plot in second row shows constraints in $H_0~-~C$ plane for $\Omega_m=0.27$ and $w_0=-1$ and the right plot shows confidence contours in $\Omega_m~-~C$ plane for $H_0=70~km~s^{-1}~Mpc^{-1}$ and $w_0=-1$ from SN observations.
}
\label{fig::jla_pan}
\end{figure*}

In Fig. \ref{fig::jla_pan}, the blue contours correspond to the pantheon data and the red contours are for JLA data by setting $w_0=-1$. The top row consists of plots of $\Omega_m$ vs $H_0$ 
for two different values of $C$. The left plot is for $C = -1$ 
and the right plot is for $C=-0.5$. 
The bottom row (left plot) the constraints are in $H_0-C$ plane 
by setting $\Omega_m=0.27$. The bottom row (right plot) the  constraints are in $\Omega_m-C$ plane by setting $H_0=70~km~s^{-1}~Mpc$. 

We find that the best fit values of the parameters 
are the approximately the same from Pantheon and JLA data. However, 
Pantheon data provides smaller range of parameters. By fixing $w_0$, we find that both the data allow the entire range of the interaction strength $C$. 

\section{Parameter constraints for $U(\phi) \sim 1/\phi^2$}
\label{sec::n=2}

\begin{figure}[!htbp]
\centering
\begin{tabular}{ccc}
\includegraphics[scale=0.32]{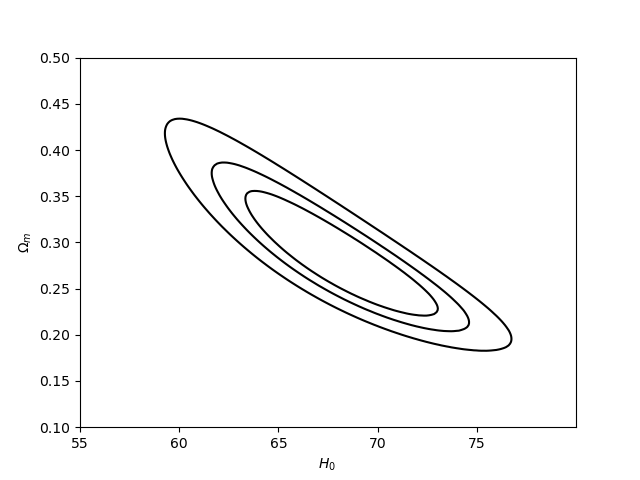}&\includegraphics[scale=0.32]{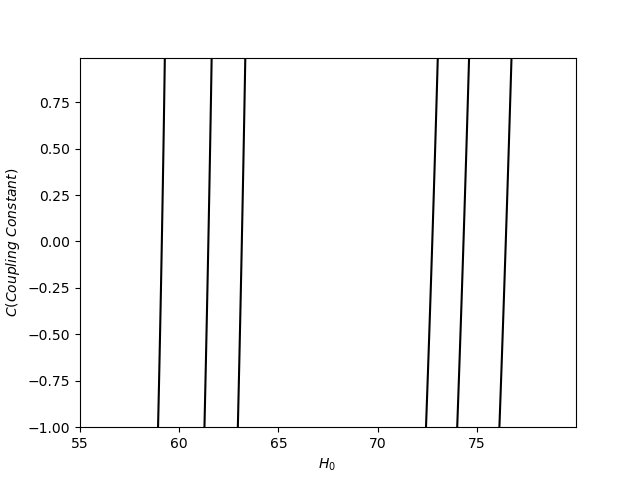}&\includegraphics[scale=0.32]{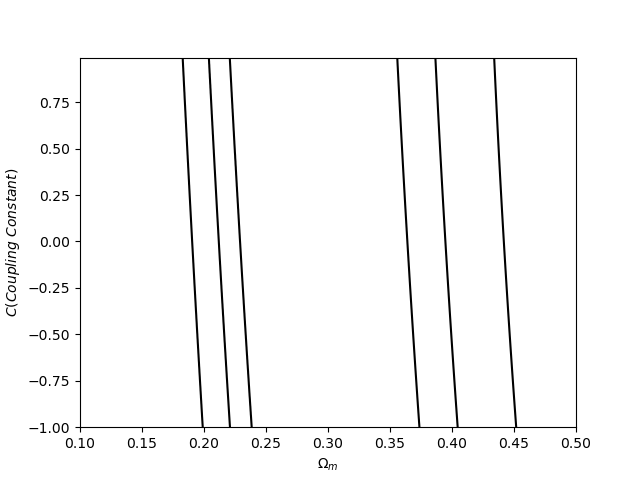}\\ 
\includegraphics[scale=0.32]{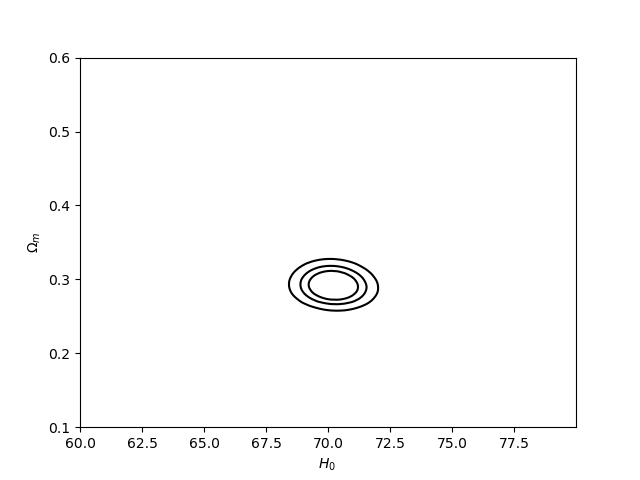}&\includegraphics[scale=0.32]{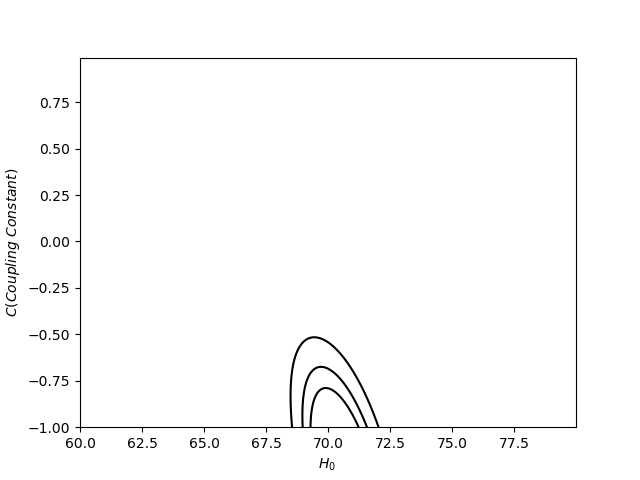}&\includegraphics[scale=0.32]{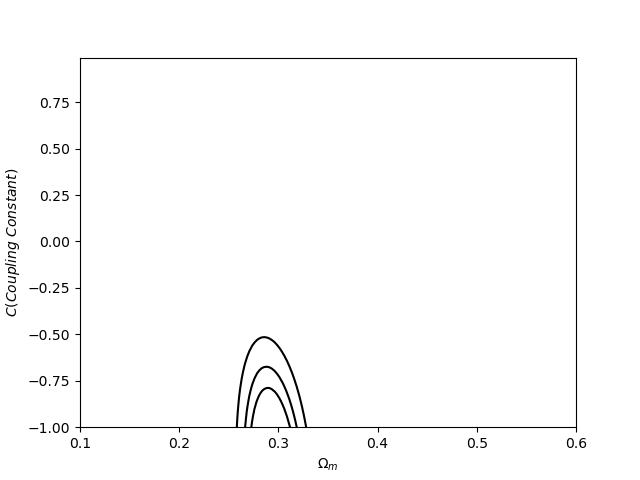}\\
\includegraphics[scale=0.32]{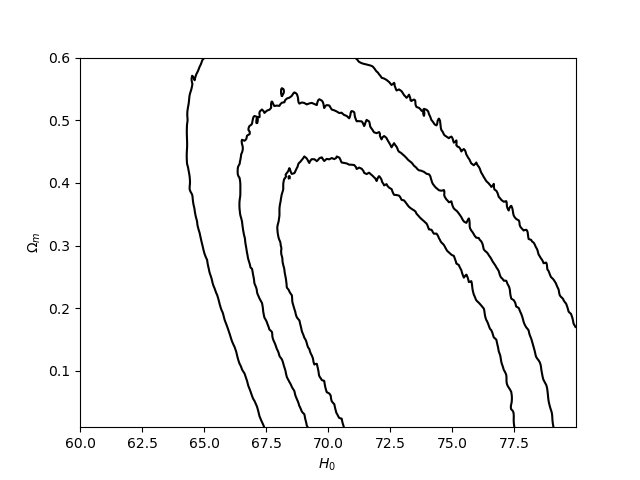}&\includegraphics[scale=0.32]{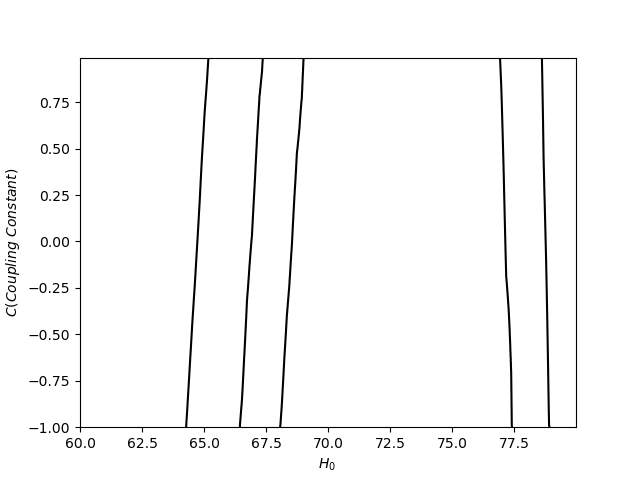}&\includegraphics[scale=0.32]{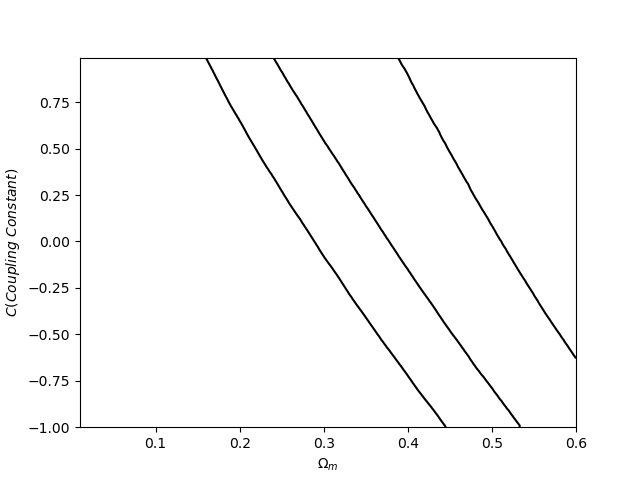}\\
\includegraphics[scale=0.32]{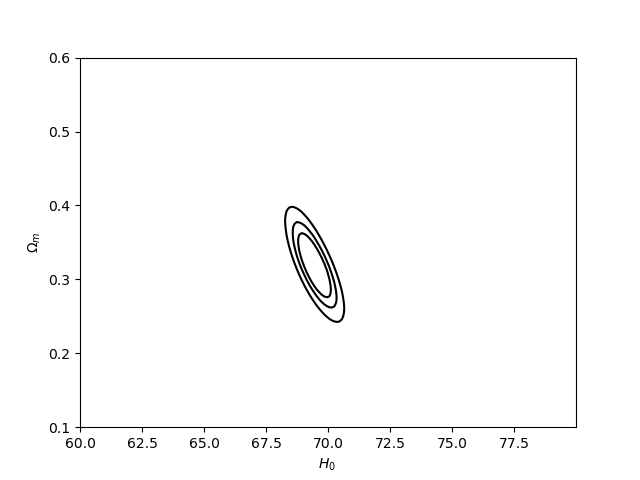}&\includegraphics[scale=0.32]{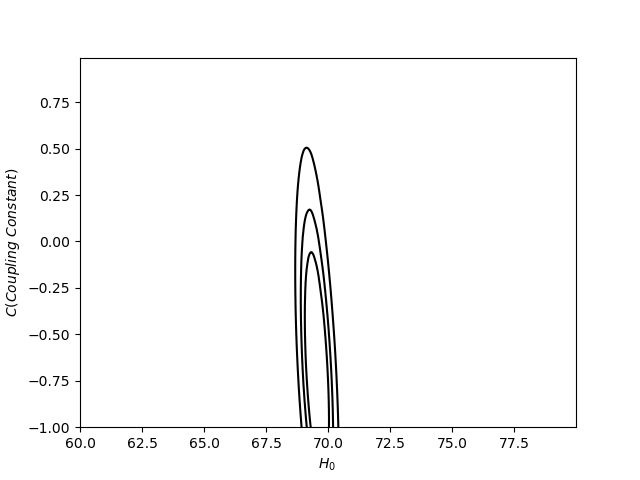}&\includegraphics[scale=0.32]{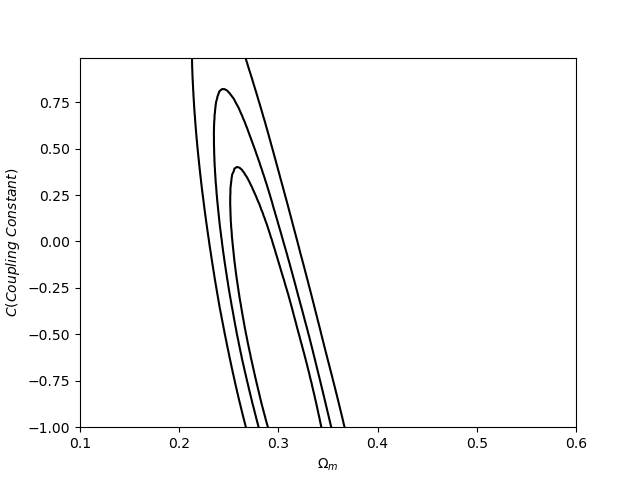}\\
\includegraphics[scale=0.32]{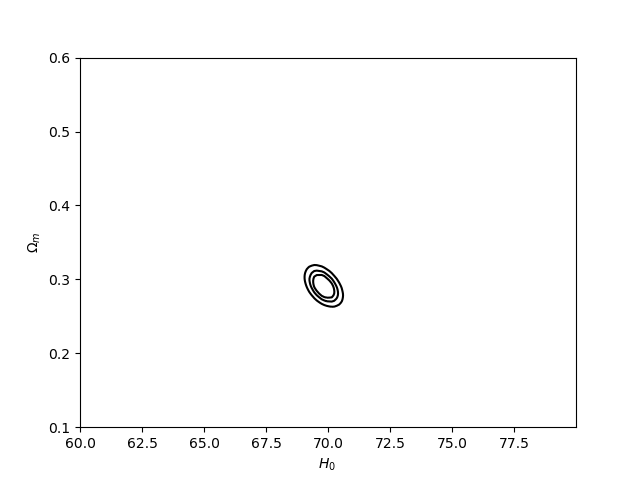}&\includegraphics[scale=0.32]{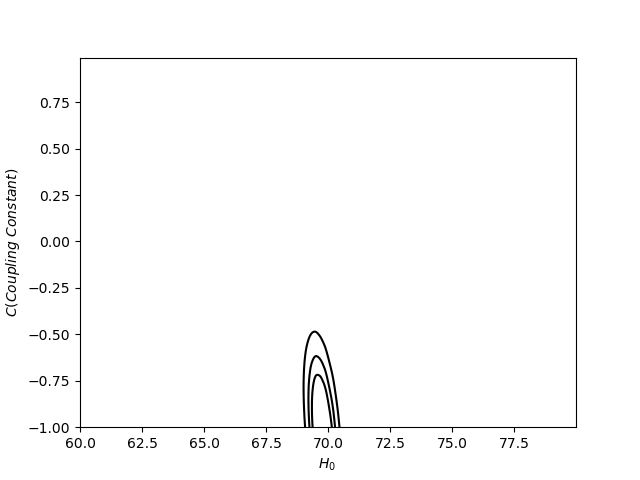}&\includegraphics[scale=0.32]{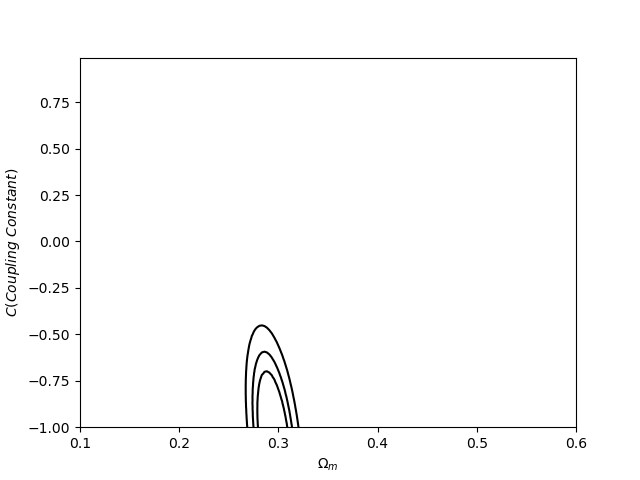}\\
\end{tabular}
\caption{1,2,3-$\sigma$ likelihood contours for Hz data (I row), BAO+Hz 
 data (II row),  HIIG data (III row), SN+Hz data (IV row) and all four data sets (V row). The two-dimensional contours are obtained by performing marginalization over other parameters.}
\label{fig::h0_om_c2}
\end{figure}

\begin{figure}[!htbp]
\centering
\begin{tabular}{cc}
\includegraphics[scale=0.45]{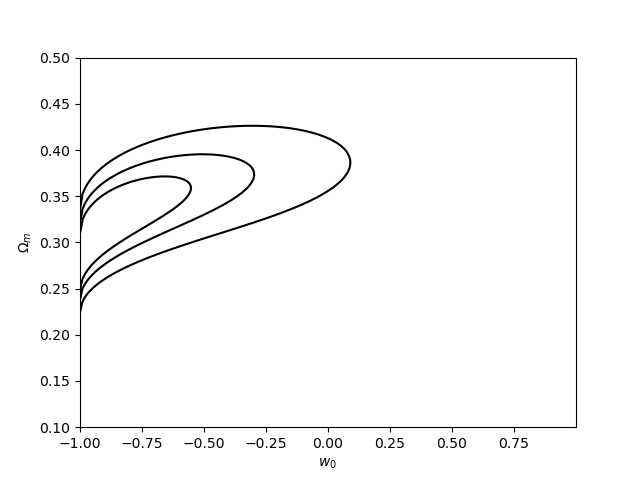}&\includegraphics[scale=0.45]{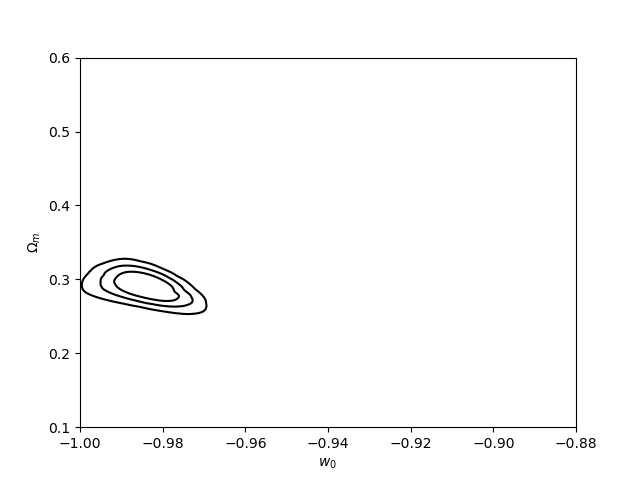}\\ 
\includegraphics[scale=0.45]{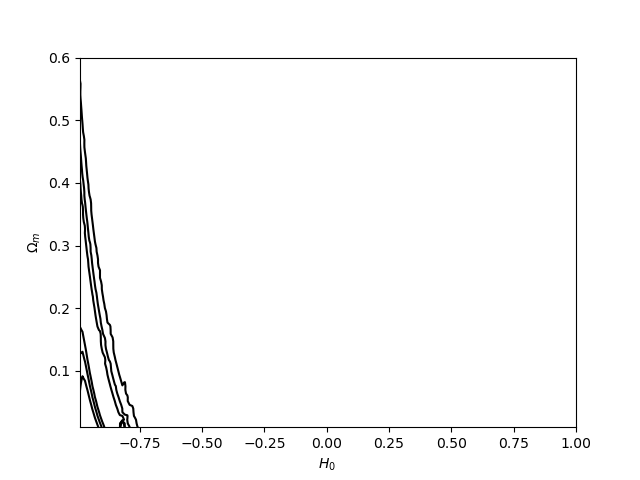}&\includegraphics[scale=0.45]{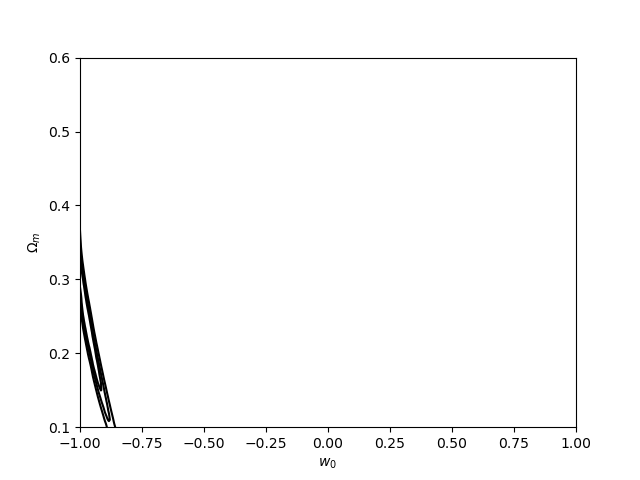}\\
\end{tabular}
\caption{1,2,3-$\sigma$ likelihood contours in `$w_0$-$\Omega_m$' plane. The top
  row shows constraints from Hz data (left) and BAO+Hz observations (right).
 The second row shows constraints from  HIIG measurements (left) and {SN+Hz} observations (right).}

\label{fig::w0_om2} 
\end{figure}

For completeness, in this Appendix, we present the constraints for 
$n = 2$ in  the quintessence potential \eqref{eq:Potentialdef}. Note that in Section \ref{sec::results}, we presented the detailed analysis for $n = 1$. As mentioned earlier, the parameter constraints are roughly the same for $n=1$ and $n=2$. 
Figures~\ref{fig::h0_om_c2} and \ref{fig::w0_om2} contain the constraints on parameters $H_0$, interaction strength $C$, and $\Omega_m$ for the four observational data sets --- Hz, BAO, HIIG, and SN.

Here are the key inferences from Figures~\ref{fig::h0_om_c2} and \ref{fig::w0_om2}: 
(i) For $n =2$, the constrains on $H_0$, $w_0$ and $C$ obtained from the data sets are almost same as for $n=1$. 
(ii) From Hz data, the minimum value of $\chi^2$ is 18.77 which  corresponds to the best fit values of the parameters are $H_0 = 69.37$ km s$^{-1}$Mpc$^{-1}$, $\Omega_m$= 0.29, $w_0$=-0.98 and the interaction strength is $C$=0.98.  
(iii) For BAO+Hz data, when it comes to the interaction strength $C$, the preference for negative value is more evident here than for $n=1$. Although for $1/\phi^2$ potential, the data does not allow for a non-accelerating universe, a larger allowed range for $w_0$ is obtained. 
(iv) For HIIG observations, $n=2$ provides a larger range of allowed values of the parameters than  $n=1$.
(v) For SN+Hz data, the $H_0$ and $\Omega_m$ constraints are as narrow as in $n=1$ case, but the observations prefer negative value for $C$.

From Figure \ref{fig::w0_om2}, we see that the four data sets do not provide a lower limit on $w_0$. Hz data provides an upper limit of -0.68 within 1$\sigma$ and $w_0$=0.03 within 3$\sigma$ region, showing that this particular model does not allow for a non-accelerating universe within 1$\sigma$ region. The allowed
ranges are almost the same as in the case $n=1$.

BAO+Hz observation does not allow for a non-accelerating Universe within the 3$\sigma$ region, and the allowed range for $w_0$ is wider as compared to the $n=1$ case. The HIIG data also allows a slightly wider range for $w_0$, with the 3$\sigma$ upper limit being $-0.749$ and allows the entire range of $\Omega_m$ considered in the analysis. The SN+Hz data also allows a wider range for $w_0$ and $\Omega_m$ as compared to $n=1$ case. Apart from Hz data, the three remaining observational data sets considered in the analysis do not allow for a non-accelerating universe for both $n=1$ and $2$. For $w_0$, Hz observations provide the widest allowed range within 3$\sigma$ confidence level.


\section{Evolution of scalar perturbations for $\phi^{-2}$ potential}
\label{app:phi2}

For completeness, in this Appendix, we present the evolution of the matter density perturbation $\delta_m$  and related quantities fo $n = 2$ in the quintessence potential \eqref{eq:Potentialdef}. Note that in Sec. \ref{sec:pertevo}, we presented the detailed analysis for $n = 1$. As mentioned earlier, the evolution of the perturbed quantities is not sensitive to $n$.  

\subsection{Evolution of the scaled interaction function $\delta q$}

\begin{figure}[!htb]
\begin{minipage}[b]{.45\textwidth}
\includegraphics[scale=0.4]{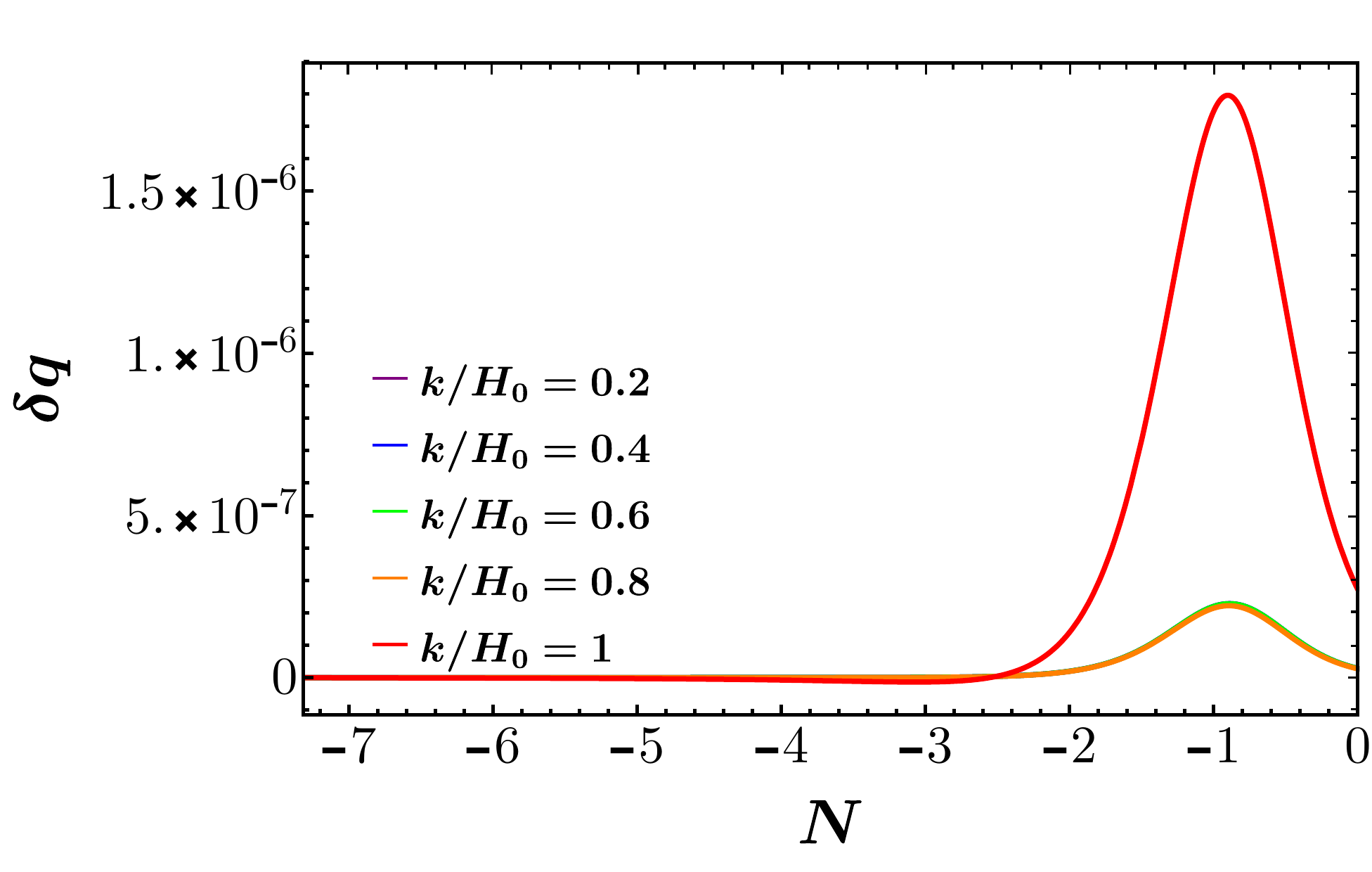}
\end{minipage}\hfill
\begin{minipage}[b]{.45\textwidth}
\includegraphics[scale=0.4]{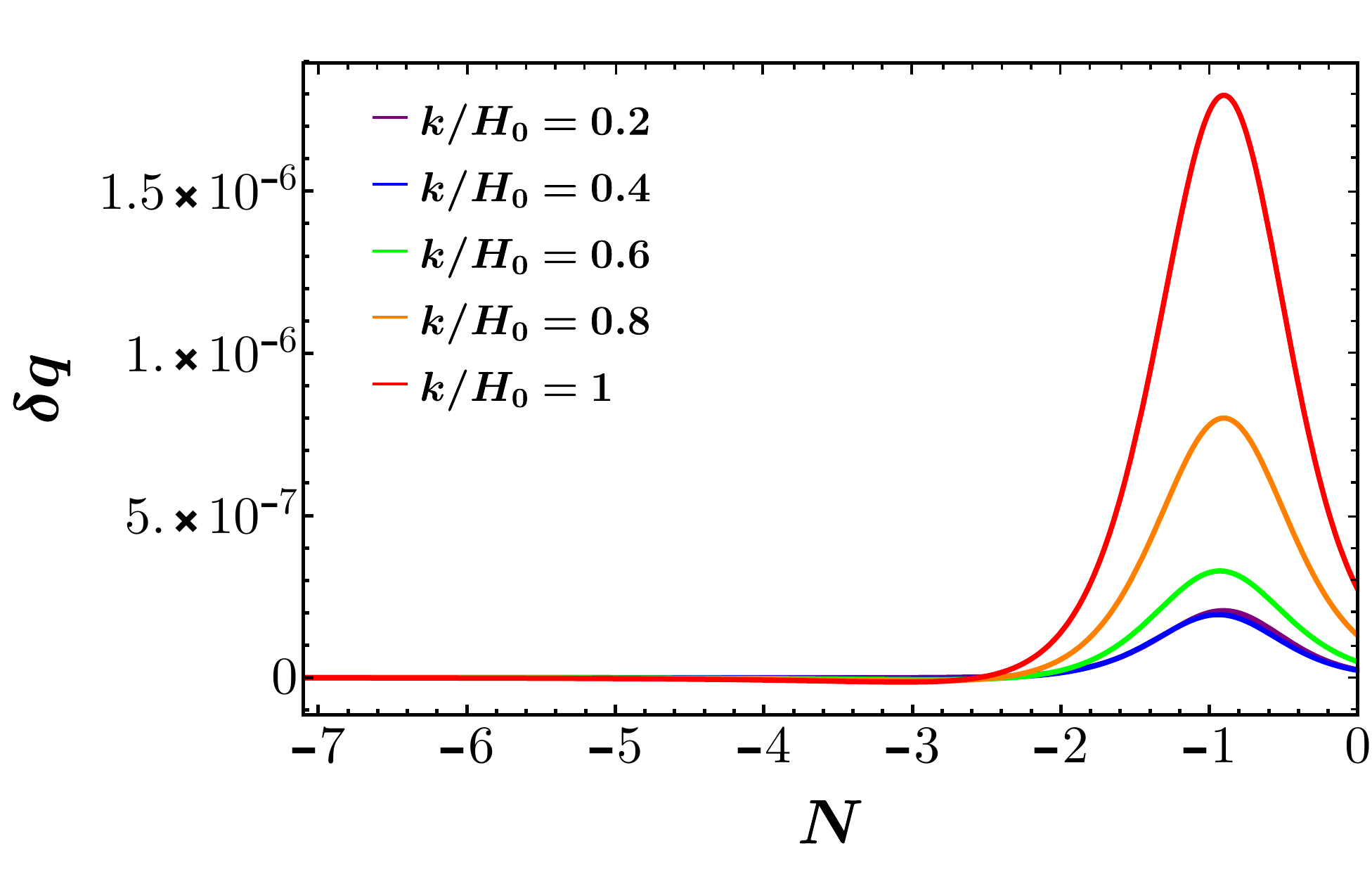}
\end{minipage}
\caption{Evolution of $\delta q$ as a function of $N$ for different values of $k$ with $C=-0.6$}
\label{fig:dqphi2-2}
\end{figure}

Figure \ref{fig:dqphi2-2} is the plot of $\delta q$ (cf. Eq. \ref{def:scaledq}) 
as a function of $N$ for different $k$ values. Comparing this plot with the plots in Figure \ref{fig:dqevophi1}, we see that evolution of the interaction function is roughly the same for the both the cases. Hence, the evolution of scaled interaction function $\delta q$ is not sensitive to $n$.

\subsection{Structure formation}

\begin{figure}[!htb]
\begin{minipage}[b]{.45\textwidth}
\includegraphics[scale=0.4]{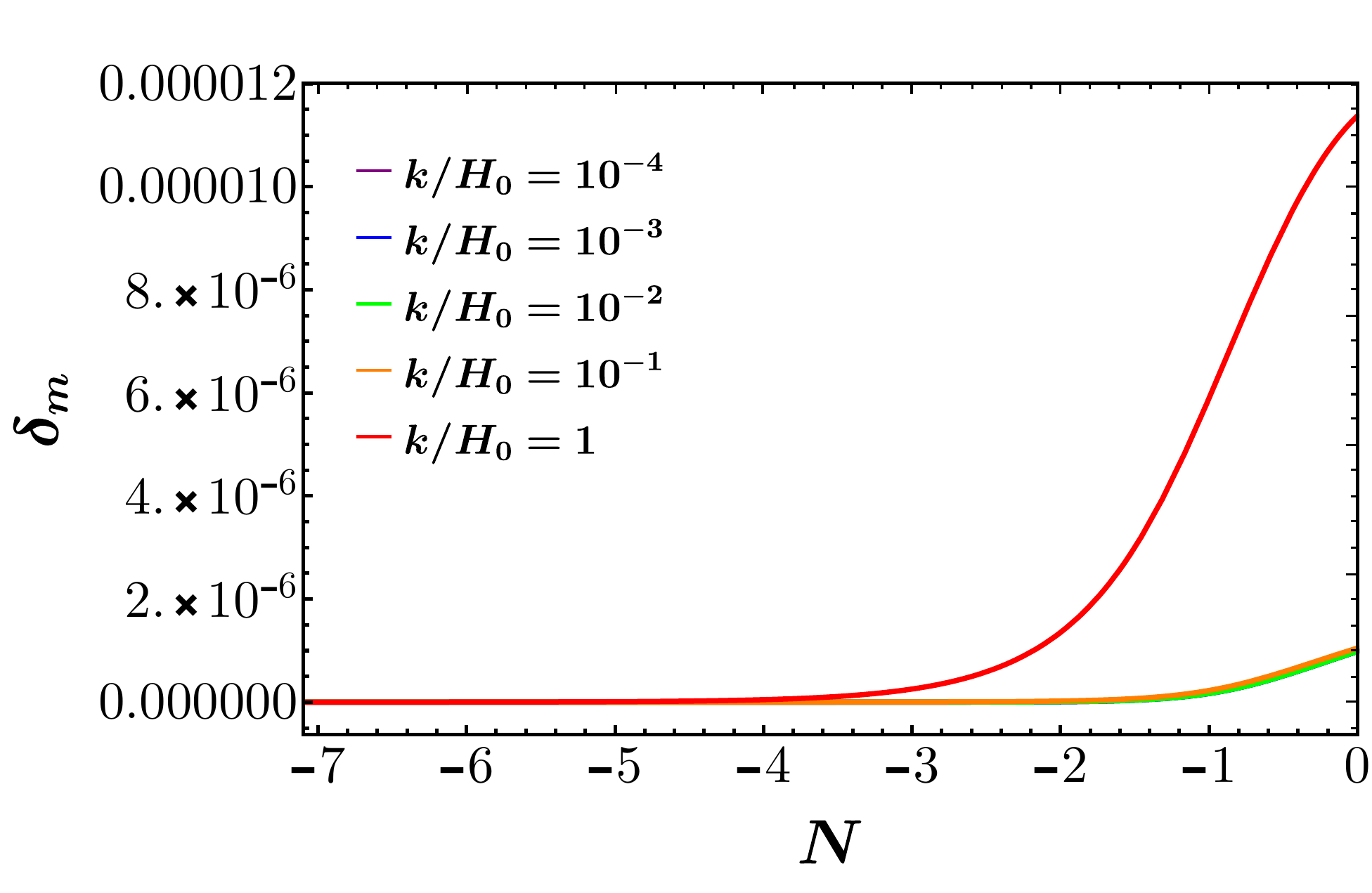}
\end{minipage}\hfill
\begin{minipage}[b]{.45\textwidth}
\includegraphics[scale=0.4]{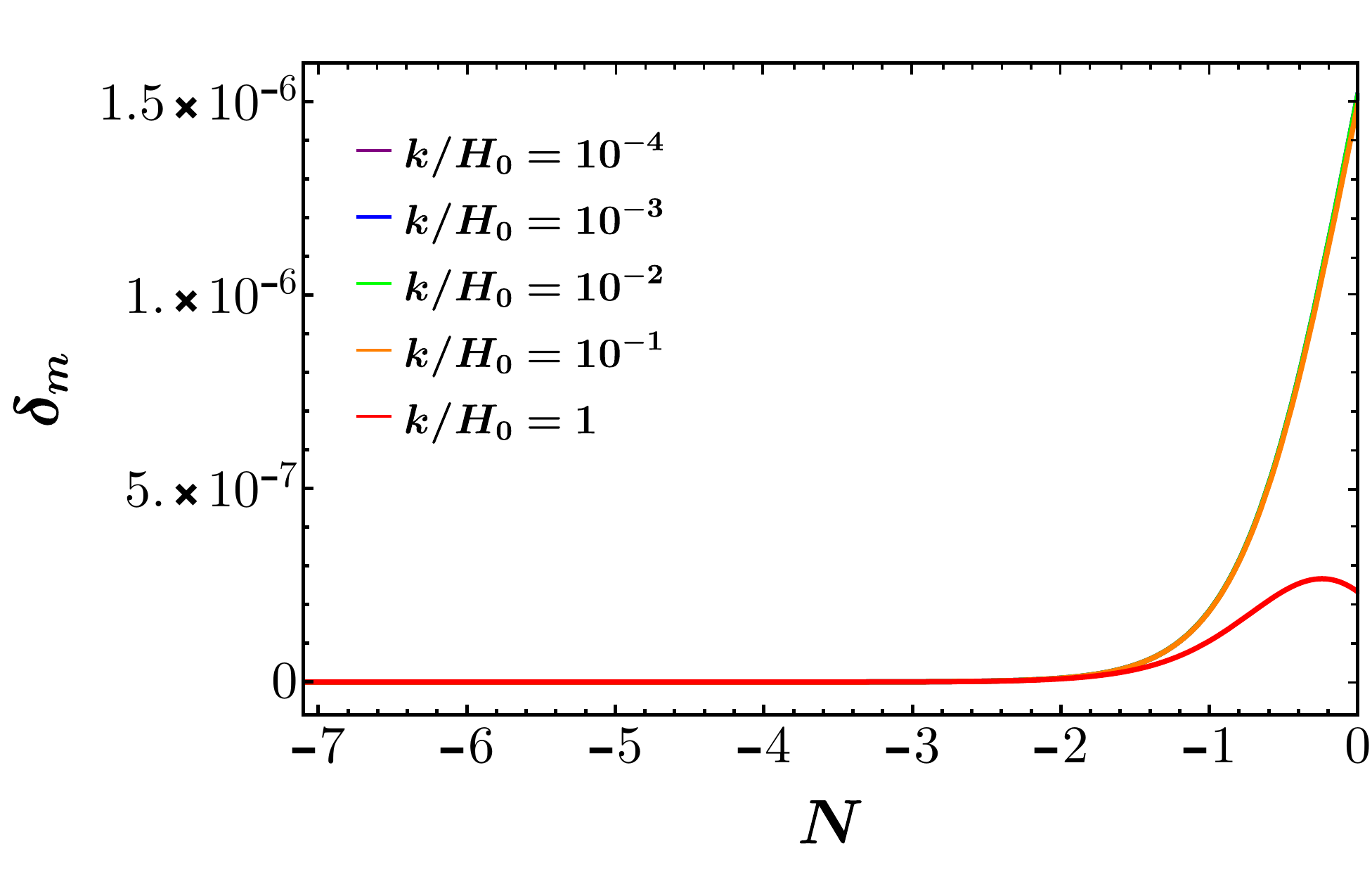}
\end{minipage}
\caption{Evolution of $\delta_m$ as a function of $N$. Left: $C=-0.6$, Right: $C=0$.}
\label{fig:deltaevophi2}
\end{figure}

\begin{figure}[!htb]
\begin{minipage}[b]{.45\textwidth}
\includegraphics[scale=0.4]{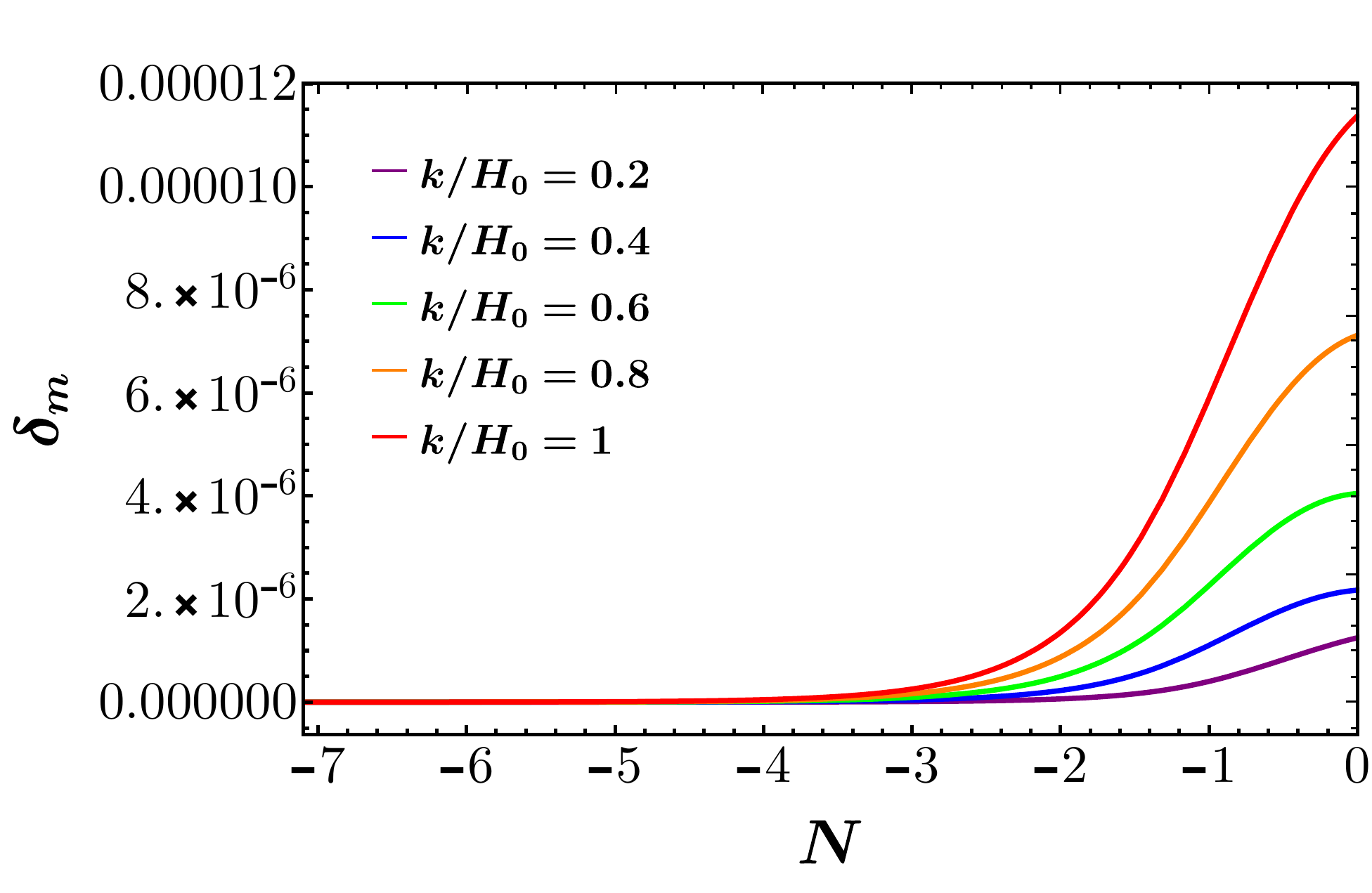}
\end{minipage}\hfill
\begin{minipage}[b]{.45\textwidth}
\includegraphics[scale=0.4]{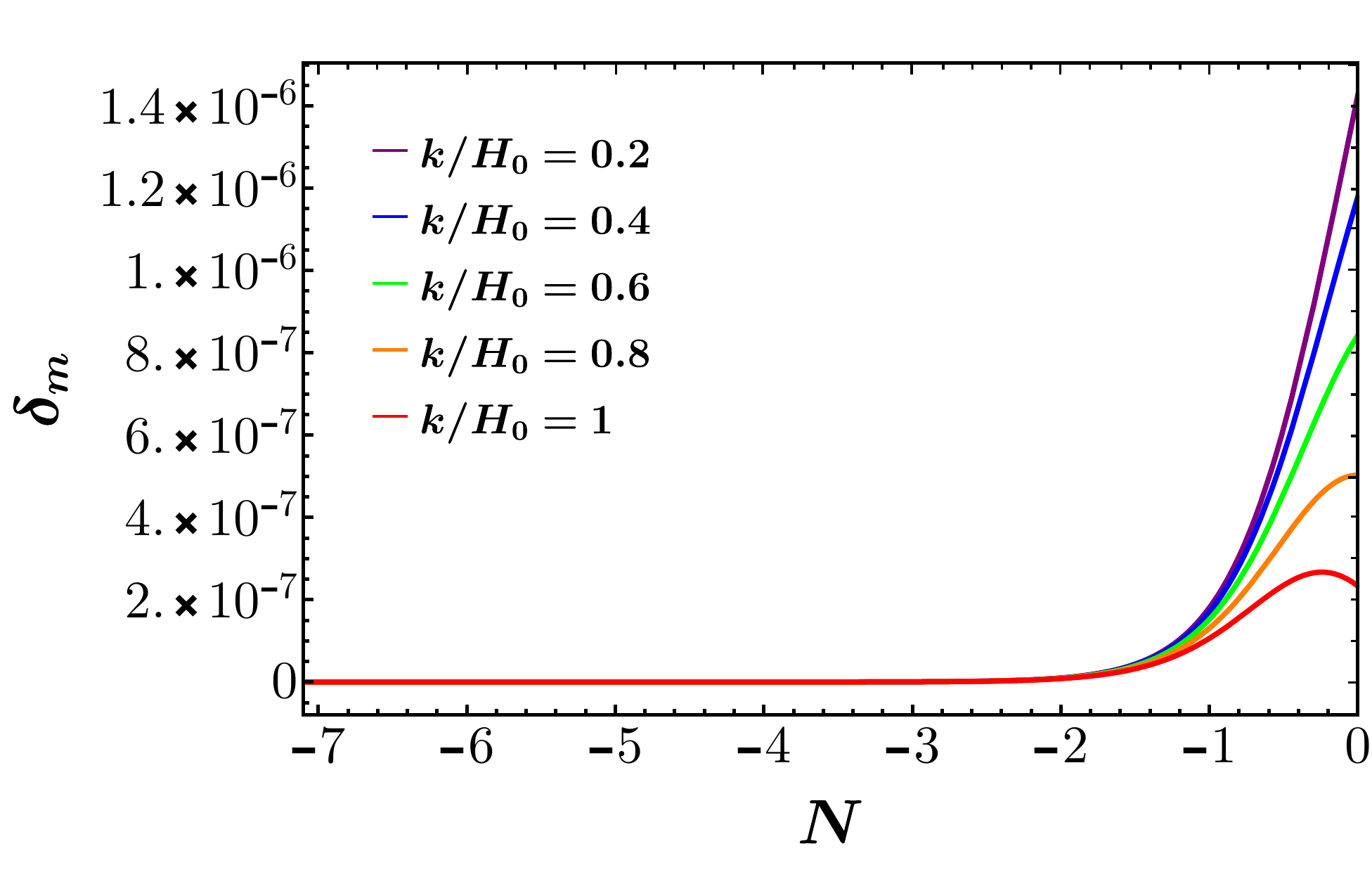}
\end{minipage}
\caption{Evolution of $\delta_m$ as a function of $N$. Left: $C=-0.6$, Right: $C=0$.}
\label{fig:deltaevozoomphi2}
\end{figure}

\begin{figure}[!htb]
\begin{minipage}[b]{.45\textwidth}
\includegraphics[scale=0.4]{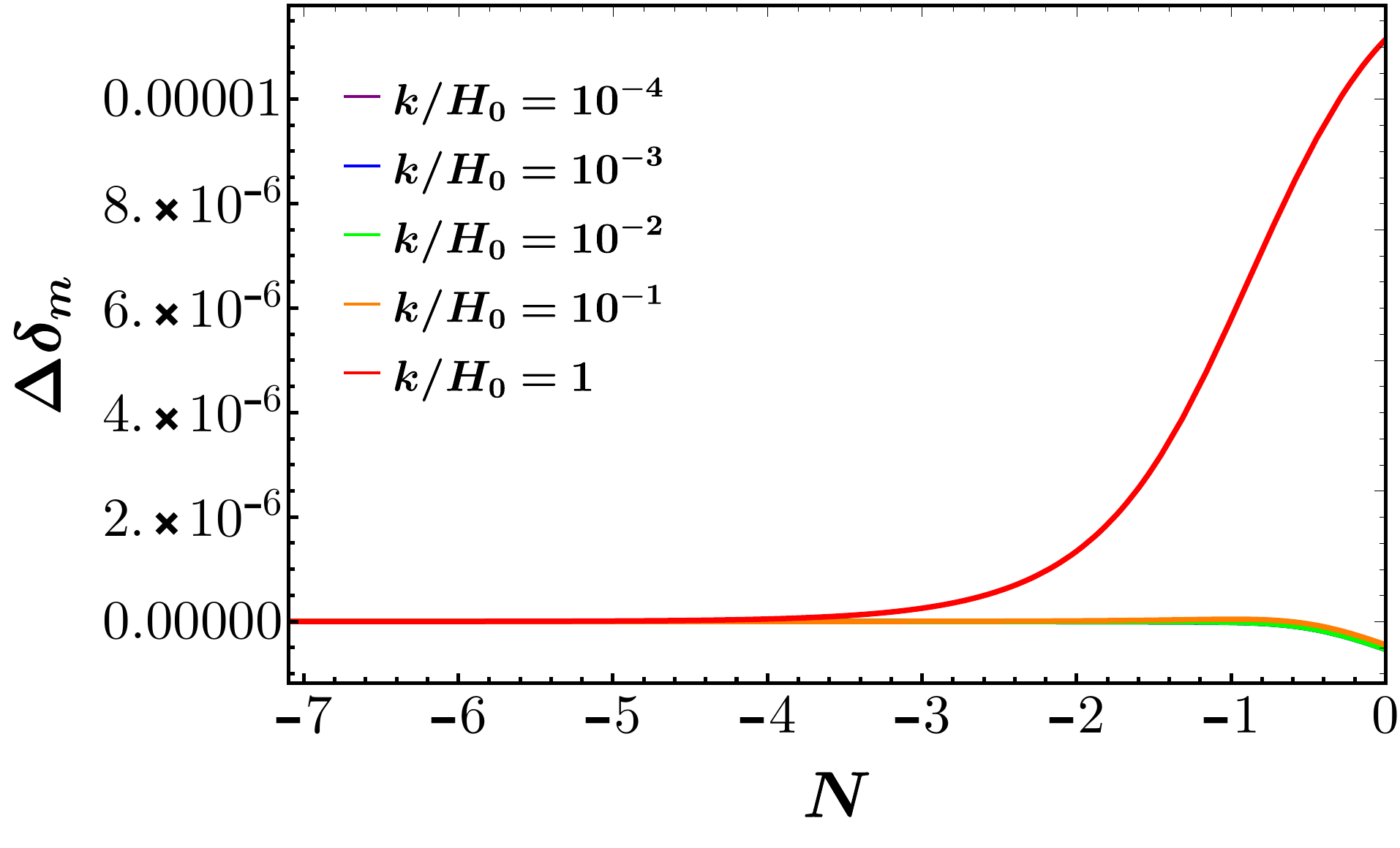}
\end{minipage}\hfill
\begin{minipage}[b]{.45\textwidth}
\includegraphics[scale=0.4]{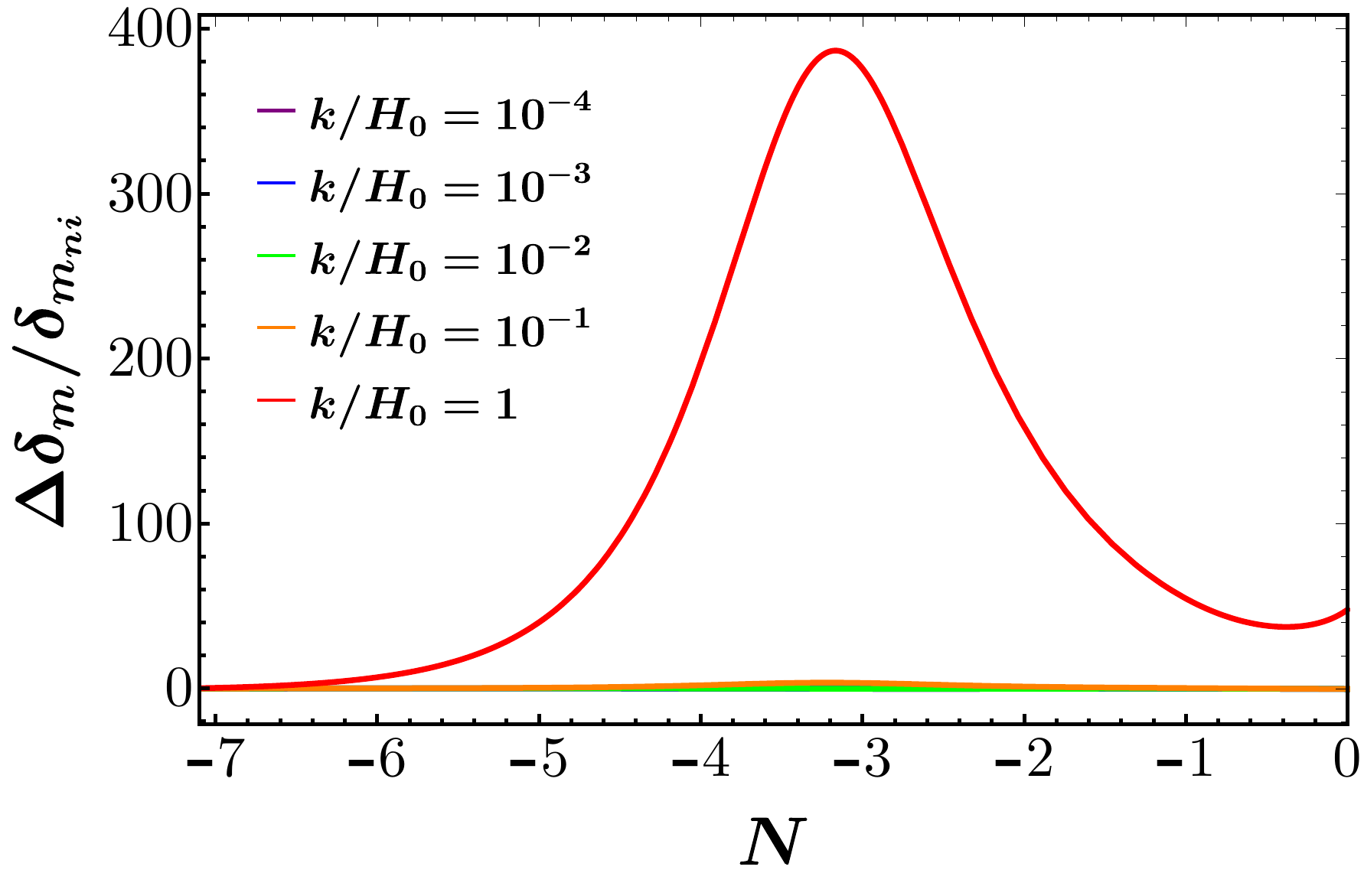}
\end{minipage}
\caption{Evolution of $\Delta \delta_m$ (left), $\Delta \delta_m/ \delta_{m_{ni}}$ (right)  as a function of $N$.}
\label{fig:ddeltaphi2}
\end{figure}

\begin{figure}[!htb]
\begin{minipage}[b]{.45\textwidth}
\includegraphics[scale=0.4]{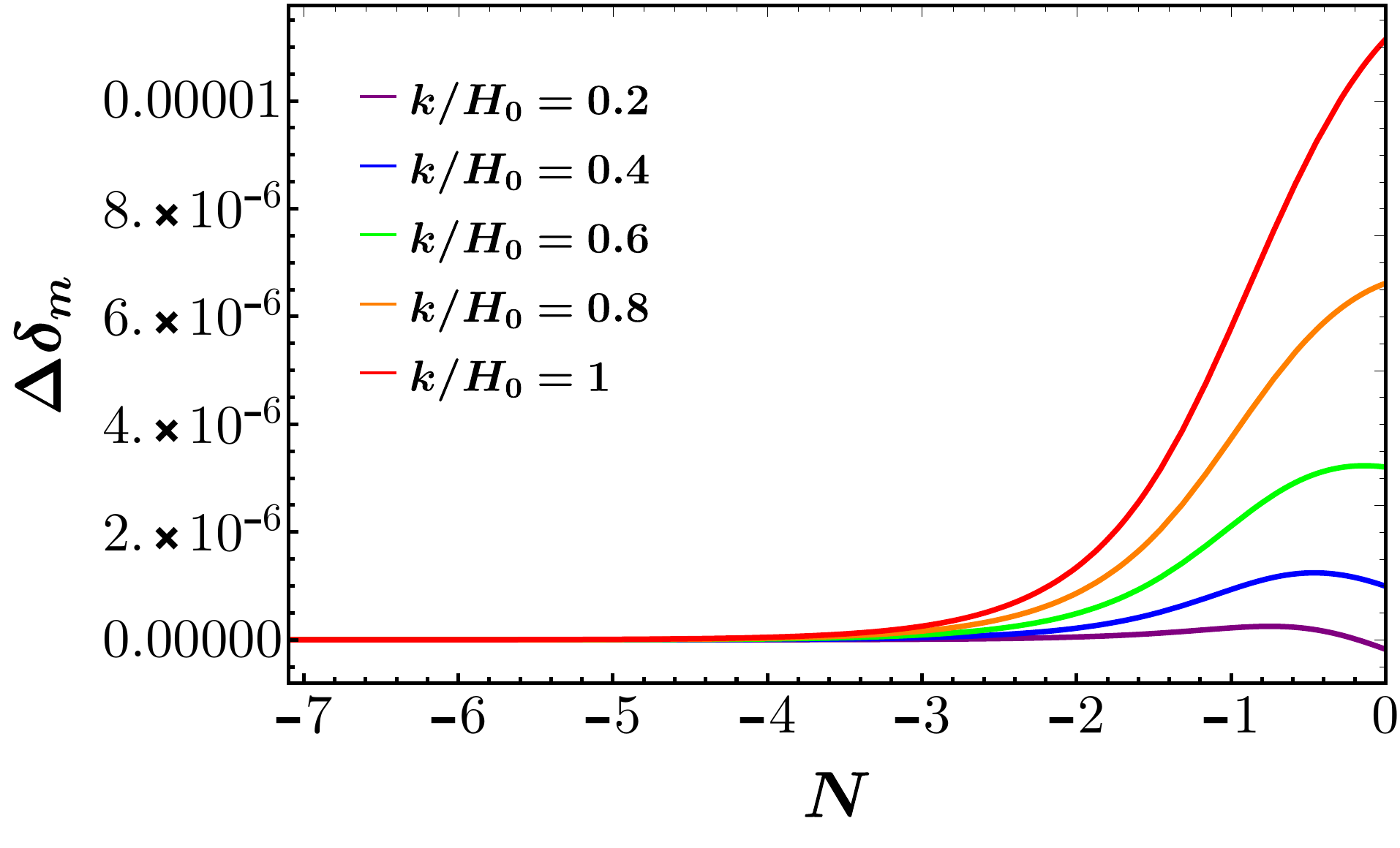}
\end{minipage}\hfill
\begin{minipage}[b]{.45\textwidth}
\includegraphics[scale=0.4]{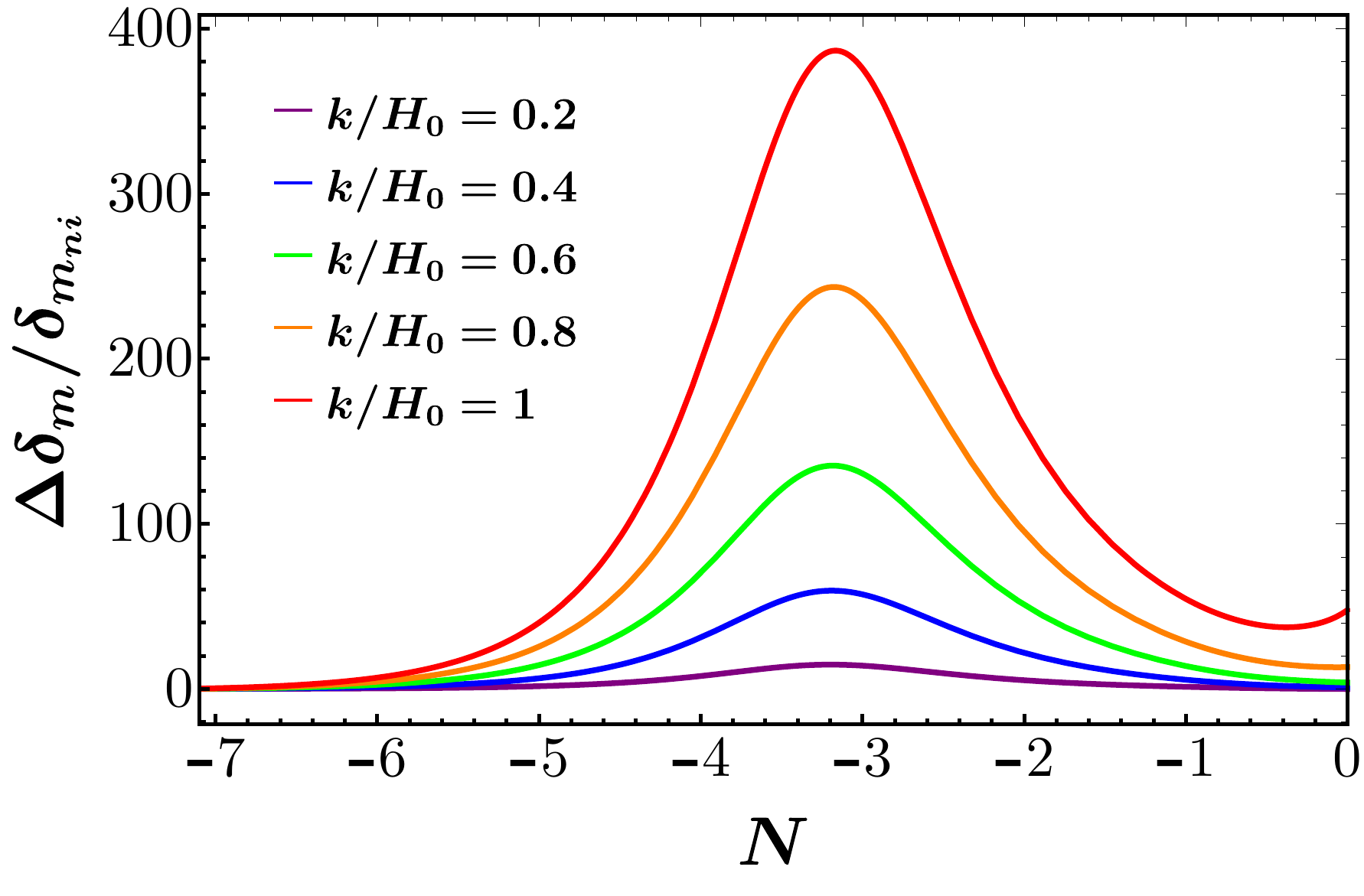}
\end{minipage}
\caption{Evolution of $\Delta \delta_m$ (left), $\Delta \delta_m/ \delta_{m_{ni}}$ (right)  as a function of $N$.}
\label{fig:ddeltazoomphi2}
\end{figure}

Figures \ref{fig:deltaevophi2} and  \ref{fig:deltaevozoomphi2} contain plots of $\delta_m$ as a function of $N$ for different length scales in interacting and non-interacting scenarios. Figures \ref{fig:ddeltaphi2}  and \ref{fig:ddeltazoomphi2} 
contain the plots of $\Delta \delta_m$ and $\Delta \delta_{m_{rel}}$ 
as a function of $N$ for different length scales, respectively.  Thus, 
we see that evolution of $\delta_m$ is roughly the same for $n=1$ and $n=2$.


\subsection{Weak gravitational lensing}

\begin{figure}[!htb]
\begin{minipage}[b]{.45\textwidth}
\includegraphics[scale=0.4]{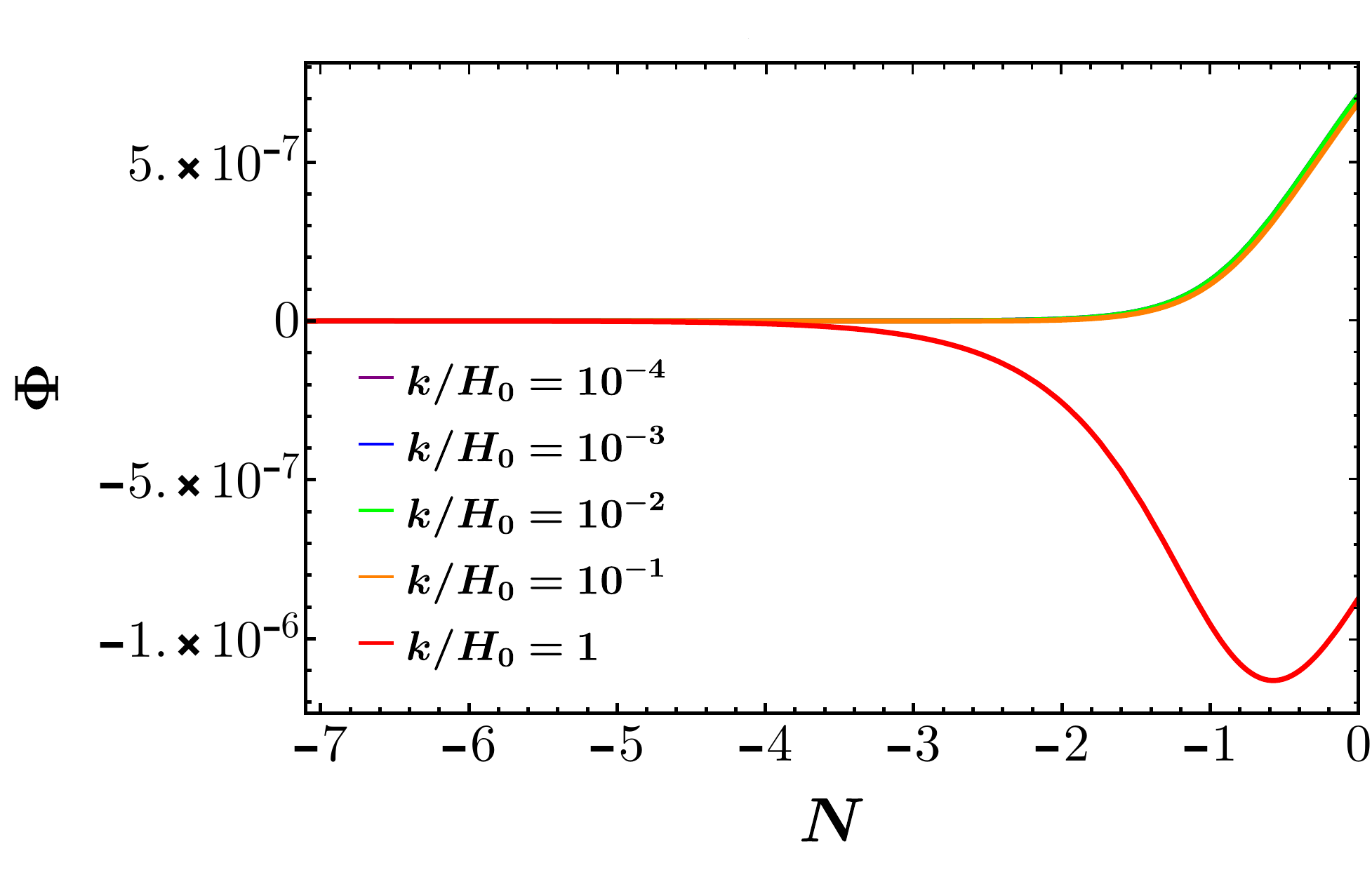}
\end{minipage}\hfill
\begin{minipage}[b]{.45\textwidth}
\includegraphics[scale=0.4]{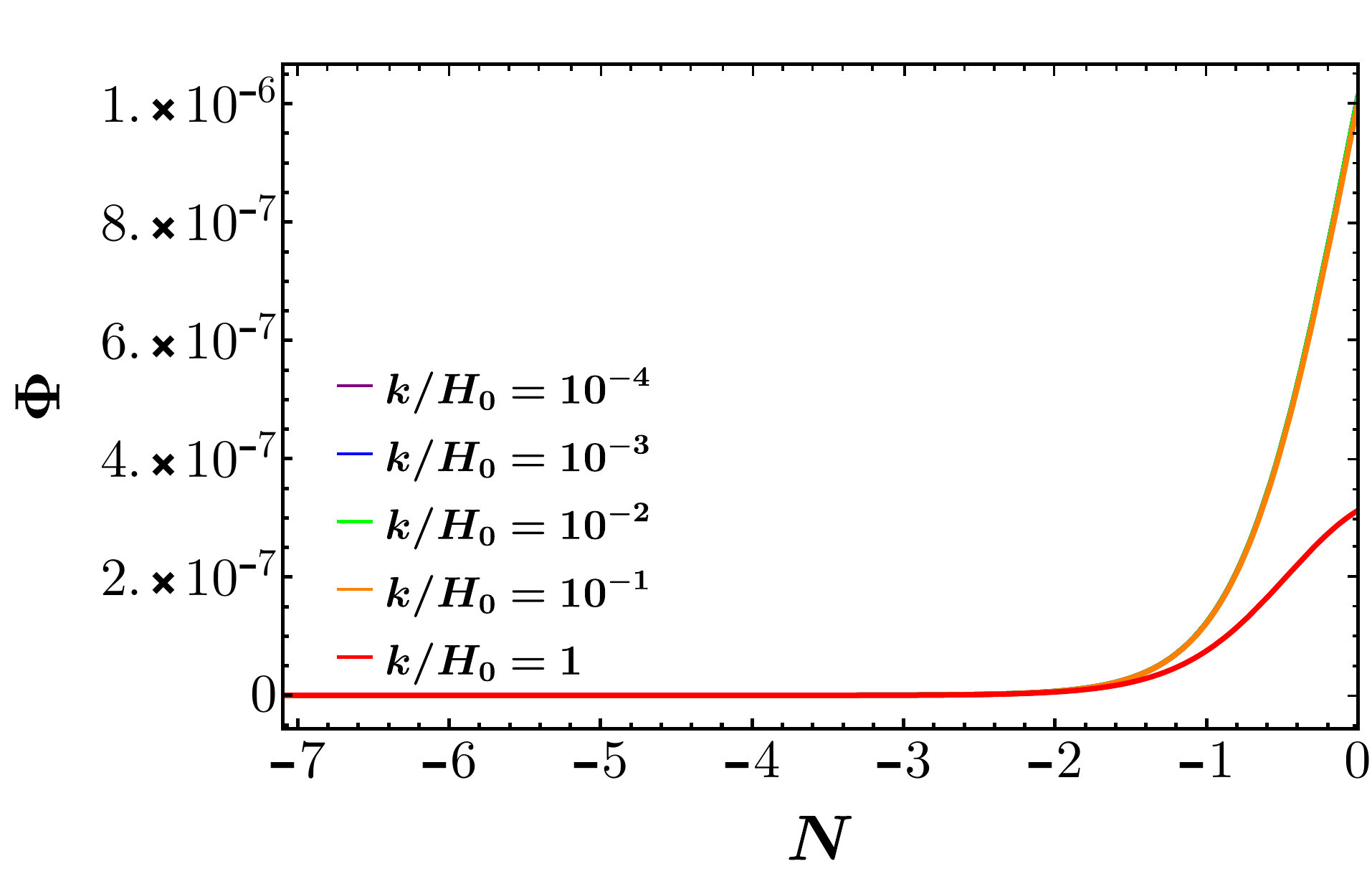}
\end{minipage}
\caption{Evolution of $\Phi$ as a function of $N$. Left: $C=-0.6$, Right: $C=0$.}
\label{fig:Phievophi2a}
\end{figure}

\begin{figure}[!htb]
\begin{minipage}[b]{.45\textwidth}
\includegraphics[scale=0.4]{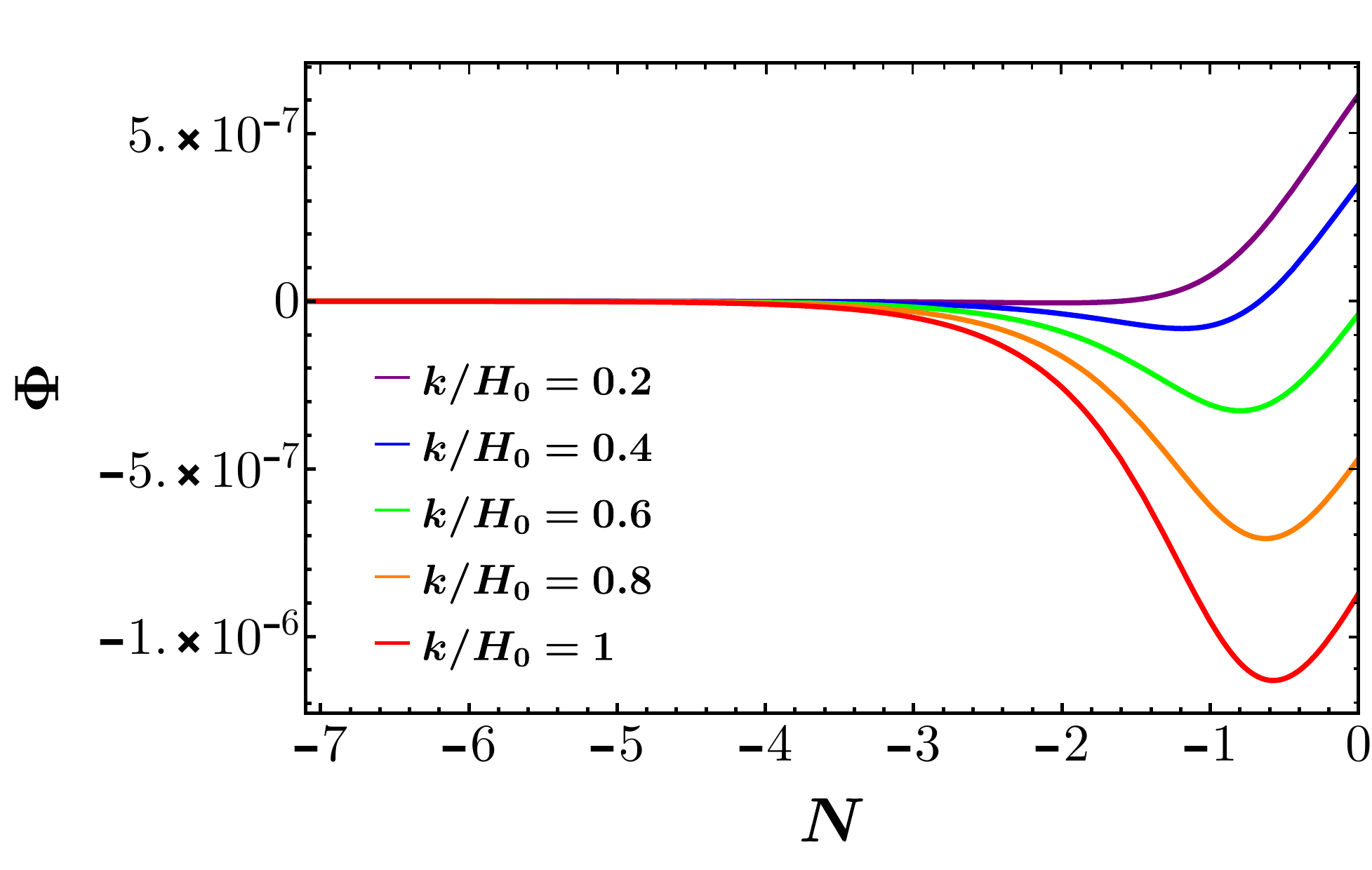}
\end{minipage}\hfill
\begin{minipage}[b]{.45\textwidth}
\includegraphics[scale=0.4]{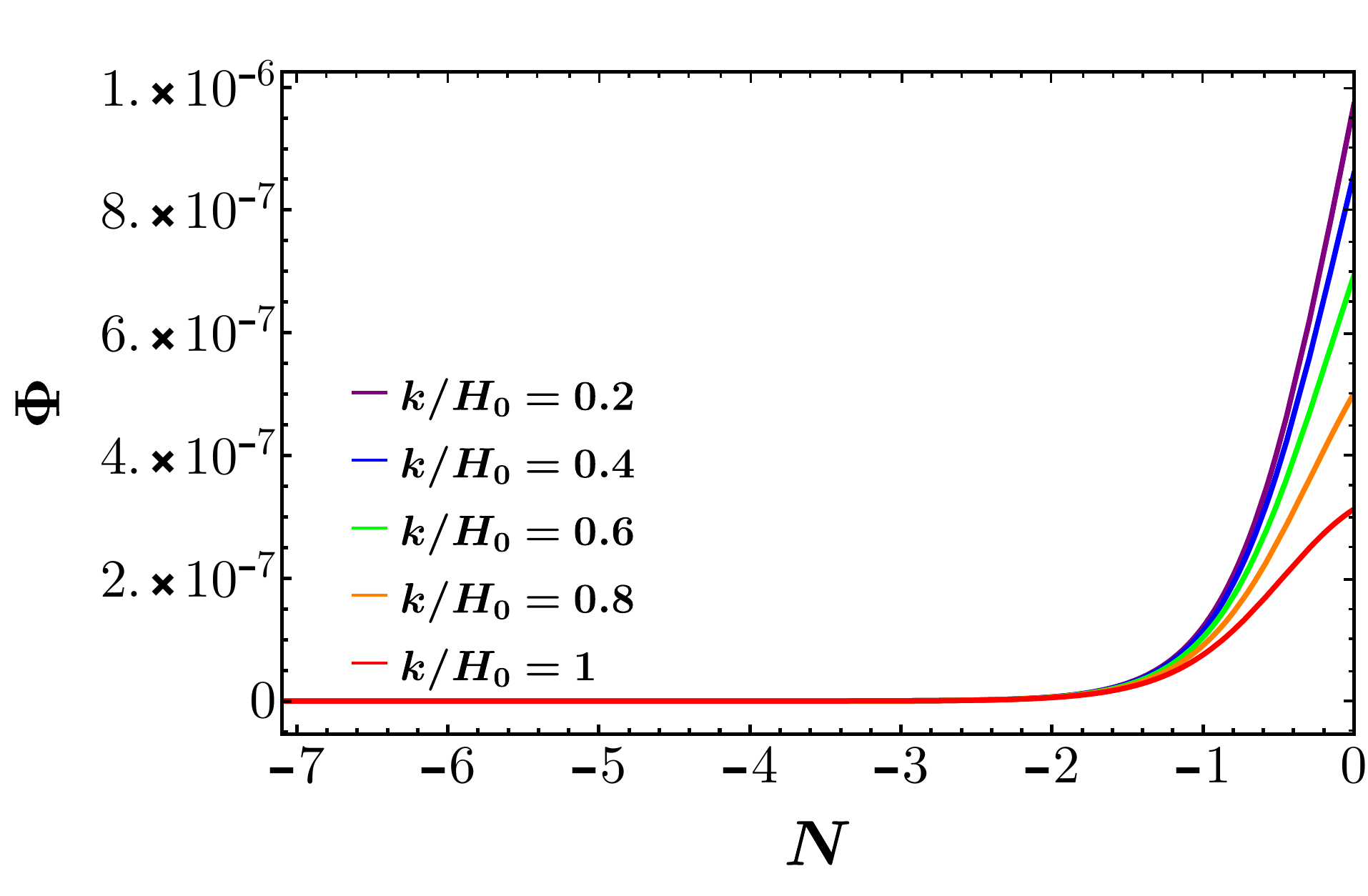}
\end{minipage}
\caption{Evolution of $\Phi$ as a function of $N$. Left: $C=-0.6$, Right: $C=0$.}
\label{fig:Phievophi2}
\end{figure}

\begin{figure}[!htb]
\begin{minipage}[b]{.45\textwidth}
\includegraphics[scale=0.4]{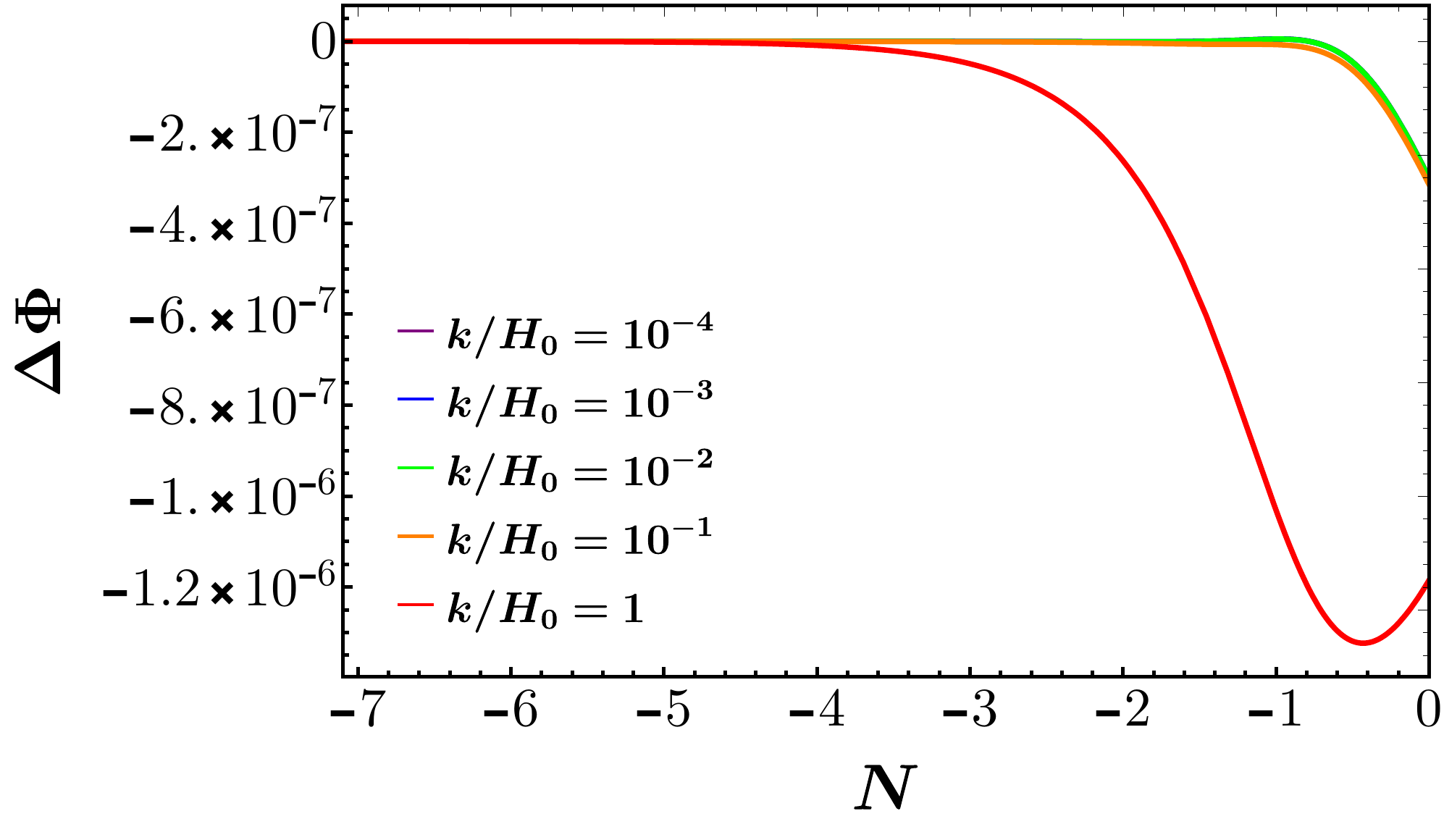}
\end{minipage}\hfill
\begin{minipage}[b]{.45\textwidth}
\includegraphics[scale=0.4]{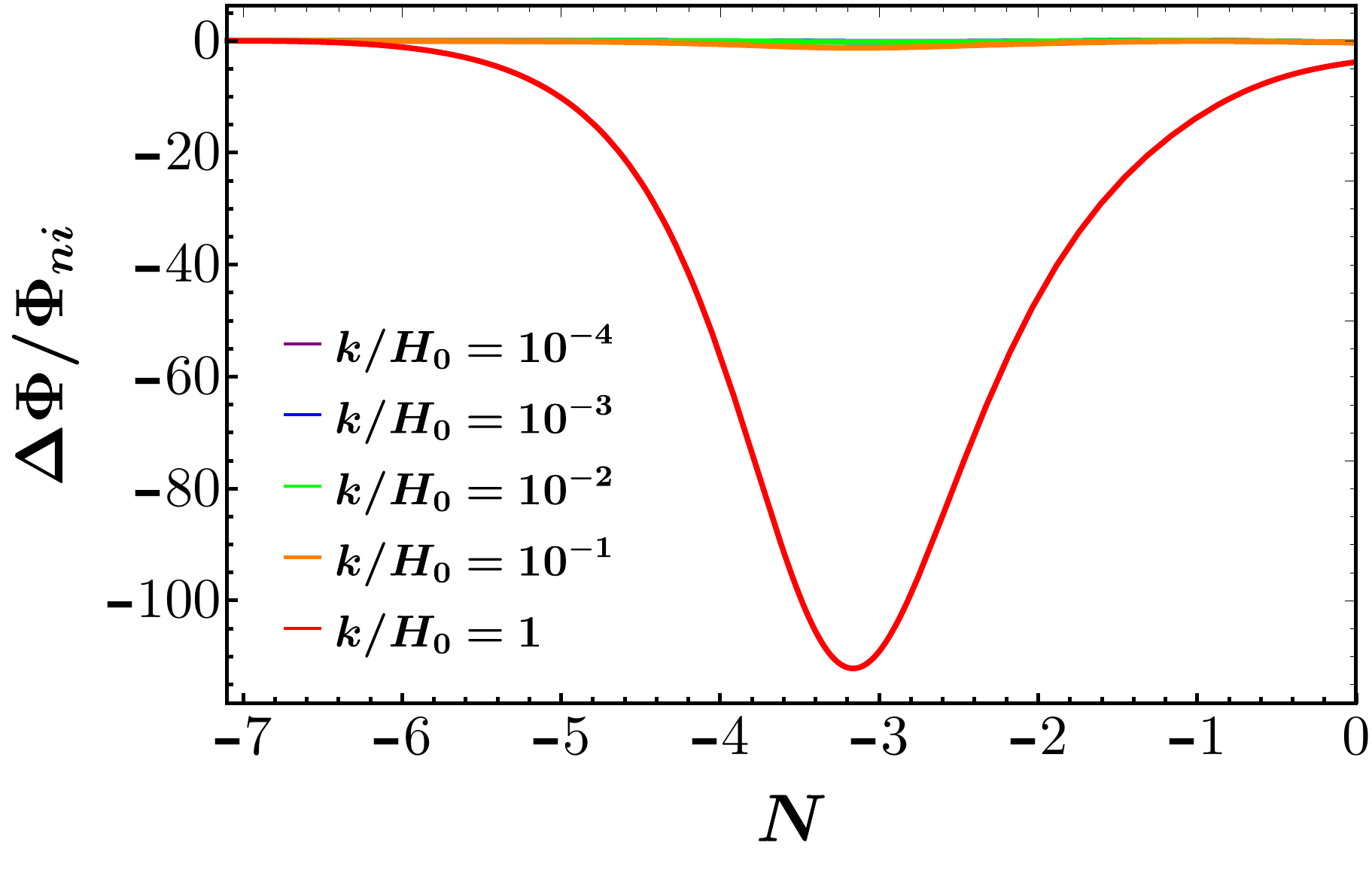}
\end{minipage}
\caption{Evolution of $\Delta \Phi$ (left), $\Delta \Phi/ \Phi_{ni}$ (right)  as a function of $N$.}
\label{fig:dlensphi2a}
\end{figure}

\begin{figure}[!htb]
\begin{minipage}[b]{.45\textwidth}
\includegraphics[scale=0.4]{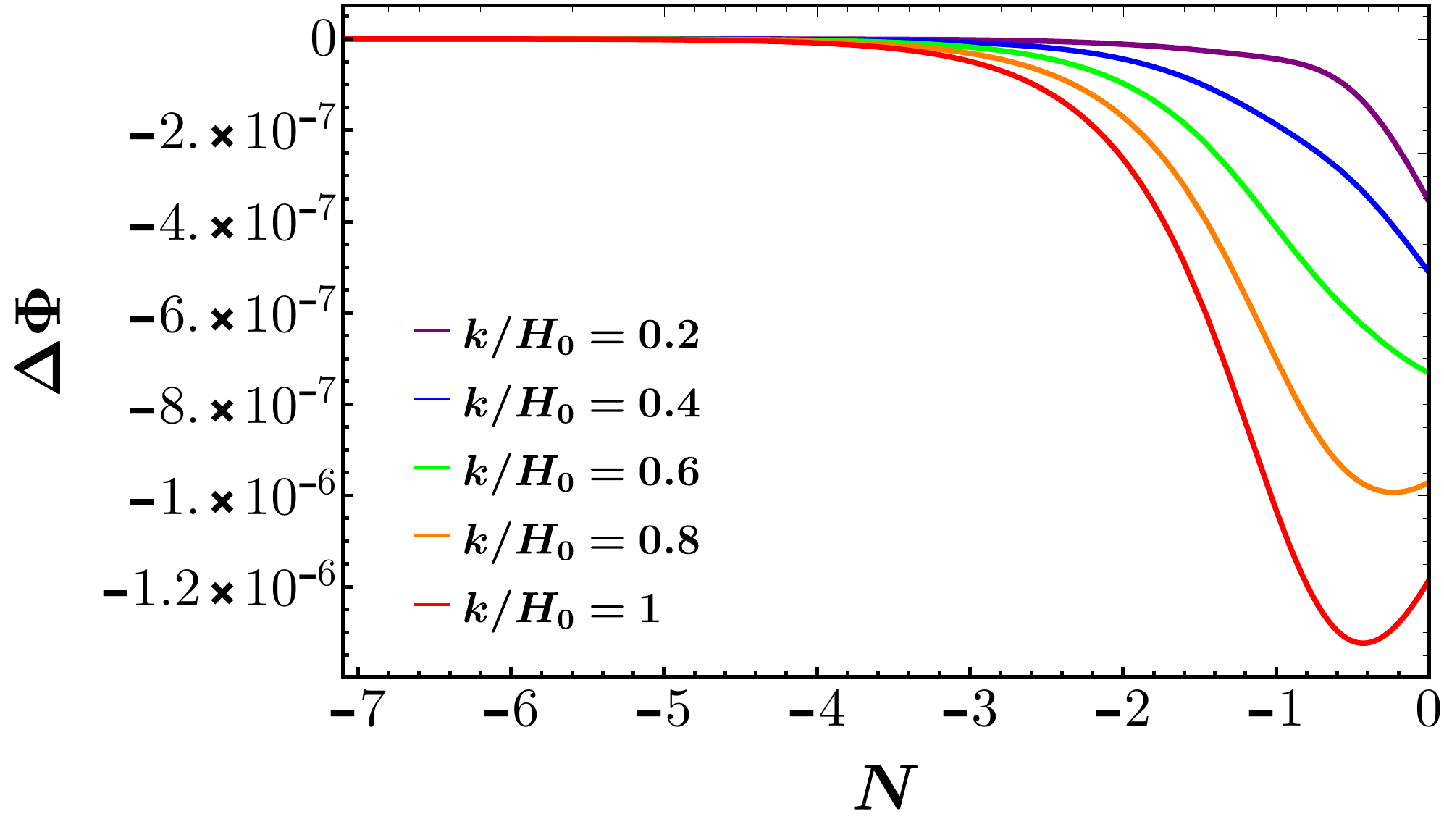}
\end{minipage}\hfill
\begin{minipage}[b]{.45\textwidth}
\includegraphics[scale=0.4]{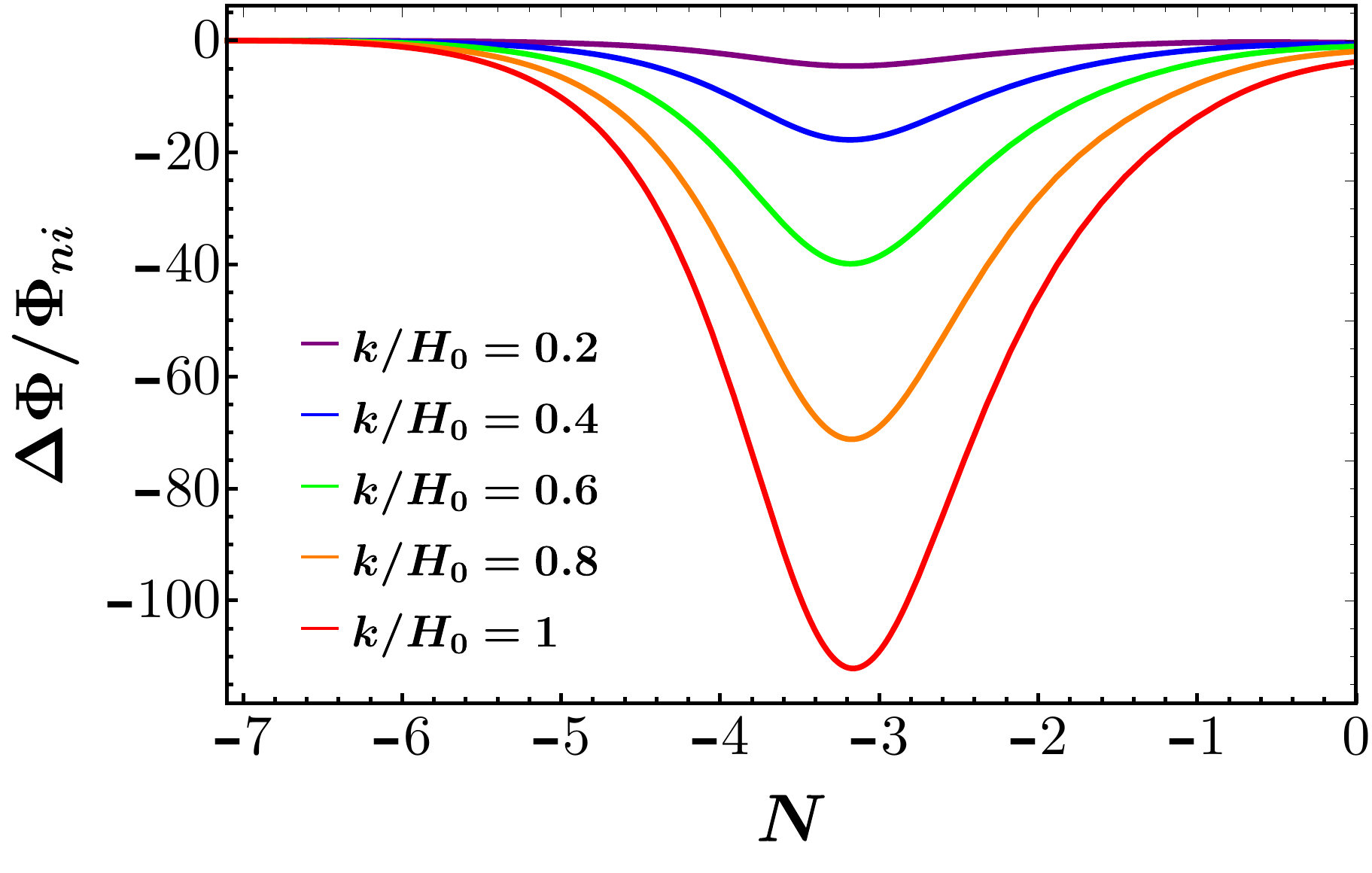}
\end{minipage}
\caption{Evolution of $\Delta \Phi$ (left), $\Delta \Phi/ \Phi_{ni}$ (right)  as a function of $N$.}
\label{fig:dlensphi2}
\end{figure}

Figures \ref{fig:Phievophi2a} and \ref{fig:Phievophi2}  contain plots of $\Phi$ as a function of $N$ for different length scales in interacting and non-interacting 
Figures \ref{fig:dlensphi2a}  and \ref{fig:dlensphi2} 
contain the plots of $\Delta \Phi$ and $\Delta \Phi/\Phi_{ni}$ 
as a function of $N$ for different length scales, respectively.  Thus, 
we see that evolution of $\Phi$ is roughly the same for the both the cases and is not sensitive to $n$.

\subsection{ISW effect}

\begin{figure}[!htb]
\begin{minipage}[b]{.45\textwidth}
\includegraphics[scale=0.4]{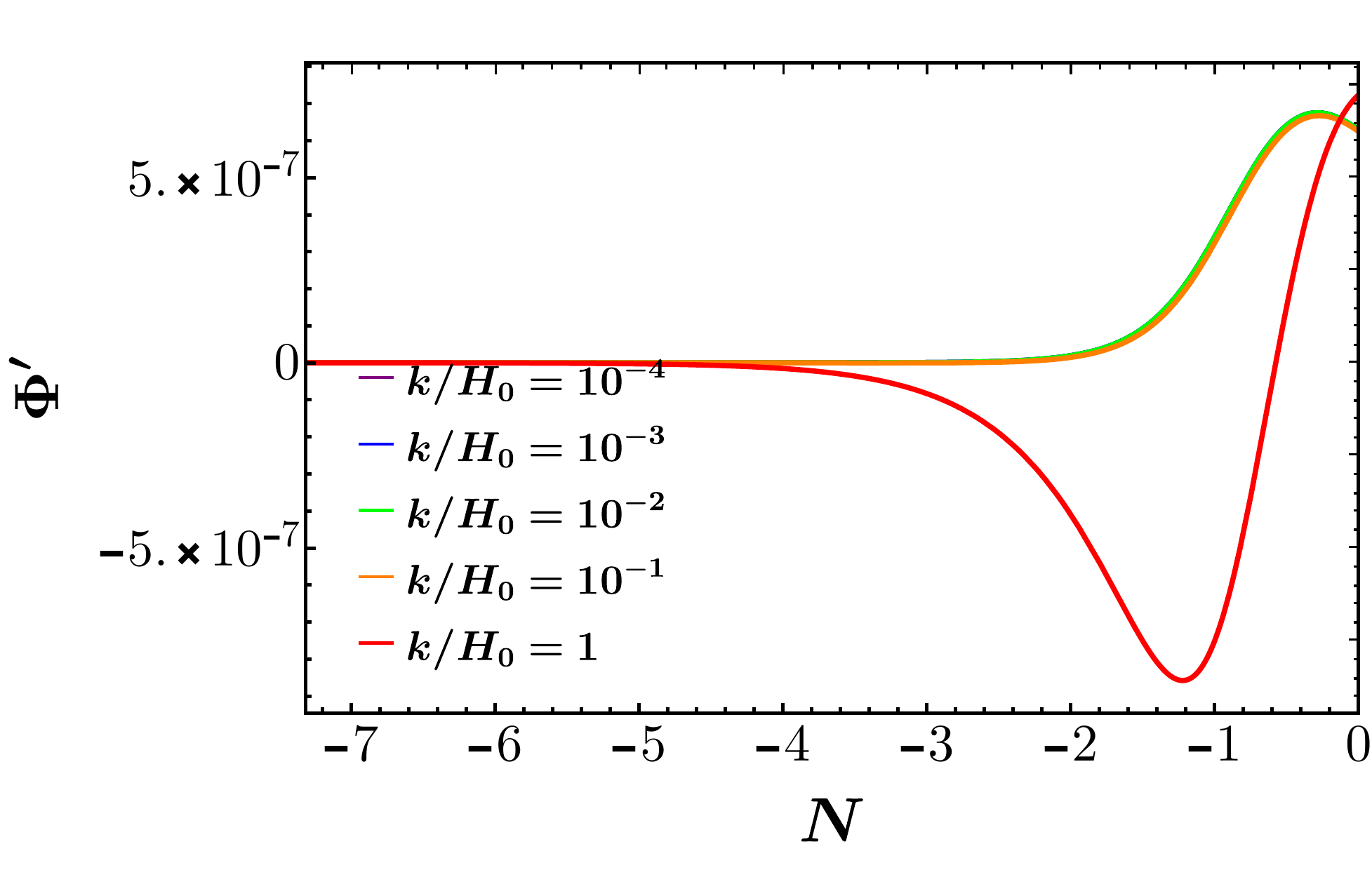}
\end{minipage}\hfill
\begin{minipage}[b]{.45\textwidth}
\includegraphics[scale=0.4]{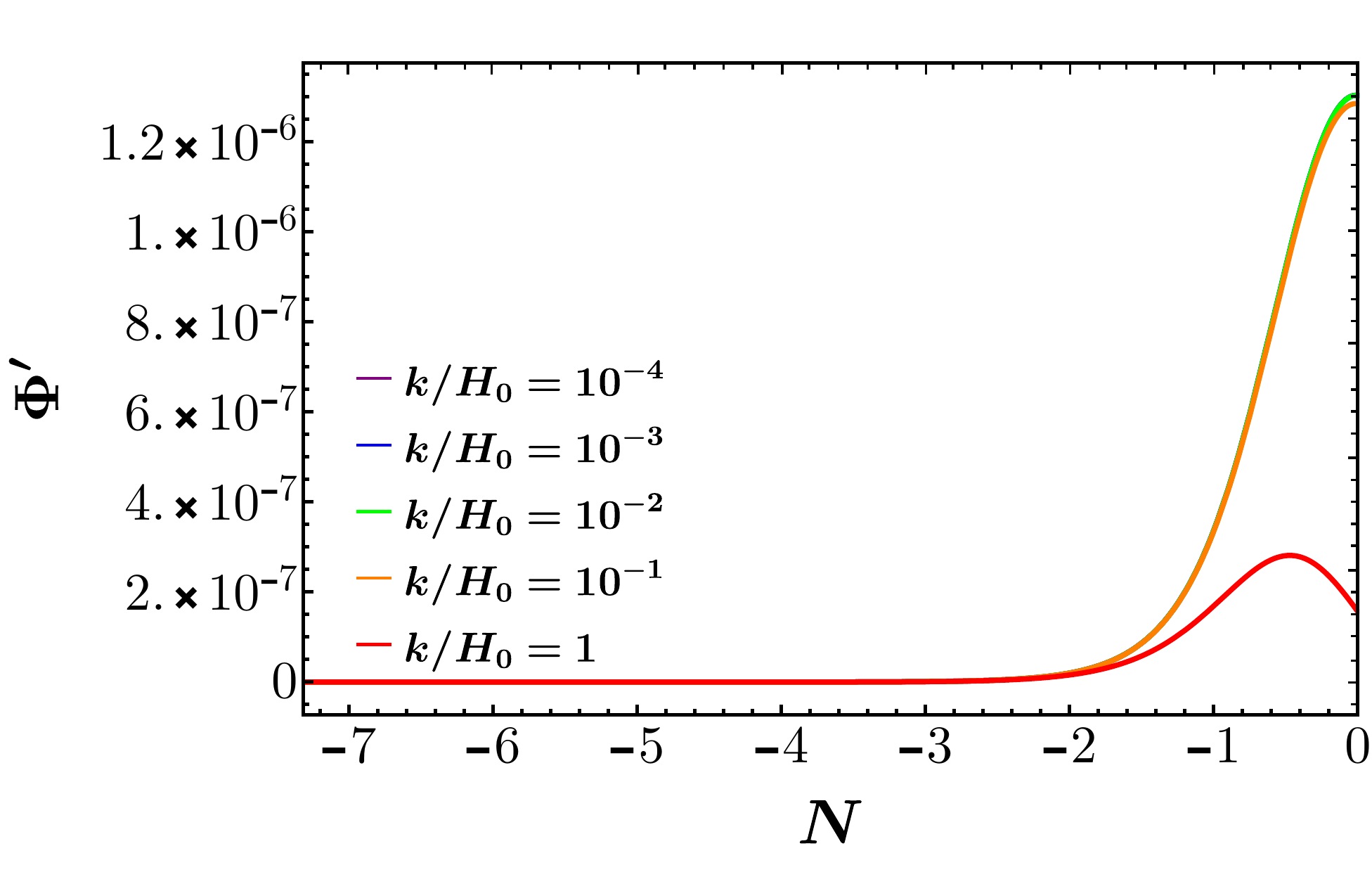}
\end{minipage}
\caption{Evolution of $\Phi'$ as a function of $N$. Left: $C=-0.6$, Right: $C=0$.}
\label{fig:iswevophi2}
\end{figure}

\begin{figure}[!htb]
\begin{minipage}[b]{.45\textwidth}
\includegraphics[scale=0.4]{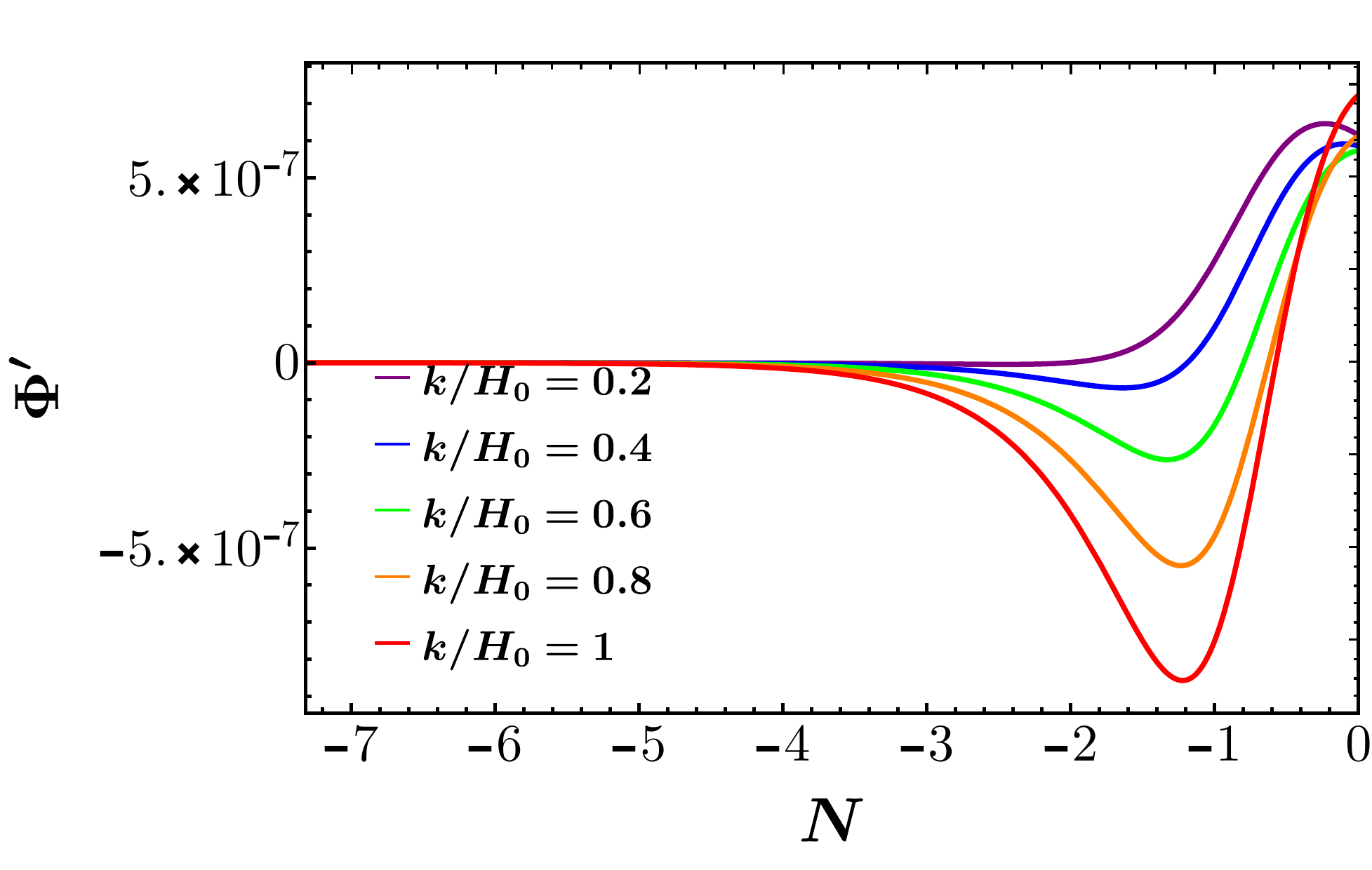}
\end{minipage}\hfill
\begin{minipage}[b]{.45\textwidth}
\includegraphics[scale=0.4]{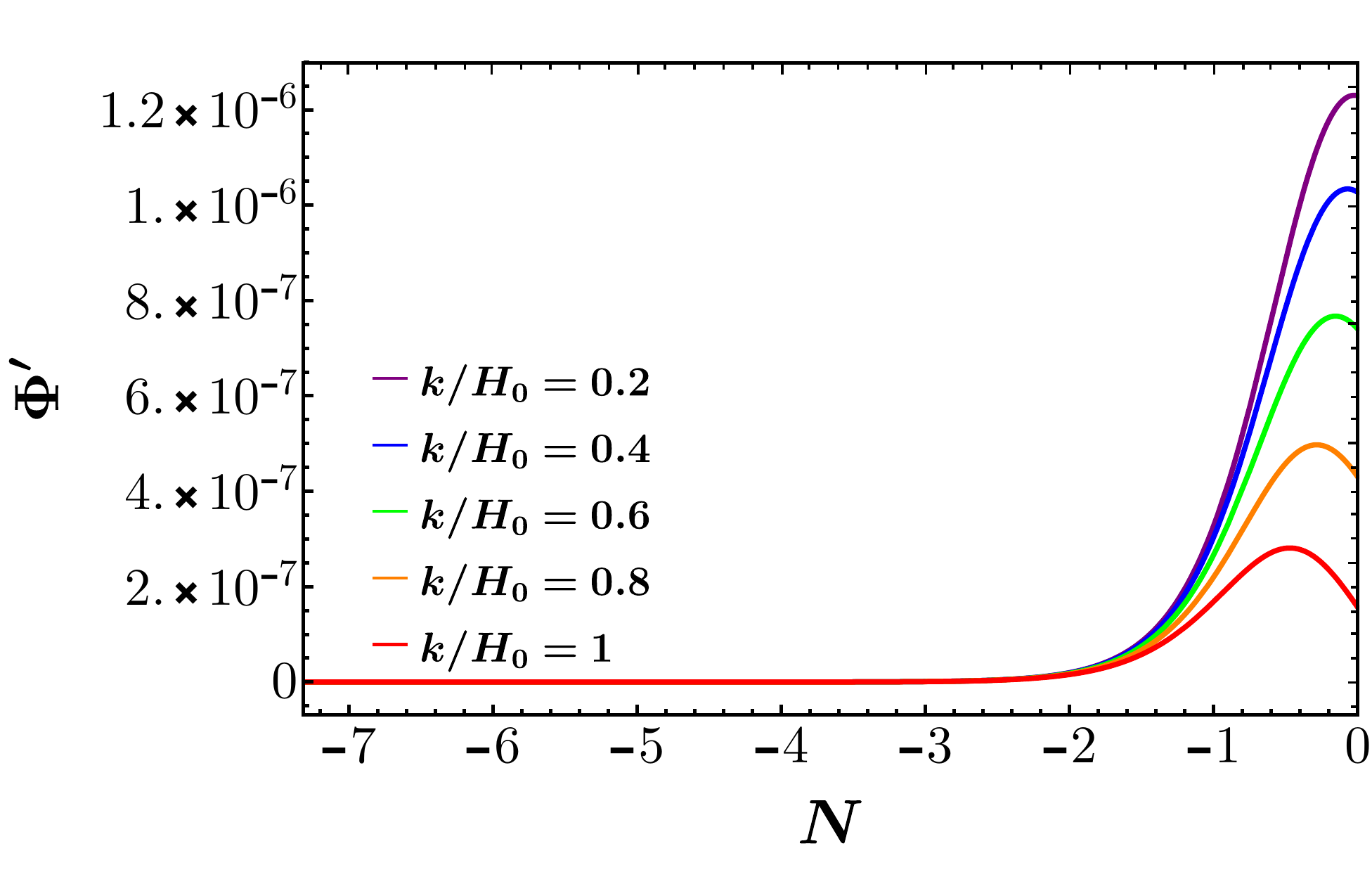}
\end{minipage}
\caption{Evolution of $\Phi'$ as a function of $N$. Left: $C=-0.6$, Right: $C=0$.}
\label{fig:iswevozoomphi2}
\end{figure}

\begin{figure}[!htb]
\begin{minipage}[b]{.45\textwidth}
\includegraphics[scale=0.4]{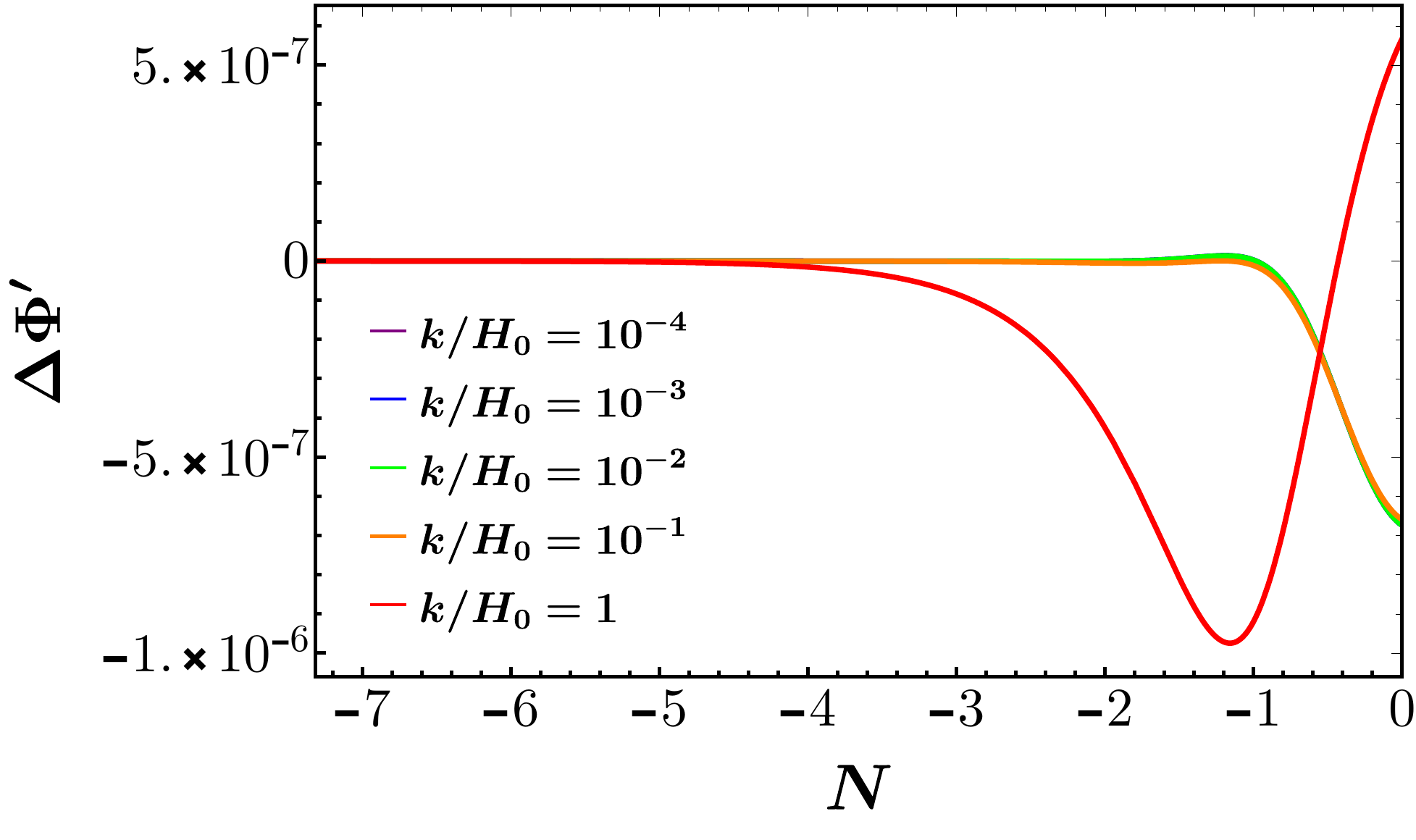}
\end{minipage}\hfill
\begin{minipage}[b]{.45\textwidth}
\includegraphics[scale=0.4]{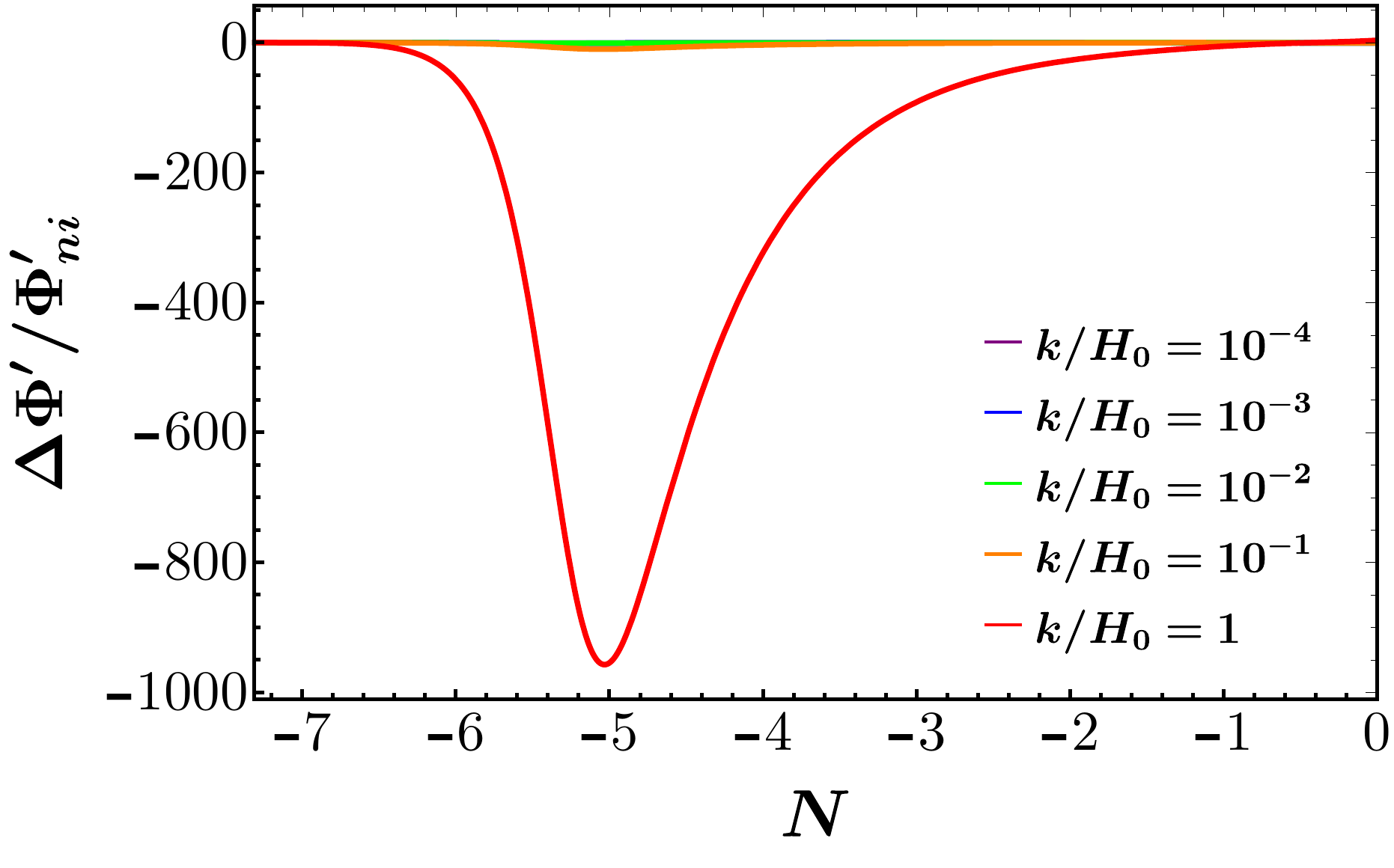}
\end{minipage}
\caption{Evolution of $\Delta \Phi'$ (left), $\Delta \Phi'/ \Phi'_{ni}$ (right)  as a function of $N$.}
\label{fig:diswphi2}
\end{figure}

\begin{figure}[!htb]
\begin{minipage}[b]{.45\textwidth}
\includegraphics[scale=0.4]{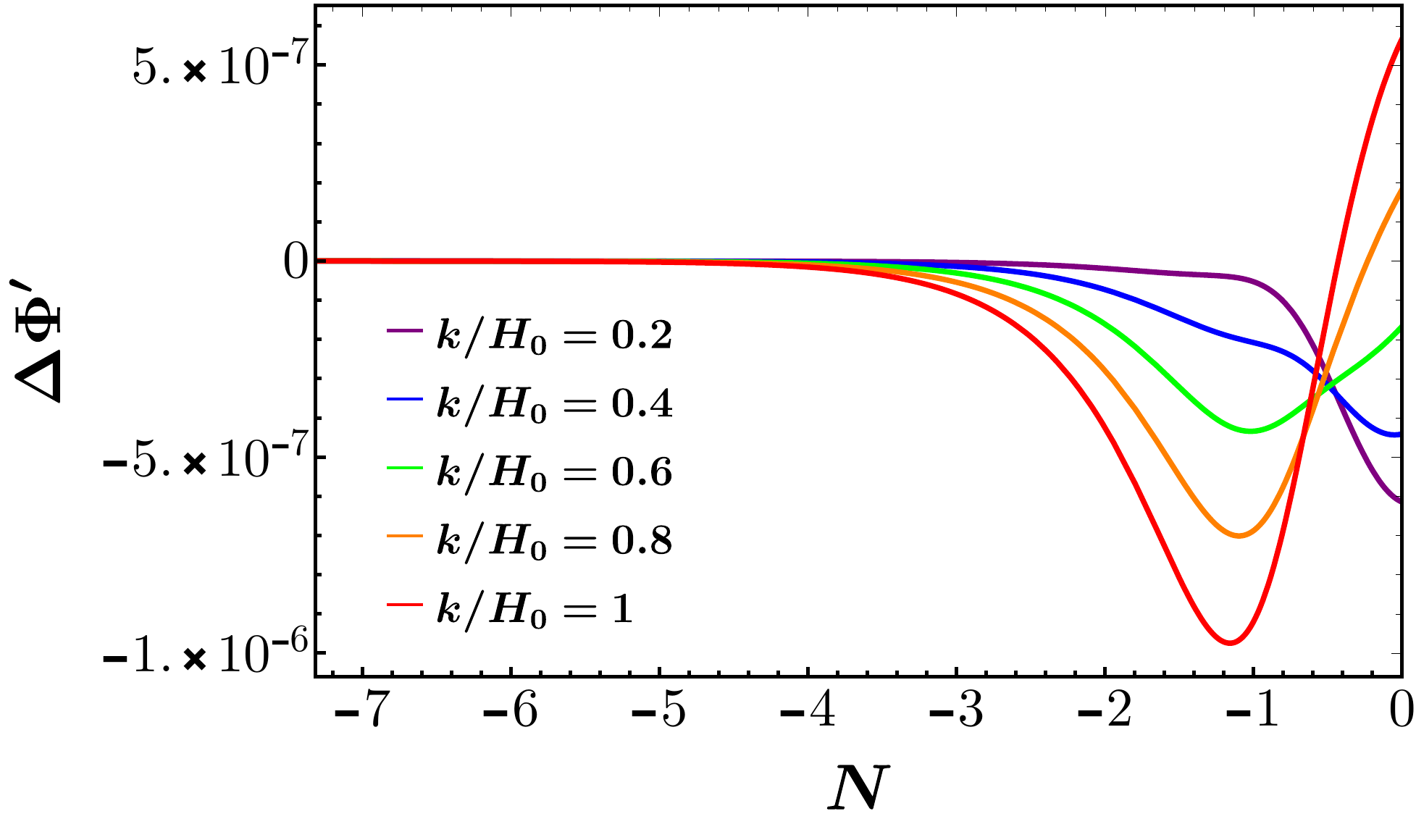}
\end{minipage}\hfill
\begin{minipage}[b]{.45\textwidth}
\includegraphics[scale=0.4]{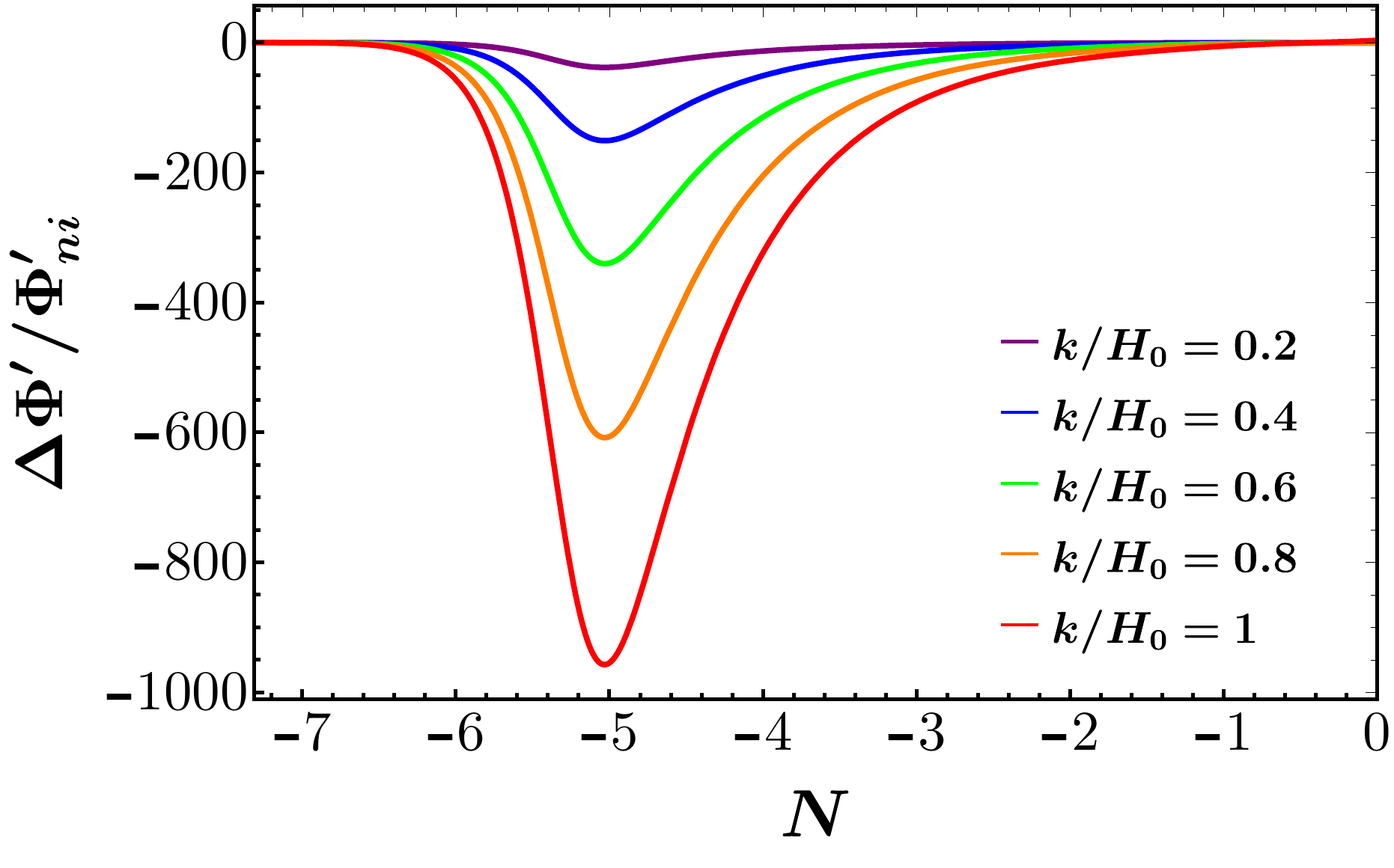}
\end{minipage}
\caption{Evolution of $\Delta \Phi'$ (left), $\Delta \Phi'/ \Phi'_{ni}$ (right)  as a function of $N$.}
\label{fig:diswzoomphi2}
\end{figure}

Figures \ref{fig:iswevophi2} and \ref{fig:iswevozoomphi2}  contain plots of $\Phi'$ as a function of $N$ for different length scales in interacting and non-interacting 
Figures \ref{fig:diswphi2}  and \ref{fig:diswzoomphi2} 
contain the plots of $\Delta \Phi'$ and $\Delta \Phi'/\Phi'_{ni}$ 
as a function of $N$ for different length scales, respectively.  Thus, 
we see that evolution of $\Phi'$ is roughly the same for the both the cases and is not sensitive to $n$.



We thus conclude that the evolution of $\delta_m$, $\Phi$, and $\Phi'$ for the inverse square potential follow a similar trend as compared to the $U(\phi) \sim 1/\phi$ case. The difference in the evolution becomes significant for $z < 20$, for all length scales. This means that cosmological observations related to the formation of large-scale structures can potentially detect the signatures of dark matter - dark energy interaction.

\section{Sound speed of the scalar field}
\label{app:Sound}

Sound speed  and adiabatic sound speed of the dark energy scalar field $(\phi)$ 
is given by ~\cite{1985-Brandenberger-Rev.Mod.Phys.}
\begin{equation}
c_s^2 = \dfrac{\delta p_{\phi}}{\delta \rho_{\phi}}, \quad c_{s_{ad}}^2 = \dfrac{\dot{\bar{p_{\phi}}}}{\dot{\bar{\rho_{\phi}}}}=-1-\dfrac{2\ddot{\bar{\phi}}}{3H\dot{\bar{\phi}}+2\alpha_{\phi}\bar{\rho}_m M_{Pl}^2}
\end{equation}
In terms of the dimensionless variables, these quantities can be expressed as
\begin{equation}
c_{s_{ad}}^2=\dfrac{h'(x^2-y^2) + h(xx' - yy')}{h'(x^2 + y^2) + h(xx' + yy')} = -1-\dfrac{2 \left(x' + \frac{h'}{h}\right)}{\sqrt{3}\left(\sqrt{12}x - \sqrt{2}\alpha \beta \Omega_m \right)}
\end{equation}
\begin{equation}
c_{s}^2=\dfrac{12 \Phi x^2 - 2\sqrt{3}x\delta \phi' - 3\sqrt{2}\lambda y^2 \delta \phi}{12 \Phi x^2 - 2\sqrt{3}x\delta\phi' + 3\sqrt{2}\lambda y^2 \delta \phi}
\end{equation}

For a quintessence model, $c_s^2=1$ in the rest frame of $\phi$~\cite{2008-Valiviita.etal-JCAP}. In this work, the perturbed quantities are evaluated in the dark matter rest frame. 



\end{document}